\documentstyle[epsfig,12pt]{article}
\newlength{\dinwidth}
\newlength{\dinmargin}
\newcommand{\optbar}[1]{\shortstack{{\tiny (\rule[.4ex]{1em}{.1mm})}
  \\ [-.7ex] $#1$}}
\catcode`@=11

\def\@citex[#1]#2{\if@filesw\immediate\write\@auxout{\string\citation{#2}}\fi
  \def\@citea{}\@cite{\@for\@citeb:=#2\do
    {\@citea\def\@citea{,\penalty\@m}\@ifundefined
      {b@\@citeb}{{\bf ?}\@warning
       {Citation `\@citeb' on page \thepage \space undefined}}%
\hbox{\csname b@\@citeb\endcsname}}}{#1}}

\def\citer{\@ifnextchar [{\@tempswatrue\@citexr}{\@tempswafalse\@citexr[]}}

\def\@citexr[#1]#2{\if@filesw\immediate\write\@auxout{\string\citation{#2}}\fi
  \def\@citea{}\@cite{\@for\@citeb:=#2\do
    {\@citea\def\@citea{--\penalty\@m}\@ifundefined
       {b@\@citeb}{{\bf ?}\@warning
       {Citation `\@citeb' on page \thepage \space undefined}}%
\hbox{\csname b@\@citeb\endcsname}}}{#1}}
\catcode`@=12

\begin{document}
\begin{flushright}  
DESY 98-056\\
hep-ph/9805yyy\\
May 1998
\end{flushright}

\begin{center}
{\large\bf  CP-Violating Asymmetries in Charmless 
Non-Leptonic Decays $B \to PP$, $PV$, $VV$ in the Factorization Approach}
\end{center}
\vspace*{1.0cm}
\centerline{\large\bf A.~Ali}
\centerline{\bf Deutsches Elektronen-Synchrotron DESY, 22607 Hamburg, 
Germany}
\vspace*{0.3cm}
\centerline{\large\bf G.~Kramer\footnote{Supported
by Bundesministerium f\"ur Bildung und
Forschung, Bonn, under Contract 057HH92P(0) and EEC Program ``Human Capital
 and Mobility'' through Network ``Physics at High Energy Colliders'' under
Contract CHRX-CT93-0357 (DG12COMA).} and 
 Cai-Dian L\"u\footnote{Alexander von Humboldt Foundation Research Fellow.}}
\centerline{\bf II. Institut f\"ur Theoretische Physik, Universit\"at 
Hamburg, 22761 Hamburg, Germany}
\date{}

\vspace*{1.0cm}
\centerline{\Large\bf Abstract}
\vspace*{1cm}

   We present estimates of the direct (in decay amplitudes) and indirect 
(mixing-induced) CP-violating asymmetries in the
 non-leptonic charmless two-body
decay rates for $B \to PP$, $B \to PV$ and $B \to VV$ decays and
their charged conjugates,
where $P(V)$ is a light
pseudoscalar (vector) meson. These estimates are based on a generalized 
factorization approach making use of next-to-leading order
perturbative QCD contributions
which generate the required strong phases. No soft final state interactions
are included. We study the dependence of the
asymmetries on a number of input parameters and show that there are at least
two (possibly three) classes of decays in which the asymmetries
are parametrically stable in this approach. The decay modes of particular
interest are:
$\optbar{B^0} \to \pi^+ \pi^-$, 
$\optbar{B^0} \to K_S^0 \pi^0$, $\optbar{B^0}  \to K_S^0 \eta^\prime$, 
$\optbar{B^0} \to K_S^0 \eta$ and $\optbar{B^0}  \to \rho^+ \rho^-$.
Likewise, the CP-violating asymmetry in the decays $\optbar{B^0} \to K_S^0
h^0$ with
$h^0=\pi^0,K_S^0,\eta,\eta^\prime$
is found to be parametrically stable and large. Measurements of
these asymmetries will lead to a determination of the phases $\sin 2
\alpha$ and $\sin 2 \beta$ and we work out the relationships 
in these modes in the
present theoretical framework. We also show the extent of 
the so-called ``penguin
pollution'' in the rate asymmetry $A_{CP}(\pi^+ \pi^-)$ and of the ``tree
shadow'' in the asymmetry $A_{CP}(K_S^0\eta^\prime)$ which will effect
the determination of  
$\sin 2 \alpha$ and $\sin 2 \beta$ from the respective  measurements. 
CP-violating asymmetries in $B^\pm$ decays depend on a model parameter
in the penguin-amplitudes and theoretical predictions require 
further experimental or theoretical input.
Of these,  CP-violating asymmetries in $B^\pm \to \pi^\pm \eta^\prime$,
 $B^\pm \to K^{*\pm} \eta$, $B^\pm \to K^{*\pm} \eta^\prime$ and $B^\pm \to
K^{*\pm}\rho^0$ are potentially interesting and are studied here.

\vspace*{0.70cm}
\centerline{(Submitted to Physical Review D)}

\newpage

\section{Introduction}

Recent measurements by the CLEO Collaboration \cite{cleo,cleobok} of a 
number 
of decays  $B\to h_1h_2$, where $h_1$ and $h_2$ are light hadrons 
such as $h_1h_2= \pi\pi, \pi K, \eta^\prime K, \omega K$, 
have lead to renewed theoretical 
interest in understanding hadronic $B$ decays \cite{lss98}.

In a recent work \cite{akl98-1} we have calculated the branching fractions of
two-body non-leptonic decays  $B \to PP$, $PV$, $VV$, where $P$ and $V$ 
are the lowest lying light pseudoscalar and vector mesons, respectively. 
The theoretical framework used was based on the next--to-leading logarithmic 
improved effective Hamiltonian and a factorization Ansatz for the 
hadronic matrix elements of the four-quark operators \cite{ag}. 
We worked out the parametric dependence of the decay rates using currently 
available information on the weak mixing matrix elements, form factors, 
decay constants and quark masses.
In total we considered seventy six decay channels with a large fraction of 
them having branching ratios of order $10^{-6}$ or higher which hopefully 
will be measured in the next round of experiments on $B$ decays. 
 The recently measured decay modes $B^0 \to K^+\pi^-$, 
$B^+ \to K^+ \eta^\prime$, $B^0 \to K^0 \eta^\prime$, $B^+ \to \pi^+ K^0 
$ and $B^+ \to \omega K^+  $ are shown to be largely in agreement with the 
estimates based on factorization \citer{akl98-1,acgk}.
This encourages us to further pursue this framework and calculate
quantities of experimental interest in two-body non-leptonic $B$ decays. 

Besides branching fractions, other observables which will help to
test the factorization approach and give information on the 
Cabibbo-Kobayashi-Maskawa (CKM) matrix \cite{CKM} are CP-violating rate
asymmetries in partial decay rates. 
In the past a large variety of ways has been proposed to observe CP 
violation in $B$ decays \cite{cpbook}.
One method is to study CP-violating  asymmetries 
in the time-dependence of the neutral $B$ meson decay rates in specific
modes, which involve an interference between two weak amplitudes. 
Asymmetries in charged $B$ decays require an interference between two 
amplitudes involving both a CKM phase  and a final state strong
interaction phase-difference. 
Such asymmetries occur also in decays of neutral $B$ mesons in which 
$B^0$ and $\bar B^0$ do not decay into common final states or where
these states are not CP-eigenstates. 
In these decays the weak phase difference arises from the superposition of 
various penguin contributions and the usual tree diagrams in case they
are present.
The strong-phase differences 
arise through the absorptive parts of perturbative penguin diagrams (hard 
final state interaction) \cite{SEW} and  non-perturbatively 
(soft final state interaction).

When a $B^0$ and $\bar B^0$ decay to a common 
final state $f$, $B^0$-$\bar B^0$ mixing plays a crucial role in determining
the CP-violating asymmetries, requiring time-dependent measurements. 
For the final states which are both CP-eigenstates and involve only one 
weak phase in the decays, the CP-violating asymmetry is independent of the
hadronic matrix elements. 
This occurs in the well studied $\optbar{B^0} \to J/ \psi K_S$ decays
 making it possible to extract
the value of $\sin 2 \beta$ with no hadronic uncertainties. 
For neutral $B$ decays into two light mesons such a direct translation of 
the CP-violating asymmetries in terms of CP-violating phases $\alpha$,
$\beta$ and 
$\gamma$ will not be possible, in general.
Hence, the predicted asymmetries are subject to hadronic uncertainties.
In principle, these uncertainties can be removed by resorting to a set
of time-dependent and time-independent measurements as suggested in the
literature \citer{GL90,DH94}. In practice, this program requires a number
of difficult measurements. We pursue here the other alternative, namely
we estimate these uncertainties in a specific model, which can be tested
experimentally in a variety of decay modes.  

 CP-violating asymmetries are expected in a large number of $B$
decays; in particular the partial rate asymmetries in all the
$B \to h_1 h_2$ decay modes and their charge conjugates studied in \cite{akl98-1} 
are potentially interesting for studying CP violation.
 We recall that CP-violating asymmetries in $B\to h_1 h_2$ decays
have been studied earlier in
the factorization framework \cite{cpbook},\citer{sw,dg97}.
With the measurement of some of the $B\to h_1 h_2$ decays
\cite{cleo,cleobok}, some selected modes have received renewed interest
in this approach \citer{dds97-1,Petrov97}. These papers, however,  make
specific assumptions about $\xi\equiv1/N_c$ (here $N_c$ is the number of
effective colors)  and certain other input parameters;
in particular, the earlier ones used CKM-parameter values which are now
strongly disfavored by recent unitarity fits \cite{aliapctp97,Parodi98}
and/or they do not include the anomaly contributions (or not quite
correctly) and the latter ones make specific
assumptions about $\xi$, which may or may not be consistent with data on
$B \to h_1 h_2$ decays. We think it is worthwhile to study again these
CP-violating asymmetries by including theoretical improvements
\cite{ag,acgk} and determine their $N_c$-and other parametric dependences.

Following our previous work \cite{akl98-1} we study this on the basis of the 
factorization approach. 
We consider the same seventy six decay channels as in \cite{akl98-1} and 
calculate 
the CP-violating asymmetries for charged and neutral $B$ decays with the
classification 
I to V as in \cite{akl98-1} to distinguish those channels which can be 
predicted with some certainty in the factorization approach. These are the
class-I and class-IV (and possibly some class-III) decays, whose decay 
amplitudes
are $N_c$-stable and which do not involve delicate cancellations among
components of the amplitudes. In our study here, we invoke two
models to  estimate the form factor dependence of the asymmetries, 
study their dependence  on the effective 
coefficients of the QCD and electroweak penguin operators in term of $N_c$,
the dependence on $k^2$, the virtuality of the gluon, photon or Z
in the penguin amplitudes decaying into
the quark-antiquark pair $q\bar q^\prime$ in $b\to q q^\prime\bar q^\prime$
and, of course, the CKM parameters. The last of these 
is the principal interest in measuring the CP-violating  asymmetries.
Our goal, therefore, is to identify, by explicit calculations, those decay
modes whose CP-violating asymmetries are relatively insensitive to the
variations of the rest of the parameters.

 In this pursuit, the sensitivity
of the asymmetries on $k^2$ is a stumbling block. As the branching
ratios are relatively insensitive to the parameter $k^2$, this
dependence can be removed only by the measurement of at least one of the
CP-violating asymmetries sensitive to it (examples of which are abundant),
enabling us to predict quite a few others. A mean value of $k^2$ can also be
estimated in specific wave function models \cite{sw} - an alternative, we
do not consider here. However, quite interestingly, we show that a number of
class-I and class-IV  (hence $N_c$-stable) decays involving $B^0/\bar{B}^0$
mesons have CP-violating asymmetries  which are also stable against
variation in $k^2$. Hence, in this limited number of decays, the asymmetries
can be reliably calculated within the factorization framework.
  We find that the CP-violating asymmetries in the
following decays are
particularly interesting and relatively stable: $\optbar{B^0} \to \pi^+
\pi^-$,
$\optbar{B^0} \to K_S^0 \pi^0$, $\optbar{B^0} \to K_S^0 \eta$,
$\optbar{B^0} \to K_S^0 \eta^\prime$ and $\optbar{B^0} \to \rho^+ \rho^-$.  
Likewise, the CP-violating asymmetry in the decays $\optbar{B^0} \to K_S^0
h^0$ with $h^0=\pi^0,K_S^0,\eta,\eta^\prime$ is
large as the individual decay modes have the same intrinsic CP-parity.
The $k^2$-dependences in  the individual asymmetries in this sum, which are
small to start with but not negligible, compensate each other resulting
in a CP-violating asymmetry which is practically independent of $k^2$.
Ideally, i.e., when only one decay amplitude dominates, the
asymmetries in the mentioned decays measure one of
the CP-violating phases $\alpha$ and $\beta$. In actual decays, many
amplitudes are present and we estimate  their contribution in the
asymmetries. To quantify this more pointedly, we work out the dependence of
the time-integrated partial rate asymmetry $A_{CP}(\pi^+\pi^-)$ in the
decays $\optbar{B^0} \to \pi^+ \pi^-$ on $\sin 2 \alpha$ and show the extent
of the so-called ``penguin pollution". Likewise, we work out the dependence 
of $A_{CP}(K_S^0 \eta^\prime)$,  $A_{CP}(K_S^0 \pi^0)$,  $A_{CP}(K_S^0
\eta)$ and  $A_{CP}(K_S^0 h^0)$ on $\sin 2 \beta$. We also study the effect
of the tree contribution - which we call a ``tree shadow" of the
penguin-dominated amplitude, on  $A_{CP}(K_S^0 \eta^\prime)$.
 The CP-violating asymmetries in $B^\pm$ decays are in general
$k^2$-dependent.
Supposing that this can be eventually fixed, as discussed above, the 
interesting
 asymmetries in $B^\pm \to h_1 h_2$ decays in our approach are: 
$B^\pm \to \pi^\pm \eta^\prime$,
 $B^\pm \to K^{*\pm} \eta$, $B^\pm \to K^{*\pm} \eta^\prime$ and $B^\pm \to
 K^{*\pm}\rho^0$. We study the asymmetries in the mentioned decays and also in 
$\optbar{B^0} \to \rho^\pm \pi^\mp$  in detail in this paper.

The effects of
soft final state interactions (SFI) may influence some (or
all) of the estimates presented here for the asymmetries. By the same token,
decay rates are also susceptible to such
non-perturbative effects \citer{DGPS96,DLWZ98}, which are, however, 
notoriously
difficult to quantify. We think that the role of SFI in $B \to h_1 h_2$
decays will be clarified already as the measurements of the branching ratios
become more precise and some more decays are measured. Based on the ``color
transparency" argument \cite{Bjorken}, we subscribe to the point of view
that the effects of SFI are subdominant in decays whose amplitudes are not
(color)-suppressed. However, it should be noted that the effects of the
so-called non-perturbative ``charm penguins" \cite{Martinellifudge}
are included here in the factorization approach in terms of the
leading power ($1/m_c^2$) corrections which contribute only to the decays
$B \to h_1 h_2$ involving an $\eta$ or $\eta^\prime$  \cite{acgk}, as
explained in the next section.

This paper is organized as follows: 
In section 2 we review the salient features of the generalized factorization
framework used in estimating the $B \to h_1 h_2$ decay rates in 
\cite{akl98-1}. In section 3
we give  the formulae  from which the various CP-violating asymmetries 
for the charged and neutral $B$ decays are calculated. 
Section 4 contains the numerical results for the CP-violating 
coefficients, required for time-dependent measurements of the 
CP-violating asymmetries in $B^0$ and $\bar{B}^0$ decays, and
time-integrated CP-violating asymmetries. The numerical results are
tabulated for three specific values of the effective number of colors 
$N_c=2,3,\infty$, varying $k^2$ in the range $k^2 =m_b^2/2 \pm 
2~\mbox{GeV}^2$, and two sets of the CKM parameters. We show the 
CKM-parametric dependence of the CP-violating asymmetries for some
representative
decays belonging to the class-I, class-III and class-IV decays, which 
have stable asymmetries and are
estimated to be measurably large in forthcoming experiments at $B$ factories
and hadron machines.
 Finally, in this section we study some decay modes which
have measurable but $k^2$-dependent CP-violating asymmetries, mostly
involving $B^\pm$ decays but also a couple of $B^0/\bar{B}^0$ decays.
Section 5 contains a summary of our results and conclusions.
\section{Generalized Factorization Approach and Classification of $B 
\to h_1 h_2$ Decays}

The calculation of the CP-violating asymmetries 
reported here is based on our work described in \cite{akl98-1}.
There, we started from the short-distance effective weak Hamiltonian 
$H_{eff}$ for $b\to s$ and $b\to d$ transitions.
We write below $H_{eff}$ for the $\Delta B=1$ transitions with five active 
quark
flavors by integrating out the top quark and the $W^\pm$ bosons:
\begin{equation}
\label{heff}
{\cal H}_{eff}
= \frac{G_{F}}{\sqrt{2}} \, \left[ V_{ub} V_{uq}^*
\left (C_1 O_1^u + C_2 O_2^u \right)
+ V_{cb} V_{cq}^*
\left (C_1 O_1^c + C_2 O_2^c \right) -
V_{tb} V_{tq}^* \,
\left(\sum_{i=3}^{10}
C_{i} \, O_i + C_g O_g \right) \right] \quad ,
\end{equation}
where $q=d,s$; $C_i$ are the Wilson coefficients evaluated at the
renormalization scale $\mu$ and $V_{ij}$ are the 
CKM matrix elements for which we shall use the Wolfenstein 
parameterization \cite{Wolfenstein83}. The operators $ O_i^u$ and $ O_i^c$ 
with $i=1,2$
are the current-current four-quark operators inducing the $b \to u      
q\bar{q}$ and $b \to c q \bar{q}$ transitions, respectively. The rest of
the operators  are the QCD penguin operators 
($O_{3},...,O_{6}$),  electroweak penguin
operators ($O_{7},...,O_{10}$), and $O_g$ represents the chromo-magnetic 
penguin operator.
The operator basis for $H_{eff}$ is given in \cite{akl98-1} together with 
the coefficients $C_1$, ..., $C_{6}$, evaluated in NLL precision, and 
$C_7$, ..., $C_{10}$ and $C_g$, evaluated in LL precision. 
Effects of weak annihilation and W-exchange diagrams have been neglected.

Working in NLL precision, the quark level matrix elements of $H_{eff}$
are treated at the one-loop level. 
They can be rewritten in terms of the tree-level matrix 
elements of the effective operators with new coefficients 
$C_1^{eff},...,C_{10}^{eff}$ 
(For details see \cite{akl98-1} and the references quoted therein.).
The effective coefficients $C_1^{eff}$, $C_2^{eff}$, $C_8^{eff}=C_8$,
and $C_{10}^{eff}=C_{10}$
 have no absorptive parts to the order we are working.
The effective coefficient $C_3^{eff}$, $C_4^{eff}$, $C_5^{eff}$, $C_6^{eff}$,
 $C_7^{eff}$ 
and $C_9^{eff}$ contain contributions of penguin diagrams with insertions of 
tree operator $O_{1,2}$, denoted by $C_t$ and $C_e$ in \cite{akl98-1} and with 
insertions of the QCD penguin operators $O_3$, $O_4$ and $O_6$ (denoted by 
$C_p$ in \cite{akl98-1}).
These penguin-like matrix elements have absorptive parts which generate the
required strong phases in the quark-level matrix elements.
The contributions $C_t$ and $C_e$ depend on the CKM matrix elements.
All three functions $C_t$, $C_p$ and $C_e$ depend on quark masses, the scale
$\mu$, and  $k^2$, and are given explicitly in eqs.~(10), (11) and (14),
respectively, of ref.~\cite{akl98-1}.

Having defined $H_{eff}$ in terms of the four-quark operators $O_i$ and 
their effective coefficients $C_i^{eff}$ the calculation of the hadronic 
matrix elements of the type $\langle h_1h_2 |O_i|B\rangle$ 
proceeds with the generalized
 factorization assumption \cite{NS97}.
The result of this calculation for the various $B\to PP$, $PV$ and $VV$ 
decays are written down in detail in \cite{akl98-1}. 
The hadronic matrix elements depend on the CKM matrix elements, which  contain 
the weak phases, the form factors and decay constants of current matrix 
elements, various quark masses and other parameters. The quantities $a_i$, 
given in terms of the effective short-distance coefficient $C_i^{eff}$,
\begin{equation}
  \label{ai}
  a_i=C_i^{eff} +\frac{1}{N_c} C_{i+1}^{eff} ~(i=odd);~~~~
  a_i=C_i^{eff} +\frac{1}{N_c} C_{i-1}^{eff} ~(i=even),
\end{equation}
where $i$ runs from $i=1,...,10$,
are of central phenomenological importance.
The terms in eq.~(\ref{ai}) proportional to $\xi=1/N_c$ originate from 
fierzing the operators $O_i$ to produce quark currents to match the quark
 content of the hadrons in the initial and final state after adopting the 
factorization assumption.
This well-known procedure results in general in matrix elements with the 
right flavor quantum number but involves both color singlet-singlet and 
color octet-octet operators. 
In the naive factorization approximation, one discards the color 
octet-octet operators.
This amounts to having $N_c=3$ in (\ref{ai}).
To compensate for these neglected octet-octet and other non-factorizing 
contribution one treats  $\xi\equiv1/N_c$ in
 eq.~(\ref{ai}) as a phenomenological 
parameter. In theory, $\xi$ can be obtained only by fully calculating the 
octet-octet and other non-factorizing contributions and can, in principle,
be different for each of the ten $a_i$.

Starting from the numerical values of the ten perturbative short distance 
coefficients $C_i^{eff}$ ($i=1,...,10$) we investigated in \cite{akl98-1} 
the $N_c$ 
dependence of the ten effective coefficients $a_i$ for the four types of 
current-current and penguin induced decays, namely $b\to s$ ($\bar b\to 
\bar s$) and  $b\to d$ ($\bar b\to \bar d$).
It was found, that $a_1$, $a_4$, $a_6$, $a_8$ and  $a_9$ are rather stable 
with respect to variations of $\xi$ in the usually adopted interval $\xi \in 
[0,1/2]$ (or $2<N_c <\infty$) for all four types of transitions, whereas 
$a_2$,  $a_3$, $a_5$, $a_7$ and $a_{10}$ depend very much on $\xi$. 

Based on this result we introduced a classification of factorized amplitudes 
which is an extension of the classification for tree decays in \cite{BSW87} 
relevant for $B$ decays involving charmed hadrons.
These classes I, II, III, IV and V are fully described in \cite{akl98-1} and 
will be used also in this work. 
The classes I, II and III in the decays $B \to h_1 h_2$ are defined as in 
previous work \cite{BSW87}. They involve dominantly (or only) current-current
transitions.
Class IV and V involve pure penguin or penguin-dominated decays. 
The classification is such, that decays in classes I and IV are  
stable against variations of $N_c$, whereas decays in classes II and V depend
strongly on $\xi=1/N_c$ and decays in class III have an intermediate status,
sometimes depending more, sometimes less on $\xi$. 
We concluded in \cite{akl98-1} that decay rates in the classes I and IV
decays  can be predicted in the factorization approximation. 
The decays in class II and V have sometimes rather small weak transition 
matrix elements, depending on the values of the effective $N_c$ and
CKM matrix elements. This introduces delicate cancellations which makes 
their amplitudes rather unstable as a function of $N_c$.
 Predicting the decay rates in these
classes involves a certain amount of theoretical fine-tuning, and hence we
are less sure about their estimates in the factorization approach.
Depending on the value of $\xi$, it is probable that other contributions 
not taken into account in the factorization approach used in \cite{akl98-1},
like annihilation, W exchange or soft final state interactions, are
important. We expect that the matrix elements of the decays in class-I and 
class-IV (and most class-III), being dominantly of $O(1)$ as far as their
$N_c$-dependence is concerned will be described, in the first approximation,
by a universal value of the parameter $\xi$. We are less sure that this will
be the case for class-II and class-V decays. As we show here, this 
$\xi$-sensitivity of the decay rates reflects itself also in estimates
of the CP-violating rate asymmetries.  

There is also an uncertainty due to the non-perturbative penguin
contributions \cite{Martinellifudge}, as we do not know how to
include their effects in the amplitudes $\langle h_1 h_2 | H_{eff}
|B\rangle$ from first principles. However, these  
effects can be calculated as an expansion in $1/m_c^2$ in the factorization
approach.
The dominant diagram contributing to the power corrections is the process
 $b \to s (c\bar{c} \to g(k_1)g(k_2))$, which was calculated in the
full theory (Standard Model) in \cite{SW90}. In the operator product
language which we are using, this contribution can be expressed as
a new induced effective Hamiltonian \cite{acgk}:
\begin{equation}
H_{eff}^{gg} = -\frac{\alpha_s}{2 \pi} a_2 \frac{G_F}{\sqrt{2}} V_{cb}
V_{cs}^* \Delta i_5\left(\frac{q^2}{m_c^2}\right)\frac{1}{k_1.k_2} O^{gg}~,
\label{heffgg}
\end{equation}
where the operator $O^{gg}$ is defined as:
\begin{equation}
O^{gg} \equiv G^{\alpha \beta}_a (D_\beta \tilde{G}_{\alpha \mu})_a ~
\bar{s} \gamma^\mu (1-\gamma_5) b~,
\label{Ogg}
\end{equation}
with $\tilde{G}_{\mu \nu, a}=1/2
\epsilon_{\alpha \beta \mu \nu} G^{\alpha \beta}_a$,
and $G^{\alpha \beta}_a$ being the QCD field strength tensor.
This formula holds for on-shell gluons
$q^2=(k_1 + k_2)^2=2 k_1.k_2$, and the sum over the color indices
is understood. The function
$\Delta i_5(z)$ is defined as \cite{acgk}:
\begin{equation}
\Delta i_5(z) = -1 + \frac{1}{z} \left[ \pi -2 \arctan 
\left(\frac{4}{z} -1\right)^{\frac{1}{2}}
\right]^2 ~~\mbox{for} ~0 < z < 4 ~.
\end{equation}  
The $H_{eff}^{gg}$ gives a non-local contribution but
one can expand the function  $\Delta i_{5}(z)$ in $z$ for
$z <1$ and the leading
term in this expansion can be represented as a higher dimensional
local  operator.
In fact, it is just the chromo-magnetic analogue of the operator
considered by Voloshin \cite{Voloshin} to calculate the power ($1/m_c^2)$
corrections in the radiative decay $B \to X_s + \gamma$. Now comes the
observation made in \cite{acgk} that in
the assumption of factorization, only the states which
have non-zero matrix elements $\langle M | \alpha_s
G^{\alpha \beta}_a (D_\beta \tilde{G}_{\alpha \mu})_a  | 0 \rangle$
contribute to the $1/m_c^2$ corrections in the decay rates for $B \to Mh$.
For $M=\eta, \eta^\prime$, this
matrix element is determined by the QCD anomaly, and $q^2$ also
gets fixed with $q^2=m_{\eta^{(\prime)}}^2$ which justifies 
the expansion.   
 For the decays
$B \to \eta^{(\prime)} K^{(*)}$, the $1/m_c^2$ effects
 were calculated in \cite{acgk}
in the decay rates. For the two-body $B \to h_1 h_2$ decays, these are
the only $1/m_c^2$ contributions in the factorization approach.
They are included here  in the estimates
of the rates and the asymmetries. Note that as the function $\Delta i_5
(m_{\eta^{(\prime)}}^2/m_c^2)$ has no absorptive part, there is no phase
generated by the anomaly contribution in $B \to \eta^{(\prime)} K^{(*)}$ 
decays.

Concerning the actual estimates of 
the $B\to h_1h_2$ matrix elements in the factorization approximation, we
note that they are 
calculated as in \cite{akl98-1} using two different theoretical approaches
to calculate the form factors.
First, we use the quark model due to Bauer, Stech and Wirbel \cite{BSW87}.
The second approach is based on lattice QCD and light-cone QCD sum 
rules. The specific values of the form factors and decay constants
 used by us and the references to
the literature are given in \cite{akl98-1}.
The implementation of the $\eta -\eta^\prime$ mixing  
follows the prescription of \cite{ag,acgk}.

Of particular importance for calculating the CP-violating asymmetries is
the choice of
the parameter $k^2$,
which appears in  the quantities $C_t$, $C_p$ and $C_e$ in the effective 
coefficients $C_i^{eff}$. 
Due to the factorization assumption any information on $k^2$ is lost when 
calculating two-body decays, except for the anomaly contribution as
discussed earlier. 
In a specific model and from simple two-body kinematics 
the average $k^2$ has been estimated to lie in the range $m_b^2/4 <k^2 < 
m_b^2/2$ \cite{sw}. In \cite{akl98-1} it was found that the branching ratios
(averaged over $B$ and $\bar B$ decays) are not sensitively dependent on $k^2$
if varied in the vicinity of $k^2=m_b^2/2$. Based on earlier work 
\cite{kps}, we do not expect the same result to hold for the asymmetries. 
Therefore, we calculated the CP-violating asymmetries by varying $k^2$ in
the range 
$k^2=m_b^2 /2 \pm 2$ GeV$^2$,
which should cover the expected range of $k^2$ in 
phenomenological models. Quite interestingly, we find that a number of decay modes
in the class-I and class-IV decays have asymmetries which are insensitive to
the variation of $k^2$. These then provide suitable avenues to test the 
assumption that strong interaction phases in these decays are dominantly
 generated perturbatively.

\section{CP-Violating Asymmetries in $B \to h_1 h_2$ Decays - Formalism}
%


For charged $B^\pm$ decays the CP-violating rate-asymmetries 
in partial decay rates are defined as follows:
\begin{equation}
  \label{acp}
  A_{CP} = \frac{  \Gamma ( B^+ \to  f^+)-\Gamma ( B^- \to  f^-) }{
\Gamma ( B^+ \to  f^+)+  \Gamma ( B^- \to  f^-) }.
\end{equation}
As these decays are all self-tagging, measurement of these CP-violating 
asymmetries
is essentially a counting experiment in well defined final states.
Their rate asymmetries require both weak and strong phase differences in 
interfering amplitudes. 
The weak phase difference arises from the superposition of amplitudes from 
various tree (current-current) and penguin diagrams. 
The strong phases,
which are needed to obtain non-zero values for $A_{CP}$ in (\ref{acp}),
are generated by final state interactions. 
For both $b\to s$ and $b \to d$ transitions, the strong phases 
are generated in our model
perturbatively by taking into account the full NLO corrections,
following earlier suggestions along these lines \cite{SEW}.

\subsection{CP-violating Asymmetries Involving $b \to s$ Transitions}

For the $b \to s$, and the charge conjugated $\bar{b} \to 
\bar{s}$, transitions, the respective  decay amplitudes ${\cal M}$ and 
$\overline{\cal M}$,  
including the weak and strong phases, can be generically written as:
\begin{eqnarray}
{\cal M}&=&  T \xi _u -P_t \xi_t  e^{i\delta_t} -P_c \xi_c  e^{i\delta_c}
-P_u \xi_u  e^{i\delta_u},\nonumber\\
\overline{\cal M} &=&  T \xi^* _u -P_t \xi_t^*  e^{i\delta_t} -P_c \xi_c^*  
e^{i\delta_c} -P_u \xi_u^*  e^{i\delta_u},\label{mm}
\end{eqnarray}
where $\xi_i=V_{ib}V^*_{is}$.
Here we denote by $T$  
the contributions from the current-current operators proportional 
to the effective coefficients $a_1$ and/or $a_2$; $P_t$, $P_c$ 
and $P_u$ denote the contributions from penguin operators proportional to
the product of the CKM matrix elements $\xi_t$, $\xi_c$ and $\xi_u$, 
respectively. The corresponding 
strong phases are denoted by $\delta_t$, $\delta_c$ and $\delta_u$, 
respectively.
Working in the standard model, 
we can use the unitarity relation $\xi_c=-\xi_u-\xi_t $
to simplify the above equation (\ref{mm}),
\begin{eqnarray}
{\cal M}&=&  T \xi _u -P_{tc} \xi_t  e^{i\delta_{tc}} 
-P_{uc} \xi_u  e^{i\delta_{uc}},\nonumber\\
\overline{\cal M} &=&  T \xi^* _u -P_{tc} \xi_t^*  e^{i\delta_{tc}}   
 -P_{uc} \xi_u^*  e^{i\delta_{uc}},\label{msim}
\end{eqnarray}
where we define
\begin{eqnarray}
P_{tc}  e^{i\delta_{tc}}&=&  P_t e^{i\delta_t} -P_c e^{i\delta_c} ,\nonumber\\
P_{uc}  e^{i\delta_{uc}}&=&  P_u   e^{i\delta_u}-P_c e^{i\delta_c}.
\end{eqnarray}
 Thus, the direct CP-violating asymmetry is
 \begin{equation}
A_{CP} \equiv a_{\epsilon^\prime}=\frac{A^-}{A^+}~,
\end{equation}
where
\begin{eqnarray}
A^- &=&\frac{1}{2}\left( |\overline{\cal M} |^2-|{\cal M}|^2 \right)\nonumber\\
&=& 2 TP_{tc} |\xi _u^* \xi_t|\sin \gamma \sin \delta_{tc}
+ 2 P_{tc} P_{uc} |\xi _u^* \xi_t|\sin \gamma \sin (\delta_{uc}-\delta_{tc}),
\label{aminus}
\end{eqnarray}
\begin{eqnarray}
A^+ &=& \frac{1}{2}\left( |{\cal M} |^2+ |\overline{\cal M}|^2 \right)
\nonumber\\
&=& (T^2+P_{uc}^2) |\xi_u|^2 + P_{tc}^2  |\xi_t|^2 
- 2 P_{tc} P_{uc} |\xi _u^* \xi_t|\cos \gamma \cos (\delta_{uc}-\delta_{tc})
\nonumber\\
&&- 2 TP_{uc} |\xi _u |^2\cos \delta_{uc}
+2 TP_{tc} |\xi _u^* \xi_t|\cos \gamma \cos \delta_{tc}.
\label{aplus}
\end{eqnarray}
In the case of $b\to s$ transitions the weak phase entering in $A^-$ is 
equal to $\gamma$, as we are using the Wolfenstein approximation
\cite{Wolfenstein83} in which $\xi_t$ has no weak phase and the phase of 
$\xi_u$ is $\gamma$. Thus, the weak phase dependence factors out in an
overall $\sin \gamma$ in $A^-$. Despite this,
the above equations for the CP-violating asymmetry $A_{CP}$ are quite
involved due to the fact
that several strong phases are present which are in general hard to
calculate except in specific models such as the ones being used here.
However, there are several small parameters involved in the numerator and
denominator given above. Expanding in these small parameters, 
much simplified forms for $A^-$ and $A^+$ and hence $A_{CP}$ can be 
obtained in specific decays. 

First, we note that  $|\xi_u| \ll | \xi_t|\simeq
|\xi_c|$, with an upper bound  $|\xi_u|/| \xi_t| \leq 0.025$.
In some channels, such as $B^+ \to K^+ \pi^0$, $K^{*+} \pi^0$,
$K^{*+} \rho^0$, $B^0 \to K^+ \pi^-$, $K^{*+} \pi^-$, $K^{*+} \rho^-$,
typical
value of the ratio $|P_{tc}/T|$ is of $O(0.1)$, with both $P_{uc}$ and $P_{tc}$
comparable with typically $|P_{uc}/P_{tc}| =O(0.3)$.
The importance of including the contributions proportional to $P_{uc}$ has 
been stressed earlier in the literature \cite{BF95} (see, also
\cite{Fleischer94,FMalpha}). 
These estimates are based on perturbation theory but the former 
inequality $|P_{tc}/T| \ll 1$ should hold generally as the top quark 
contribution is genuinely short-distance. The other  inequality can
be influenced by non-perturbative penguin contributions. However, also in
this case, for the mentioned transitions, we expect that $|P_{uc}/T| \ll 1$
 should hold. Using these approximations, eq.~(\ref{aminus},
\ref{aplus}) become simplified:
\begin{eqnarray}
A^- &  \simeq & 2 T P_{tc} |\xi _u^* \xi_t|\sin \gamma \sin \delta_{tc},
\end{eqnarray}
\begin{eqnarray}
A^+ & \simeq &  P_{tc}^2  |\xi_t|^2 + T^2  |\xi_u|^2 
 +2 TP_{tc} |\xi _u^* \xi_t|\cos \gamma \cos \delta_{tc}.
\end{eqnarray}
The CP-violating asymmetry in this case is 
\begin{equation}
  \label{cps1}
  A_{CP} \simeq \frac{2z_{12} \sin \delta_{tc} \sin \gamma}
{1+ 2 z_{12} \cos \delta_{tc} \cos \gamma +z_{12}^2},\label{app1}
\end{equation}
where $z_{12}=|\xi_u/\xi_t|\times T/P_{tc}$, where we use the notation
used in \cite{akl98-1}.
This relation was suggested in the context of the decay $B \to K\pi$ by
Fleischer and Mannel \cite{FM97}.
Due to the circumstance that the suppression due to $|\xi_u/\xi_t|$ is
 stronger than the enhancement
due to $T/P_{tc}$, restricting the value of $z_{12}$,  the CP-violating 
asymmetry for these kinds of decays are
O(10\%).
To check the quality of the approximation made in eq.~(\ref{app1}),
we have calculated the CP-violating asymmetry using this formula for $B^0
\to K^+ \pi^-$, 
which yields $A_{CP}=-7.1\%$ at $N_c=2$, very close to the value $-7.7\%$ in 
Table 5 calculated using the full formula, with $\rho=0.12$, $\eta=0.34$ and
$k^2=m_b^2/2$ in both cases. The results for other values of $N_c$ are similar.
Thus, we conclude that 
eq.~(\ref{app1}) holds to a good approximation in the factorization
framework for the decays mentioned earlier on.
However, the CP-violating asymmetries $A_{CP}$ in the mentioned decays
are found to depend on $k^2$, making their theoretical predictions
considerably uncertain. These can be seen in the various tables for
$A_{CP}$.
Of course, the relation (\ref{cps1}) given above, and others given below,
can
be modified through  SFI - a possibility we are not entertaining here.

There are also some decays with vanishing tree contributions, such as 
$B^+ \to \pi^+ K_S^0$, $\pi^+ K^{*0}$, $\rho^+ K^{*0}$. For these decays,
$T=0$, and $|\xi_{u}|\ll  |\xi_{t}|$, then for these decays
\begin{eqnarray}
A^- & = & 2 P_{tc} P_{uc} |\xi _u^* \xi_t|\sin \gamma 
 \sin (\delta_{uc}-\delta_{tc}) ,
\end{eqnarray}
\begin{eqnarray}
A^+ & \simeq &  P_{tc}^2  |\xi_t|^2 
- 2 P_{tc} P_{uc} |\xi _u^* \xi_t|\cos \gamma \cos (\delta_{uc}-\delta_{tc})\\
& \simeq & P_{tc}^2  |\xi_t|^2 .
\end{eqnarray}
The CP-violating asymmetry is 
\begin{equation}
  \label{cps2}
  A_{CP} \simeq 2\frac{P_{uc}}{P_{tc}}\left |\frac{\xi_u}{\xi_t}\right |
 \sin (\delta_{uc}-\delta_{tc}) \sin \gamma.
\end{equation}
Without the $T$ contribution, the suppression due to both $P_{uc}/P_{tc}$ and
$|\xi_u/\xi_t|$ is much stronger and the CP-violating asymmetries are only
around 
$-$($1$-$2$)\%. This is borne out by the numerical results obtained with
the complete contributions, which can be seen in the Tables.

\subsection{CP-violating Asymmetries Involving $b \to d$ Transitions}

For  $b\to d$ transitions, we have 
\begin{eqnarray}
{\cal M} &=&  T \zeta _u -P_{t} \zeta_t  e^{i\delta_{t}} -P_c \zeta_c  
e^{i\delta_c} -P_{u} \zeta_u  e^{i\delta_{u}},\nonumber\\
\overline{\cal M} &=&  T \zeta^* _u -P_{t} \zeta_t^*  e^{i\delta_{t}} -P_c 
\zeta_c^*  e^{i\delta_c}
-P_{u} \zeta_u^*  e^{i\delta_{u}},
\end{eqnarray}
where $\zeta_i=V_{ib}V^*_{id}$, and again using CKM unitarity relation
$\zeta_c = -\zeta_t-\zeta_u $, we have
\begin{eqnarray}
{\cal M} &=&  T \zeta _u -P_{tc} \zeta_t  e^{i\delta_{tc}}
 -P_{uc} \zeta_u  e^{i\delta_{uc}},\nonumber\\
\overline{\cal M} &=&  T \zeta^* _u -P_{tc} \zeta_t^*  e^{i\delta_{tc}} 
 -P_{uc} \zeta_u^*  e^{i\delta_{uc}},
\end{eqnarray}
\begin{eqnarray}
A^- &=& -2 TP_{tc} |\zeta _u^* \zeta_t|\sin \alpha \sin \delta_{tc}
- 2 P_{tc} P_{uc} |\zeta _u^* \zeta_t|\sin \alpha \sin (\delta_{uc}
-\delta_{tc}),
\label{aminusd}
\end{eqnarray}
\begin{eqnarray}
A^+ &=& (T^2+P_{uc}^2) |\zeta_u|^2 + P_{tc}^2  |\zeta_t|^2 
- 2 P_{tc} P_{uc} |\zeta _u^* \zeta_t|\cos \alpha \cos 
(\delta_{uc}-\delta_{tc})
\nonumber\\
&&- 2 TP_{uc} |\zeta _u |^2\cos \delta_{uc}
+2 TP_{tc} |\zeta _u^* \zeta_t|\cos \alpha \cos \delta_{tc}.
\end{eqnarray}
For the tree-dominated decays involving $b\to d $ transitions, such as 
$B^+\to \pi^+ \eta ^{(\prime)}$, $\rho^+ \eta ^{(\prime)}$, $\rho^+ \omega$,
the relation $P_{uc} < P_{tc} \ll T$ holds. This makes the formulae 
simpler, yielding 
 \begin{eqnarray}
A^- &\simeq & -2 TP_{tc} |\zeta _u^* \zeta_t|\sin \alpha \sin \delta_{tc},
\end{eqnarray}
\begin{eqnarray}
A^+ &\simeq & T^2 |\zeta_u|^2 - 2 TP_{uc} |\zeta _u |^2\cos \delta_{uc}
+2 TP_{tc} |\zeta _u^* \zeta_t|\cos \alpha \cos \delta_{tc}\nonumber\\
 &\simeq & T'^2 |\zeta_u|^2
+2 TP_{tc} |\zeta _u^* \zeta_t|\cos \alpha \cos \delta_{tc},
\end{eqnarray}
with $T'^2\equiv T^2  - 2 TP_{uc}\cos \delta_{uc}$.
The CP-violating asymmetry is now approximately given by
\begin{equation}
 A_{CP} \simeq \frac{-2z_{1} \sin \delta_{tc} \sin \alpha}
{1+ 2 z_{1} \cos \delta_{tc} \cos \alpha},\label{app3}
\end{equation}
with $z_{1}=|\zeta_t/\zeta_u|\times TP_{tc}/T'^2$.
Note, the CP-violating asymmetry is approximately proportional to
$\sin\alpha$ in this 
case.
Here the suppression due to $P_{tc}T/T^{\prime 2}$ is accompanied with some 
enhancement from
$|\zeta_t/\zeta_u|$ (the central value of this quantity is about 3 
\cite{aliapctp97}),
 making the CP-violating asymmetry in this  kind of decays to have a
value  $A_{CP}=(10$-$20)\%$.
We have calculated the CP-violating asymmetry of $B^\pm\to \rho^\pm 
\omega$ using the
approximate
 formula (\ref{app3}). The number we got for $N_c=2$ is $A_{CP}=9.2\%$, 
which is very close to the value $A_{CP}=8.9\%$ in Table 11 calculated 
using the exact formula, with $\rho=0.12$, $\eta=0.34$ and $k^2=m_b^2/2$. 

For the decays with a vanishing tree contribution, such as 
$B^+ \to K^+  K_S^0$, $K^+ \bar K^{*0}$, $K^{*+} \bar K^{*0}$, we have 
$T=0$. Thus,
 \begin{eqnarray}
A^- &= &- 2 P_{tc} P_{uc} |\zeta _u^* \zeta_t|\sin \alpha \sin 
(\delta_{uc}-\delta_{tc}),
\end{eqnarray}
\begin{eqnarray}
A^+ &= & P_{tc}^2  |\zeta_t|^2 +P_{uc}^2 |\zeta_u|^2
- 2 P_{tc} P_{uc} |\zeta _u^* \zeta_t|\cos \alpha \cos(\delta_{uc}
-\delta_{tc}).
\end{eqnarray}
The CP-violating asymmetry is approximately  proportional to  $\sin\alpha$
again,
\begin{equation}
 A_{CP} = \frac{-2z_{3} \sin (\delta_{uc}-\delta_{tc}) \sin \alpha}
{1- 2 z_{3} \cos (\delta_{uc}-\delta_{tc}) \cos \alpha+z_3^2},\label{app4}
\end{equation}
with $z_3=|\zeta_u/\zeta_t|\times P_{uc}/P_{tc}$.
As the  suppressions from  
$|\zeta_u/\zeta_t|$ and $|P_{uc}/P_{tc}|$  
are not very big, the CP-violating asymmetry can again be
of order $(10$-$20)\%$. However, being direct CP-violating
asymmetries, the mentioned asymmetries in the specific $B^\pm \to
(h_1 h_2)^\pm$ modes depend on $k^2$ and are uncertain.

\subsection{CP-violating Asymmetries in Neutral $B^0$ Decays}

For the neutral $B^0(\bar{B}^0)$ decays, there is an additional
complication due to  $B^0$ - $\overline{B^0}$ mixing.
These CP-asymmetries may require time-dependent
measurements, as discussed in the literature 
\cite{cpbook},\citer{gr89,PW95}. 
Defining the time-dependent asymmetries as
\begin{equation}
  \label{ae}
  A_{CP}(t)= \frac{ \Gamma (B^0(t) \to f) -\Gamma (\overline  B^0(t) \to  
\bar f)}{
 \Gamma (B^0(t) \to f) +\Gamma ( \overline  B^0(t) \to  \bar  f)},
\end{equation}
there are four cases that one encounters for neutral $B^0(\bar{B}^0)$ 
decays:
 \begin{itemize}
\item case (i): $B^0 \to f $, $\bar{B}^0 \to \bar{f}$, where $f$ or $\bar{f}$
is not a common final state
of $B^0$ and $\bar{B}^0$, for example ${B}^0 \to K^+\pi^-$.
\item case (ii): $B^0 \to (f=\bar{f}) \leftarrow \bar{B}^0$ with 
$f^{CP}=\pm f$, involving final
states which are CP eigenstates, i.e., decays such as $\bar{B}^0 (B^0) \to 
\pi^+ \pi^-, \pi^0 \pi^0, K_S^0 \pi^0$ etc.
\item case (iii): $B^0 \to (f=\bar{f}) \leftarrow \bar{B}^0$ with
$f$, involving final states which are not CP eigenstates. They include
decays such as $B^0 \to (VV)^0$, as the $VV$ states are not CP-eigenstates.  
\item case (iv): $B^0 \to (f \& \bar{f}) \leftarrow 
\bar{B}^0$ with $f^{CP} \neq f$, i.e.,  both $f$ and  $\bar{f}$ are common 
final states of $B^0$ and $\bar{B}^0$, but they are not  CP eigenstates.
 Decays ${B}^0  \to \rho^+ \pi^-$, $\rho^- \pi^+$ and $B^0\to K^{*0} 
 K_S^0$, $\bar K^{*0} K_S^0$ are two examples of interest for us.
\end{itemize}
Here case (i) is very similar to the charged $B^\pm$ decays. For case (ii),
and (iii), $A_{CP}(t)$ would involve $B^0$ - $\overline{B^0}$ mixing.
Assuming $|\Delta \Gamma | \ll |\Delta m |$
and $|\Delta \Gamma/\Gamma | \ll 1$, which hold in the standard model for
the mass and width differences $\Delta m $ and
$\Delta \Gamma$ in the neutral $B$-sector, one can express
$A_{CP}(t)$ in a simplified form:
\begin{equation}
\label{acpeps}
A_{CP}(t) \simeq a_{\epsilon^\prime} \cos (\Delta m t) + a_{\epsilon + 
\epsilon^\prime} \sin (\Delta m t) ~.
\end{equation}
The quantities $a_{\epsilon^\prime}$ and $a_{\epsilon + 
\epsilon^\prime}$, for which we follow the definitions given in
\cite{PW95}, depend on 
the hadronic matrix elements which we have calculated in our model.
\begin{equation}\label{a1}
a_{\epsilon^\prime} = \frac{ 1- |\lambda_{CP}|^2}{1+ |\lambda_{CP}|^2},
\end{equation}
\begin{equation}\label{a2}
a_{\epsilon+\epsilon^\prime }= \frac{ - 2Im(\lambda_{CP})}
{1+ |\lambda_{CP}|^2} ,
\end{equation}
where
\begin{equation}
\lambda_{CP} = \frac{V_{tb}^* V_{td}}{V_{tb} V_{td}^*}
\frac{\langle f|H_{eff}|\bar B^0\rangle  }{\langle f|H_{eff}| B^0\rangle  }. 
\end{equation}

For case (i) decays, the coefficient $a_{\epsilon^\prime}$ determines
$A_{CP}(t)$, and since no mixing is involved for these decays, the
CP-violating asymmetry is independent of time. We shall call these,
together with the
CP-asymmetries in charged $B^\pm$ decays, CP-class (i) decays. 
For case (ii) and (iii), 
one has to separate the $\sin(\Delta m t)$ and  $\cos (\Delta m t)$
terms to get the CP-violating asymmetry $A_{CP}(t)$.
The time-integrated asymmetries are:
\begin{equation}
A_{CP} =\frac{1}{1+x^2} a_{\epsilon^\prime }
+\frac{x}{1+x^2} a_{\epsilon+\epsilon^\prime },\label{cpint}
\end{equation}
with $x=\Delta m /\Gamma \simeq 0.73$ for the $B^0$ - $\overline{B^0}$
case \cite{PDG96}.

Case (iv) also involves mixing but requires additional formulae. Here one 
studies the four 
time-dependent decay widths for $B^0(t) \to f$, $\bar B^0(t) \to \bar f$, 
$B^0(t) \to \bar f$ and $\bar B^0(t) \to  f$ \cite{gr89,adkd91,PW95}. 
These time-dependent widths can be expressed by four basic matrix elements
\begin{equation}
\begin{array}{ll}
g=\langle f|H_{eff} |B^0\rangle  ,& h=\langle f|H_{eff}|\bar B^0\rangle ,\\
\bar g=\langle \bar f|H_{eff} |\bar B^0\rangle  ,
& \bar h=\langle \bar f|H_{eff}| B^0\rangle ,
\end{array}
\end{equation}
which determine the decay matrix elements of $B^0\to f \& \bar f$ and 
of $\bar B^0 \to \bar f \& f$ at $t=0$.
For example, when $f=\rho^-\pi^+$ the matrix element $h$ is given in 
appendix B of \cite{akl98-1} in eq.~(99) and $\bar g$ for the decay $\bar 
B^0\to
\rho^+\pi^-$ is written down in eq.~(100) in appendix B of \cite{akl98-1}.
The matrix elements $\bar h$ and $g$ are obtained from $h$ and $\bar g$ by
changing the signs of the weak phases contained in the products of the
CKM matrix elements. We also need to know  the CP-violating parameter
coming from the $B^0$ - $\bar{B}^0$ mixing. Defining:
\begin{eqnarray}
B_1 &=& p |B^0\rangle + q | \bar{B}^0 \rangle, \nonumber\\
B_2 &=& p |B^0\rangle - q | \bar{B}^0 \rangle,
\label{B12}
\end{eqnarray}
with $\vert p \vert^2 + \vert q \vert ^2=1$ and $q/p=\sqrt{H_{21}/H_{12}}$,
 with $H_{ij}= M_{ij} - i/2\Gamma_{ij}$ representing the 
 $\vert \Delta B \vert =2$ and $\Delta Q=0$ Hamiltonian \cite{cpbook}. 
 For the decays of $B^0$ and $\bar{B}^0$, we use,
as before, 
\begin{equation}
\frac{q}{p} = \frac{V_{tb}^* V_{td}}{V_{tb}V_{td}^*} = e ^{-2 i \beta}.
\label{qpdef}
\end{equation}
So, $|q/p|=1$, and this ratio has only a phase given by $-2 \beta$.
Then, the four time-dependent widths 
are given by the following formulae (we follow the notation of \cite{PW95}):
\begin{eqnarray}
  \Gamma (B^0 (t) \to f)& =& e^{-\Gamma t} \frac{1}{2} ( |g|^2 + |h|^2 )
\left \{ 1+ a_{\epsilon '} \cos \Delta m t + a_{\epsilon +\epsilon '} 
\sin \Delta m t \right \},\nonumber \\
 \Gamma (\bar B^0 (t) \to \bar f)& =& e^{-\Gamma t} \frac{1}{2} ( |\bar g|^2 
+ |\bar h|^2 ) \left \{ 1- a_{\bar \epsilon '} \cos \Delta m t 
- a_{\epsilon +\bar \epsilon '} 
\sin \Delta m t \right \},\nonumber \\
 \Gamma ( B^0 (t) \to \bar f) &= &e^{-\Gamma t} \frac{1}{2} ( |\bar g|^2 
+ |\bar h|^2 )\left \{ 1+ a_{\bar \epsilon '}
 \cos \Delta m t + a_{\epsilon +\bar \epsilon '} 
\sin \Delta m t \right \},\nonumber \\
 \Gamma (\bar B^0 (t) \to f) &=& e^{-\Gamma t} \frac{1}{2} ( | g|^2 
+ | h|^2 )
\left \{ 1- a_{\epsilon '} \cos \Delta m t - a_{\epsilon +\epsilon '} 
\sin \Delta m t \right \},  \label{rate}
\end{eqnarray}
where
\begin{equation}
  \label{aepsilon}
  \begin{array}{ll}
a_{\epsilon '} = \displaystyle \frac{ |g|^2 -|h|^2}{ |g|^2 +|h|^2}, &
a_{\epsilon +\epsilon '} = \displaystyle 
\frac{-2Im \left( \frac{q}{p}\frac{h}{g}\right)}
{1+|h/g|^2},\\
a_{\bar \epsilon '} = \displaystyle 
\frac{ |\bar h|^2 -|\bar g|^2}{ |\bar h|^2 +|\bar g|^2},
 &
a_{\epsilon +\bar \epsilon '} = \displaystyle 
\frac{-2Im \left( \frac{q}{p}\frac{\bar g}
{\bar h}\right)}
{1+|\bar g/\bar h |^2}.
   \end{array}
\end{equation}
By measuring the time-dependent spectrum of the decay rates of $B^0$ and 
$\bar B^0$, one can find the coefficients of the two functions $\cos\Delta mt$
and $\sin \Delta m t$ and extract the quantities $a_{\epsilon '}$,
$a_{\epsilon +\epsilon '}$, $|g|^2+|h|^2$, $a_{\bar \epsilon '}$, 
$a_{\epsilon +\bar \epsilon '}$ and $|\bar g|^2+|\bar h|^2$ as well as 
$\Delta m$ and $\Gamma$, which, however, are already well measured
\cite{PDG96}. The signature of CP violation is
$\Gamma (B^0 (t) \to f) \neq \Gamma (\bar B^0 (t) \to \bar f)$ and 
$\Gamma (\bar B^0 (t) \to f) \neq \Gamma ( B^0 (t) \to \bar f)$ which means,
that $a_{\epsilon '} \neq - a_{\bar \epsilon '}$ and/or 
$a_{\epsilon +\epsilon '} \neq - a_{\epsilon +\bar \epsilon '}$.
In the two examples, $f=\rho^+\pi^-$ and $f=K^{*0} K_S^0$, 
the amplitudes $g$ and $h$ contain contributions of several terms similar 
to what we have written down above for the charged $B$ decays. They have weak 
and strong phases with the consequence that 
$|g|\neq |\bar g|$ and $|h| \neq |\bar h|$.

%

\begin{table}
\caption{ CP-violating asymmetry parameters $a_{\epsilon^\prime}$ and 
$a_{\epsilon +\epsilon^\prime}$ (in percent) for the decays 
$\protect\optbar{B^0} 
\to h_1h_2$ using $\rho=0.12$, $\eta=0.34$ and $N_c=2,3,\infty$,
for $k^2=m_b^2/2\pm 2$ GeV$^2$.
}
\label{cppa}
\begin{tabular}{||l||r|r|r||r|r|r||}
\hline
\hline
 &   \multicolumn{3}{|c||}{$a_{\epsilon^\prime}$} 
&\multicolumn{3}{|c||}{$a_{\epsilon +\epsilon^\prime}$} \\
\hline
Channel &  $N_c=2$ &  $N_c=3$ & $N_c=\infty$
&  $N_c=2$ &  $N_c=3$ & $N_c=\infty$ \\
\hline
\hline
$\optbar{B^0} \to \pi^+ \pi^-$  &
$ 6.9^{+1.6}_{-3.5}$ &$ 7.0^{+1.6}_{-3.6}$&$ 7.0^{+1.7}_{-3.6}$
&$ 35.3^{-1.6}_{+2.2}$& $ 35.0^{-1.6}_{+2.2}$&$ 34.5^{-1.7}_{+2.2}$\\
$\optbar{B^0} \to \pi^0 \pi^0$     &
$ 1.0^{-2.2}_{+4.6}$&$ 17.6^{-2.0}_{+3.9}$ &$ 10.1^{+0.9}_{-2.0}$&
$-89.7^{+2.7}_{-2.9}$&$-55.8^{-4.0}_{+5.6}$&$ 81.8^{-2.9}_{+3.7}$\\

$ \optbar{B^0} \to \eta^\prime \eta^\prime$  &
$ 26.0^{+5.6}_{-11.6}$ &$ 38.1^{+3.9}_{-6.1}$ &$-17.2^{-7.2}_{+16.0}$ &
$ 62.8^{-4.8}_{+16.0}$ &$ 78.0^{-2.3}_{+3.6}$ &$-85.7^{+4.5}_{-5.1}$
\\
$ \optbar{B^0} \to \eta \eta^\prime$    &
$ 22.9^{+4.3}_{-8.5}$ &$ 23.3^{+0.5}_{-0.4}$ &$-13.3^{-6.6}_{+14.3}$ &
$ 88.5^{-2.5}_{+3.2}$ &$ 62.7^{-1.0}_{+2.1}$ &$-96.5^{+1.4}_{-2.3}$
\\
$ \optbar{B^0} \to \eta \eta$    & 
$ 19.3^{+3.0}_{-5.9}$ &$ 16.1^{-0.6}_{+1.5}$ &$-10.1^{-5.9}_{+12.5}$ &
$ 97.7^{-0.9}_{+1.2}$ &$ 50.6^{-0.9}_{+1.8}$ &$-99.5^{+0.9}_{-0.3}$
\\
$ \optbar{B^0} \to \pi^0 \eta^\prime$    &
$ 31.3^{+0.7}_{-0.8}$&$ 22.9^{-3.0}_{+5.1}$ &$ 9.2^{-7.3}_{+12.6}$&
$ 59.1^{-2.3}_{+4.0}$&$ 29.9^{-3.2}_{+5.6}$ &$-20.1^{-4.3}_{+7.6}$\\
$ \optbar{B^0} \to \pi^0 \eta$    
&$ 17.2^{-1.2}_{+2.4} $&$ 13.8^{-2.6}_{+5.0}$&$ 7.4^{-4.8}_{+9.0}$&
$ 43.1^{-1.4}_{+2.6}$&$ 21.8^{-2.3}_{+3.8}$&$-15.7^{-3.3}_{+5.4}$\\

$\optbar{B^0} \to K_S^0 \pi^0$        &
$ 0.4^{+0.6}_{-1.3}$ &$-1.2^{+0.0}_{-0.2}$ &$-3.8^{-0.9}_{+1.4}$ &
$ 75.1^{+0.2}_{-0.3}$ &$ 69.1^{+0.1}_{-0.2}$ &$ 58.1^{-0.3}_{+0.3}$
\\
$\optbar{B^0} \to K_S^0 \eta^\prime$       &
$-2.4^{-0.3}_{+0.5}$ &$-1.8^{-0.1}_{+0.2}$ &$-0.9^{+0.2}_{-0.4}$ &
$ 64.7^{-0.0}_{+0.1}$ &$ 66.9^{-0.0}_{+0.0}$ &$ 70.2^{+0.1}_{-0.2}$
\\
$\optbar{B^0} \to K_S^0 \eta$         &
$ 1.1^{+0.9}_{-1.6}$ &$-1.0^{+0.1}_{-0.3}$ &$-4.3^{-1.1}_{+1.8}$ &
$ 78.0^{+0.2}_{-0.5}$ &$ 69.7^{+0.1}_{-0.1}$ &$ 54.1^{-0.4}_{+0.5}$
\\
$\optbar{B^0}\to K^0\bar K^0$ &
$ 12.5^{-2.9}_{+5.5}$ &$ 12.3^{-2.9}_{+5.5}$ &$ 12.0^{-2.8}_{+5.5}$ &
$ 15.7^{-2.4}_{+4.0} $ &$ 15.6^{-2.4}_{+3.9} $ &$ 15.3^{-2.3}_{+3.9} $ 
\\
$\optbar{B^0} \to \rho^0 \pi^0$    &
$-6.4^{-4.6}_{+9.8}$ &$-3.1^{-17.0}_{+30.0}$ &$ 7.8^{+3.4}_{-7.0}$ &
$-40.6^{+3.6}_{-4.7}$ &$-99.5^{+1.6}_{+5.1}$ &$ 36.0^{-3.2}_{+4.5}$
\\
$\optbar{B^0} \to \omega \pi^0$    &
$ 26.2^{+2.6}_{-4.3}$ &$ 23.4^{-0.6}_{+1.5}$ &$ 1.0^{+0.2}_{-0.6}$ &
$ 84.7^{-0.5}_{+1.1}$ &$ 50.1^{-1.7}_{+3.3}$ &$ 49.8^{-0.2}_{+0.3}$ 
\\
$ \optbar{B^0} \to \rho^0 \eta $   & 
$-19.8^{-3.2}_{+26.7}$ &$ 12.9^{-8.8}_{+14.5}$ &$ 30.1^{+5.4}_{-9.9}$ &
$-97.9^{+3.6}_{-0.7}$ &$-15.9^{-4.9}_{+9.4}$ &$ 93.9^{-2.4}_{+3.1}$
\\
$ \optbar{B^0} \to \rho^0 \eta^\prime$  & $-52.7^{-6.6}_{+26.3}$ &
$-55.0^{-1.9}_{+79.7}$ &$ 38.3^{+8.9}_{-19.1}$ &$ 37.8^{+10.0}_{-15.7}$ &
$-43.5^{+35.2}_{-50.4}$ &$ 31.8^{-9.4}_{+13.2}$
\\
$ \optbar{B^0} \to \omega \eta $   &$ 16.3^{+3.3}_{-6.8}$ &
$ 25.1^{+3.4}_{-6.1}$ &$ 1.8^{+0.5}_{-1.0}$ &$ 74.6^{-2.4}_{+3.3}$ &
$ 94.8^{-0.6}_{+1.0}$ &$ 9.5^{-0.4}_{+0.7}$
\\
$ \optbar{B^0} \to \omega \eta^\prime$  & $ 17.7^{+4.0}_{-8.5}$ &
$ 43.5^{+9.5}_{-19.2}$ &$ 1.9^{+0.4}_{-1.0}$ &$ 46.0^{-3.9}_{+5.2}$ &
$ 55.6^{-9.1}_{+12.3}$ &$ 37.8^{-0.5}_{+0.6}$
\\
$ \optbar{B^0} \to \phi \pi^0$   &$ 16.2^{-3.4}_{+6.2}$ 
&$ 1.0^{-0.4}_{+0.7}$ &$ 10.5^{-2.6}_{+5.1}$ &
$ 19.0^{-2.9}_{+5.0} $ &$ 1.6^{-0.3}_{+0.4}$ &$ 13.8^{-2.1}_{+3.5}$
\\ 
$ \optbar{B^0} \to \phi \eta$   &  
$ 16.2^{-3.4}_{+6.2}$ &$ 1.0^{-0.4}_{+0.7}$ &$ 10.5^{-2.6}_{+5.1}$ &
$ 19.0^{-2.9}_{+5.0} $ &$ 1.6^{-0.3}_{+0.4}$ &$ 13.8^{-2.1}_{+3.5}$
\\ 
$ \optbar{B^0} \to \phi \eta^\prime$  & 
$ 16.2^{-3.4}_{+6.2}$ &$ 1.0^{-0.4}_{+0.7}$ &$ 10.5^{-2.6}_{+5.1}$ &
$ 19.0^{-2.9}_{+5.0} $ &$ 1.6^{-0.3}_{+0.4}$ &$ 13.8^{-2.1}_{+3.5}$
\\ 
$\optbar{B^0} \to \rho^0 K_S^0$        & 
$ 2.1^{+0.5}_{-1.4}$ &$ 0.9^{+0.1}_{-0.4}$ &$-2.0^{-0.8}_{+1.8}$ &
$ 18.7^{+0.6}_{-0.9}$ &$ 58.0^{+0.2}_{-0.2}$ &$ 98.6^{-0.2}_{+0.2}$
\\
$ \optbar{B^0} \to  \phi K_S^0$         &
$-1.7^{-0.1}_{+0.1}$ &$-1.8^{-0.1}_{+0.1}$ &$-2.7^{-0.1}_{+0.1}$ &
$ 67.5^{+0.0}_{-0.1}$ &$ 67.5^{+0.0}_{-0.1}$ &$ 67.9^{+0.2}_{-0.3}$
\\
$\optbar{B^0} \to \omega K_S^0$   & $-5.3^{-1.5}_{+2.4}$ &
$-24.0^{-7.9}_{+13.3}$ &$-3.8^{-0.9}_{+1.6}$ &$ 50.7^{-0.5}_{+0.7}$ &
$ 19.2^{-0.0}_{+1.1}$ &$ 54.8^{-0.3}_{+0.5}$
\\
$\optbar{B^0}\to\rho^+ \rho^-$ & 
$ 4.1^{+1.0}_{-2.2}$ &$ 4.2^{+1.0}_{-2.3}$ &$ 4.2^{+1.0}_{-2.3}$ &
$ 17.1^{-1.1}_{+1.5}$ &$ 16.9^{-1.1}_{+1.5}$ &$ 16.5^{-1.1}_{+1.5}$
\\
$\optbar{B^0} \to \rho^0 \rho^0$   & 
$-8.0^{-3.9}_{+8.5}$ &$ 1.9^{-4.0}_{+7.8}$ &$ 12.1^{+2.2}_{-4.5}$ &
$-97.0^{+1.2}_{-1.3}$ &$-41.4^{-2.6}_{+4.1}$ &$ 88.2^{-1.2}_{+1.6}$\\
$\optbar{ B^0} \to \omega \omega$  &
$ 20.8^{+3.6}_{-6.9}$ &$ 22.0^{+1.6}_{-2.5}$ &$ 4.7^{+1.1}_{-2.5}$ &
$ 95.2^{-1.5}_{+1.9}$ &$ 78.6^{-0.3}_{+0.8}$ &$ 24.4^{-1.1}_{+1.6}$
\\
$\optbar{B^0}\to K^{*0}\bar K^{*0}$ &
$ 15.2^{-3.3}_{+6.1}$ &$ 14.5^{-3.2}_{+5.9}$ &$ 13.4^{-3.1}_{+5.7}$ &
$ 18.1^{-2.8}_{+4.7}$ &$ 17.5^{-2.7}_{+4.5} $ &$ 16.5^{-2.5}_{+4.2} $ \\
$ \optbar{B^0} \to \rho^0 \omega$   &
$ 8.2^{-6.7}_{+12.0}$ &$ 4.5^{-0.8}_{+1.6}$ &$ 8.3^{-3.4}_{+6.5}$ &
$-21.4^{-4.1}_{+7.1}$ &$ 22.8^{-0.6}_{+0.9}$ &$-2.1^{-2.6}_{+4.1}$ \\
$\optbar{B^0} \to \rho^{0} \phi$    &
$ 16.2^{-3.4}_{+6.2}$ &$ 1.0^{-0.4}_{+0.7}$ &$ 10.5^{-2.6}_{+5.1}$ &
$ 19.0 ^{-2.9}_{+5.0} $ &$ 1.6^{-0.3}_{+0.4} $ &$ 13.8^{-2.1}_{+3.5}$ \\
$\optbar{B^0} \to \omega \phi$     &
$ 16.2^{-3.4}_{+6.2}$ &$ 1.0^{-0.4}_{+0.7}$ &$ 10.5^{-2.6}_{+5.1}$ &
$ 19.0 ^{-2.9}_{+5.0} $ &$ 1.6^{-0.3}_{+0.4} $ &$ 13.8^{-2.1}_{+3.5}$ \\
\hline
\hline
\end{tabular}
\end{table}
\begin{table}
\caption{CP-violating asymmetry parameters $a_{\epsilon^\prime}$ and 
$a_{\epsilon +\epsilon^\prime}$ (in percent) for the decays 
$\protect\optbar{B^0} 
 \to h_1h_2$ using $\rho=0.23$, $\eta=0.42$ and $N_c=2,3,\infty$,
for $k^2=m_b^2/2$.
}
\label{cppa2}
\begin{tabular}{||l||r|r|r||r|r|r||}
\hline
\hline
 &   \multicolumn{3}{|c||}{$a_{\epsilon^\prime}$} 
&\multicolumn{3}{|c||}{$a_{\epsilon +\epsilon^\prime}$} \\
\hline
Channel &  $N_c=2$ &  $N_c=3$ & $N_c=\infty$
&  $N_c=2$ &  $N_c=3$ & $N_c=\infty$ \\
\hline
\hline
$\optbar{B^0} \to \pi^+ \pi^-$  &$ 4.9$ & $ 4.9$ & $ 5.0$ & $ 29.3$ & $ 29.1$ 
& $ 28.7$\\
$\optbar{B^0} \to \pi^0 \pi^0$     &$-0.7$ & $ 20.3$ & $ 8.8$ & $-72.2$ & 
$-70.7$ & $ 68.2$\\
$ \optbar{B^0} \to \eta^\prime \eta^\prime$  &$ 19.3$ & $ 42.3$ & $-14.0$ & 
$ 50.8$ & $ 88.3$ & $-66.3$
\\
$ \optbar{B^0} \to \eta \eta^\prime$    &$ 18.6$ & $ 28.8$ & $-11.8$ & $ 75.4$
 & $ 78.7$ & $-82.2$\\
$ \optbar{B^0} \to \eta \eta$    & $ 17.2$ & $ 21.0$ & $-9.7$ & $ 90.2$ & 
$ 66.7$ & $-92.3$\\
$ \optbar{B^0} \to \pi^0 \eta^\prime$    &$ 38.9$ & $ 31.2$ & $ 13.0$ & 
$ 74.2$ & $ 41.1$ & $-28.4$
\\
$ \optbar{B^0} \to \pi^0 \eta$  &$ 22.8$ & $ 19.2$ & $ 10.5$ & $ 57.9$ & 
$ 30.5$ & $-22.3$ \\
$\optbar{B^0} \to K_S^0 \pi^0$        &$ 0.4$ & $-1.5$ & $-4.9$ & $ 90.7$ & 
$ 85.7$ & $ 75.1$\\
$\optbar{B^0} \to K_S^0 \eta^\prime$       &$-3.0$ & $-2.2$ & $-1.1$ & 
$ 81.8$ & $ 83.8$ & $ 86.6$
\\
$\optbar{B^0} \to K_S^0 \eta$         &$ 1.3$ & $-1.2$ & $-5.6$ & $ 92.7$ & 
$ 86.3$ & $ 70.9$\\
$\optbar{B^0}\to K^0\bar K^0$ &$ 17.5$ & $ 17.3$ & $ 16.9$ & $ 22.2$ & $ 22.0$ 
& $ 21.7$\\
$\optbar{B^0} \to \rho^0 \pi^0$    &$-3.6$ & $ 1.9$ & $ 4.8$ & $-26.1$ & 
$-97.0$ & $ 30.7$
\\
$\optbar{B^0} \to \omega \pi^0$    &$ 28.7$ & $ 30.4$ & $ 0.7$ & $ 94.5$ &
 $ 65.6$ & $ 40.1$
\\
$ \optbar{B^0} \to \rho^0 \eta $   & $-19.3$ & $ 18.3$ & $ 26.3$ & $-92.3$ 
& $-22.5$ & $ 85.2$
\\
$ \optbar{B^0} \to \rho^0 \eta^\prime$  & $-39.0$ & $-69.2$ & $ 27.6$ & 
$ 32.1$ & $-53.6$ & $ 27.2$
\\
$ \optbar{B^0} \to \omega \eta $   &$ 12.4$ & $ 24.7$ & $ 1.3$ & $ 60.6$ & 
$ 96.0$ & $ 11.1$
\\
$ \optbar{B^0} \to \omega \eta^\prime$  &$ 12.7$ & $ 32.6$ & $ 1.3$ & $ 37.3$
 & $ 45.7$ & $ 31.1$
\\
$ \optbar{B^0} \to \phi \pi^0$   &$ 22.7$ & $ 1.4$ & $ 14.8$ & $ 26.7$ & 
$ 2.3$ & $ 19.6$
\\ 
$ \optbar{B^0} \to \phi \eta$   &  $ 22.7$ & $ 1.4$ & $ 14.8$ & $ 26.7$ & 
$ 2.3$ & $ 19.6$
\\ 
$ \optbar{B^0} \to \phi \eta^\prime$  & $ 22.7$ & $ 1.4$ & $ 14.8$ & $ 26.7$
 & $ 2.3$ & $ 19.6$
\\ 
$\optbar{B^0} \to \rho^0 K_S^0$        & $ 2.9$ & $ 1.1$ & $-2.0$ & $ 26.3$
 & $75.1$ & $ 98.7$
\\
$ \optbar{B^0} \to  \phi K_S^0$         &$-2.1$ & $-2.2$ & $-3.4$ & $ 84.3$ 
& $84.3$ & $ 84.6$
\\
$\optbar{B^0} \to \omega K_S^0$   & $-7.0$ & $-33.3$ & $-4.9$ & $ 67.1$ & 
$ 26.8$ & $ 71.7$ \\
$B^0\to\rho^+ \rho^-$ & $ 2.9$ & $ 2.9$ & $ 3.0$ & $ 16.4$ & $ 16.2$ & $ 15.9$
\\
$\optbar{B^0} \to \rho^0 \rho^0$   & $-7.1$ & $ 2.6$ & $ 9.7$ & $-82.5$ & 
$-56.8$ & $ 74.2$
\\
$\optbar{ B^0} \to \omega \omega$  &$ 17.8$ & $ 25.3$ & $ 3.3$ & $ 84.8$ & 
$ 91.7$ & $ 21.5$
\\
$\optbar{B^0}\to K^{*0}\bar K^{*0}$ &$ 21.3$ & $ 20.3$ & $ 18.8$ & $ 25.4$ & 
$ 24.6$ & $ 23.3$
 \\
$\optbar{ B^0} \to \rho^0 \omega$   &$ 11.6$ & $ 6.3$ & $ 11.8$ & $-30.4$ & 
$ 32.1$ & $-3.0$
 \\
$\optbar{B^0} \to \rho^{0} \phi$    &$ 22.7$ & $ 1.4$ & $ 14.8$ & $ 26.7$ & 
$ 2.3$ & $ 19.6$
 \\
$\optbar{B^0} \to \omega \phi$     &$ 22.7$ & $ 1.4$ & $ 14.8$ & $ 26.7$ & 
$ 2.3$ & $ 19.6$
 \\
\hline
\hline
\end{tabular}
\end{table}

\section{Numerical Results for CP-Violating Coefficients and $A_{CP}$}

Given the amplitudes ${\cal M}$ and $\overline{\cal M}$, one can 
calculate the CP-violating asymmetry $A_{CP}$  
for all the $B \to PP$, $B \to PV$ and $B \to VV$ decay 
modes and their charged conjugates whose branching ratios were calculated
by us recently in 
the factorization approach \cite{akl98-1}. The asymmetries
depend on several variables, such as the CKM parameters,
$N_c$, the virtuality $k^2$ discussed earlier, and the scale $\mu$.
The scale dependence of $A_{CP}$ is important in only a few decays and we
shall estimate it by varying $\mu$ between $\mu=m_b/2$ and $\mu=m_b$
at the end of this section for these decays and fix the scale at 
$\mu=m_b/2$. The dependence on the
rest of the parameters is worked out explicitly. We show the results for
$N_c=2,3,\infty$, for two representative choices of the CKM parameters
in the tables:
%
%
\begin{table}
\begin{center}
\caption{CP-violating asymmetry parameters
$a_{\epsilon^\prime}$, $a_{\bar \epsilon^\prime}$, 
$a_{\epsilon+\epsilon^\prime}$, $a_{\epsilon+\bar \epsilon^\prime}$
defined in eq.~(\ref{aepsilon}) for the decays $\protect\optbar{B^0} \to
\rho^-\pi^+$, 
$\protect\optbar{B^0} \to \rho^+\pi^-$, and
 $\protect\optbar{B^0} \to \bar{K}^{*0}K_S^0$, 
$\protect\optbar{B^0} \to K^{*0} {K}^0_S$, (in percent), using
$\rho=0.12$,
$\eta=0.34$ and $k^2=m_b^2/2\pm 2$ GeV$^2$. }
\label{aeep}
\begin{tabular} {|l|l|r|r|r|r|}
\hline
Channel&$N_c$&
$a_{\epsilon^\prime}$& $a_{\bar \epsilon^\prime}$& $a_{\epsilon+\epsilon 
^\prime}$& $a_{\epsilon+\bar \epsilon^\prime}$\\
\hline
  &$N_c=2$ & $-54.9^{+0.6}_{-1.3}$ &  $ 58.6^{+0.4}_{-0.8}$ 
&$ 6.0^{-0.4}_{+0.4}$ & $ 6.2^{-0.7}_{+1.2}$ \\
$\optbar{B^0} \to\rho^-\pi^+, \rho^+ \pi^- $ &$N_c=3$ &
$-54.9^{+0.6}_{-1.3}$ & $58.7^{+0.3}_{-0.9}$
 &$ 5.8^{-0.4}_{+0.5}$ & $6.0^{-0.7}_{+1.3}$  \\
 &   $N_c=\infty$ & $-54.9^{+0.6}_{-1.3}$ &  $58.7^{+0.3}_{-0.9}$ 
&$ 5.6^{-0.5}_{+0.4}$ & $5.8^{-0.8}_{+1.2}$ \\
\hline
 & $N_c=2$ & $99.3^{-0.2}_{+0.2}$ & $-99.1^{+0.4}_{-0.5}$ 
&$ -5.3^{-2.9}_{+4.8}$ &  $ 10.0^{+2.2}_{-3.9}$ \\
$ \optbar{B^0} \to \bar K^{*0}  K_S^0, K^{*0}{K}_S^0$ &$N_c=3$ &$
99.9^{-0.1}_{+0.0}$ & 
$ -99.6^{+0.1}_{-0.2}$ & $-3.2^{-2.3}_{+3.9}$ & $8.8^{+1.5}_{-2.6} $\\
 & $N_c=\infty$ &  $ 99.8^{-0.1}_{+0.0}$ & $-99.1^{-0.1}_{+0.2}$ 
& $ -0.4^{-1.5}_{+2.7}$ & $7.2^{+0.5}_{-0.9}$\\
\hline
\end{tabular}\end{center}
\end{table}
\begin{table}
\begin{center}
\caption{CP-violating asymmetry parameters
$a_{\epsilon^\prime}$, $a_{\bar \epsilon^\prime}$, 
$a_{\epsilon+\epsilon^\prime}$, $a_{\epsilon+\bar \epsilon^\prime}$
defined in eq.~(\ref{aepsilon}) for the decays $\protect\optbar{B^0} \to
\rho^-\pi^+$,
$\protect\optbar{B^0} \to \rho^+\pi^-$, and
 $\protect\optbar{B^0} \to \bar{K}^{*0}K_S^0$,
$\protect\optbar{B^0} \to K^{*0} {K}_S^0$, (in percent), 
using $\rho=0.23$, $\eta=0.42$ and $k^2=m_b^2/2$. }
\label{aeep2}
\begin{tabular} {|l|l|r|r|r|r|}
\hline
Channel&$N_c$&
$a_{\epsilon '}$& $a_{\bar \epsilon^\prime}$& 
$a_{\epsilon+\epsilon^\prime}$& $a_{\epsilon+\bar \epsilon^\prime}$\\
\hline
  &$N_c=2$ & $-55.5$ &  $ 58.1$ &$ 7.8$ & $ 8.1$ \\
$\optbar{B^0} \to\rho^-\pi^+, \rho^+ \pi^- $ &
 $N_c=3$ & $-55.5$ & $58.1$ &$ 7.6$ & $8.0$  \\
 &   $N_c=\infty$ & $-55.5$ &  $58.1$ &$ 7.5$ & $7.8$ \\
\hline
 & $N_c=2$ & $99.4$ & $-99.0$ &$-4.4$ &  $ 11.1$ \\
$\optbar{B^0} \to \bar K^{*0}  K_S^0, K^{*0}{K}_S^0$
&$N_c=3$ &$ 99.9$ & $-99.5$ & $-2.2$ & $10.3 $\\
 & $N_c=\infty$ &  $ 99.8$ & $-98.8$ & $ 0.8$ & $9.0$\\
\hline
\end{tabular}\end{center}
\end{table}
\begin{itemize}
\item Central values emerging from the CKM unitarity fits of the existing
data, yielding: $\rho =0.12, \eta=0.34$ \cite{aliapctp97}.
\item For values of $\rho$ and $\eta$ which correspond to their
central values $+1\sigma$, yielding $\rho=0.23$ and $\eta=0.42$.
\cite{aliapctp97}.
\end{itemize}
For each decay mode and given a value of $N_c$, the errors shown on 
the numbers in the tables reflect the uncertainties due to the variation of
$k^2$ in the range $k^2=(m_b^2/2~ \pm 2 )$ GeV$^2$.
For some selected CP-asymmetries, we show
in figures, however, the dependence
on the CKM parameters for a wider range of $\rho$ and $\eta$ which
are allowed by the present $95\%$ C.L. unitarity fits \cite{aliapctp97}.

%
%
%

%
%
\begin{table}
\begin{center}
\caption{CP-violating asymmetries $A_{CP}$ in $\protect\optbar{B} \to PP$ 
decays (in percent) using 
$\rho=0.12$, $\eta=0.34$ and $N_c=2,3,\infty$,
for $k^2=m_b^2/2\pm 2$ GeV$^2$.}
\label{cpp1}
\begin{tabular}{|l|c|c|r|r|r|}
\hline
Channel & Class & CP-Class &  $N_c=2$ &  $N_c=3$ & $N_c=\infty$ \\
\hline
$\optbar{B^0} \to \pi^+ \pi^-$ &I & (ii) &$ 21.3^{+0.3}_{-1.2}$  
& $ 21.2^{+0.3}_{-1.2}$ &$ 21.0^{+0.3}_{-1.3}$\\
$\optbar{B^0} \to \pi^0 \pi^0$     &II & (ii)&$-42.0^{-0.2}_{+1.6}$  
& $-15.1^{-3.1}_{+5.3} $ &$ 45.5^{-0.8}_{+0.4}$\\

$ \optbar{B^0} \to \eta^\prime \eta^\prime$  &II & (ii)&$
46.9^{+1.4}_{-4.5}$    
&$ 62.0^{+1.4}_{-2.3}$ &$-52.0^{-2.6}_{+8.0}$\\
$\optbar{ B^0} \to \eta \eta^\prime$    &II & (ii)&$ 57.1^{+1.6}_{-4.1}$
&$ 45.1^{-0.2}_{+0.7}$ &$-54.6^{-3.1}_{+8.3}$\\
$\optbar{ B^0} \to \eta \eta$    &II & (ii)& $ 59.0^{+1.6}_{-3.2}$
&$ 34.6^{-0.8}_{+1.8}$ &$-53.9^{-3.5}_{+8.0}$\\

$ B^\pm \to \pi^\pm \pi^0$ &III & (i)&$ 0.1^{+0.0}_{-0.1}$ 
& $0.0^{+0.1}_{-0.0}$& $0.0^{+0.0}_{-0.0}$\\

$ B^\pm \to \pi^\pm \eta^\prime$ &III & (i)&$ 12.0^{+2.6}_{-5.9}$ 
&$ 14.5^{+3.2}_{-6.7}$ &$ 21.3^{+4.2}_{-8.4}$\\
$ B^\pm \to \pi^\pm \eta$  &III & (i)&$ 11.8^{+2.4}_{-5.3}$ 
&$ 14.0^{+2.8}_{-5.9}$ &$ 19.1^{+3.3}_{-6.4}$\\

$ \optbar{B^0} \to \pi^0 \eta^\prime$    &V & (ii)&$ 48.6^{-0.7}_{+1.4}$ 
&$ 29.2^{-3.5}_{+6.0}$ &$-3.6^{-6.7}_{+11.9}$\\
$\optbar{ B^0 }\to \pi^0 \eta$    &V & (ii)& $ 31.7^{-1.4}_{+2.9}$ 
&$ 19.4^{-2.8}_{+5.0}$ &$-2.6^{-4.8}_{+8.4}$\\

$ B^\pm \to K^\pm \pi^0$ &IV & (i)&$-7.1^{-2.1}_{+3.7}$    
& $-6.3^{-1.8}_{+3.2}$& $-4.9^{-1.3}_{+2.3}$\\
$\optbar{B^0} \to K^\pm \pi^\mp$  &IV & (i)&$-7.7^{-2.3}_{+4.0}$    
&$-7.9^{-2.3}_{+4.2}$ & $-8.2^{-2.4}_{+4.4}$\\
$\optbar{B^0} \to K_S^0 \pi^0$        &IV & (ii)&$ 36.0^{+0.5}_{-1.0}$  
&$ 32.0^{+0.1}_{-0.2}$ & $ 25.1^{-0.7}_{+1.1}$\\
$ B^\pm \to K^\pm \eta^\prime$ &IV & (i)&$-4.9^{-1.2}_{+2.1}$  &
$ -4.1^{-1.0}_{+1.6}$ &$-3.0^{-0.5}_{+1.0}$\\
$\optbar{B^0} \to K_S^0 \eta^\prime$       &IV & (ii)&$ 29.2^{-0.2}_{+0.4}$ 
 &$ 30.7^{-0.1}_{+0.0}$& $ 32.8^{+0.1}_{-0.3}$\\
$ B^\pm \to  K^\pm \eta$ &IV & (i)&$ 8.5^{+3.4}_{-6.3}$ &$
6.2^{+2.6}_{-4.8}$
& $ 2.8^{+1.4}_{-2.6}$\\
$\optbar{B^0} \to K_S^0 \eta$         &IV & (ii)&$ 37.8^{+0.7}_{-1.3}$  
&$ 32.5^{+0.1}_{-0.3}$ & $ 22.9^{-0.9}_{+1.5}$\\
$ B^\pm \to \pi^\pm K_S^0$   &IV & (i)&$-1.4 ^{-0.1}_{+0.0}$ 
&$-1.4^{-0.1}_{+0.0}$  & $-1.4^{-0.0}_{+0.1}$\\
$ B^\pm \to K^\pm K_S^0$ &IV & (i)&$ 12.5^{-2.9}_{+5.5}$ 
&$ 12.3^{-2.9}_{+5.5}$ & $ 12.0^{-2.8}_{+5.5}$\\
$\optbar{B^0}\to K^0\bar K^0$ &IV & (ii)&$ 15.6^{-3.0}_{+5.6}$ 
 &$ 15.5^{-3.1}_{+5.4}$ & $ 15.1^{-2.9}_{+5.5}$\\
\hline
\end{tabular}\end{center}
\end{table}

\begin{table}[htb]
\begin{center}
\caption{CP-violating asymmetries $A_{CP}$ in $\protect\optbar{B} \to PP$
decays  (in percent) using
$\rho=0.23$, 
$\eta=0.42$ and $N_c=2,3,\infty$ for $k^2=m_b^2/2$.
}
\label{cpp2}
\begin{tabular}{|l|c|c|r|r|r|}
\hline
Channel & Class & CP-Class &  $N_c=2$ &  $N_c=3$ & $N_c=\infty$ \\
\hline
$\optbar{B^0} \to \pi^+ \pi^-$ &I & (ii) &$ 17.2$ & $ 17.1$ & $ 16.9$\\
$\optbar{B^0} \to \pi^0 \pi^0$     &II & (ii)& $-34.8$ & $-20.4$ & $ 38.2$\\

$\optbar{ B^0} \to \eta^\prime \eta^\prime$  &II & (ii)&$ 36.8$ & $ 69.7$ &
$-40.7$\\
$\optbar{ B^0} \to \eta \eta^\prime$    &II & (ii)&$ 48.0$ & $ 56.3$ &
$-46.8$\\
$\optbar{ B^0} \to \eta \eta$    &II & (ii)& $ 54.2$ & $ 45.4$ & $-50.2$\\

$ B^\pm \to \pi^\pm \pi^0$ &III & (i)&$0.0$ & $0.0$ & $0.0$\\

$ B^\pm \to \pi^\pm \eta^\prime$ &III & (i)&$ 8.5$ & $ 10.5$ & $ 16.7$\\
$ B^\pm \to \pi^\pm \eta$  &III & (i)&$ 8.7$ &$ 10.6$ & $ 16.2$\\

$ \optbar{B^0} \to \pi^0 \eta^\prime$    &V & (ii)&$ 60.7$ & $ 39.9$ &
$-5.0$\\
$\optbar{ B^0} \to \pi^0 \eta$    &V & (ii)& $ 42.5$ & $ 27.1$ & $-3.7$\\

$ B^\pm \to K^\pm \pi^0$ &IV & (i)&$-9.8$ & $-8.6$ & $-6.5$\\
$\optbar{B^0} \to K^\pm \pi^\mp$  &IV & (i)&$-10.5$ & $-10.8$ & $-11.2$\\
$\optbar{B^0} \to K_S^0 \pi^0$        &IV & (ii)&$ 43.4$ & $ 39.8$ & $
32.5$\\
$ B^\pm \to K^\pm \eta^\prime$ &IV & (i)&$-6.3$ & $-5.3$ & $-3.8$ \\
$\optbar{B^0} \to K_S^0 \eta^\prime$       &IV & (ii)&$ 36.9$ & $ 38.4$ & $
40.5$\\
$ B^\pm \to  K^\pm \eta$ &IV & (i)&$ 8.4$ & $ 6.4$ & $ 3.1$ \\
$\optbar{B^0} \to K_S^0 \eta$         &IV & (ii)&$ 45.0$ & $ 40.2$ & $ 30.0$
\\
$ B^\pm \to \pi^\pm K_S^0$   &IV & (i)&$-1.8$ & $-1.7$ & $-1.7$ \\
$ B^\pm \to K^\pm  K_S^0$ &IV & (i)& $ 17.5$ & $ 17.3$ & $ 16.9$\\
$\optbar{B^0}\to K^0\bar K^0$ &IV & (ii)& $ 22.1$ & $ 21.8$ & $ 21.4$\\
\hline
\end{tabular}\end{center}
\end{table}

 The results are presented taking into account the following considerations.
 For decays belonging to the CP class-(i), the CP-asymmetry is 
time-independent.
Hence for this class, $A_{CP}=a_{\epsilon^\prime}$. For the CP class-(ii) 
and CP class-(iii) decays, the measurement of $A_{CP}$ will be done in
terms of the coefficients $a_{\epsilon^\prime}$ and 
$a_{\epsilon+\epsilon^\prime }$, which are the measures of direct and 
indirect (or mixing-induced) CP-violation, respectively. In view of this,
we give in Tables~\ref{cppa} and \ref{cppa2} these coefficients for the
thirty decay modes of the $B^0$ and $\bar{B}^0$ mesons, which belong
to these classes, for the two sets of CKM parameters given above. 
For the four decays belonging to the CP class-(iv) decays, one would
measure by time-dependent decay rates the quantities $a_{\epsilon^\prime}$,
$a_{\epsilon+\epsilon^\prime }$, $a_{\bar{\epsilon}^\prime}$ and 
$a_{\epsilon+\bar{\epsilon}^\prime }$. They are given in 
Tables~\ref{aeep} and \ref{aeep2} for the two sets of CKM parameters, 
respectively. Having worked out these quantities, we then give the
numerical values of the CP-violating asymmetries for all the seventy six
decays $B \to PP$, $B \to
PV~(b \to d$ transition), $B \to PV~(b \to s$ transition) and $B \to VV$
in Tables \ref{cpp1} - \ref{cvv2}.

\begin{table}[htb]\begin{center}
\caption{CP-violating asymmetries $A_{CP}$ in $\protect\optbar{B}\to PV$ 
decays ($b\to d$ transition) 
(in percent) using $\rho=0.12$, $\eta=0.34$ and 
$N_c=2,3,\infty$ for $k^2=m_b^2/2\pm 2$ GeV$^2$. }
\label{cpv1}
\begin{tabular} {|l|c|c|r|r|r|}
\hline
Channel & Class&   CP-Class&  $N_c=2$ &  $N_c=3$ & $N_c=\infty$ \\
\hline
$B^0/\bar{B}^0 \to\rho^-\pi^+/\rho^+\pi^- $ &I & (iv)&$ 3.6^{-0.7}_{+1.2}$
&$ 3.5^{-0.7}_{+1.2}$  &$ 3.3^{-0.7}_{+1.2}$  \\
$B^0/\bar{B}^0 \to \rho^+  \pi^-/\rho^-\pi^+ $&I & (iv) &$
6.0^{+0.7}_{-1.9}$&
$ 5.9^{+0.8}_{-1.8}$& $ 5.9^{+0.7}_{-1.9}$\\

$\optbar{B^0} \to \rho^0 \pi^0$  &II & (ii)  &$-23.5^{-1.3}_{+4.2}$
& $-49.4^{-10.3}_{+22.1}$& $ 22.2^{+0.7}_{-2.4}$\\
$\optbar{B^0} \to \omega \pi^0 $  &II & (ii)  &$ 57.5^{+1.4}_{-2.3}$ 
&$ 39.2^{-1.3}_{+2.5}$ &$ 24.3^{+0.1}_{-0.2}$\\

$\optbar{ B^0} \to \rho^0 \eta $ &II & (ii)  & $-59.5^{-7.0}_{+17.1}$ 
&$ 0.9^{-8.1}_{+14.0}$ &$ 64.4^{+2.4}_{-5.0}$ \\
$\optbar{ B^0} \to \rho^0 \eta^\prime$ &II & (ii) & $-16.5^{+0.4}_{+9.7}$ 
&$-56.7^{+15.5}_{+28.2}$ &$ 40.2^{+1.4}_{-6.2}$ \\
$\optbar{ B^0} \to \omega \eta $ &II & (ii)  & $ 46.2^{+0.9}_{-2.9}$  & 
$ 61.5^{+1.9}_{-3.5}$& $ 5.7^{+0.1}_{-0.3}$ \\
$\optbar{ B^0} \to \omega \eta^\prime$&II & (ii)  & $ 33.5^{+0.8}_{-3.1}$ & 
$ 54.9^{+1.9}_{-6.7}$& $ 19.2^{+0.1}_{-0.4}$ \\

$B^\pm \to \rho^0\pi^\pm$ &III & (i)&$-3.9^{-1.1}_{+2.6} $
&$-5.2^{-1.5}_{+3.5}$
&$-11.0^{-3.8}_{+8.8}$\\
$B^\pm\to \rho^\pm \pi^0$ &III & (i)&$ 2.7^{+0.6}_{-1.5}$ &$
3.0^{+0.7}_{-1.6}$
&$ 3.6^{+0.9}_{-1.9}$ \\
$B^\pm \to \omega\pi^\pm$ &III & (i)&$ 9.8^{+2.2}_{-4.8}$ &$
7.9^{+1.9}_{-4.0}$
&$-1.8^{-0.6}_{+1.2}$ \\

$B^\pm \to \rho^\pm \eta$ &III & (i)&$ 3.9^{+0.9}_{-2.2}$ &$
4.4^{+1.1}_{-2.4}$
&$ 5.7^{+1.4}_{-3.0}$ \\
$B^\pm \to \rho^\pm\eta^\prime$ &III & (i)&$ 3.8^{+1.0}_{-2.2}$ 
&$ 4.3^{+1.2}_{-2.5}$ & $ 5.6^{+1.4}_{-3.2}$ \\

$ \optbar{B^0} \to \bar K^{*0}  K_S^0/ K^{*0} K_S^0$ &IV  & (iv)&  $
15.9^{-3.4}_{+6.2}$ 
&$ 15.3^{-3.3}_{+6.0}$ &$ 14.3^{-3.2}_{+5.9}$\\
$\optbar{ B^0} \to  K^{*0}   K_S^0/\bar K^{*0} K_S^0$ & V & (iv)&
$-12.2^{+3.0}_{-5.7}$ 
&$-10.6^{+2.6}_{-5.2}$  &$-8.2^{+2.2}_{-4.3}$ \\

$ B^\pm \to K^\pm  \optbar{K^{*0}}$ &IV & (i)& $ 15.2^{-3.3}_{+6.1}$ 
&$ 14.5^{-3.2}_{+5.9}$ &$ 13.4^{-3.1}_{+5.7}$\\

$ B^\pm \to K^{*\pm}   K_S^0$ & V & (i)&  $-1.2^{+5.6}_{-32.8}$ 
&$ 46.8^{-13.2}_{-3.3}$ &$ 48.1^{-5.6}_{+4.8}$\\
$ B^\pm \to \phi \pi^\pm$  & V & (i)& $ 16.2^{-3.4}_{+6.2}$ 
&$ 1.0^{-0.4}_{+0.7}$ &$ 10.5^{-2.6}_{+5.1}$\\
$\optbar{B^0} \to \phi \pi^0$  & V & (ii)& $ 19.6^{-3.6}_{+6.5}$   
&$ 1.4^{-0.3}_{+0.7}$ &$ 13.4^{-2.7}_{+5.0}$\\ 
$\optbar{B^0} \to \phi \eta$  & V & (ii)&    $ 19.6^{-3.6}_{+6.5}$   
&$ 1.4^{-0.3}_{+0.7}$ &$ 13.4^{-2.7}_{+5.0}$\\ 
$\optbar{B^0} \to \phi \eta^\prime$  & V & (ii)&   $ 19.6^{-3.6}_{+6.5}$   
&$ 1.4^{-0.3}_{+0.7}$ &$ 13.4^{-2.7}_{+5.0}$\\ 
\hline
\end{tabular}\end{center}
\end{table}

\begin{table}[htb]\begin{center}
\caption{CP-violating asymmetries $A_{CP}$ in $\protect\optbar{B} \to PV$ 
decays ($b\to d$ transition)
(in percent) 
using $\rho=0.23$, $\eta=0.42$ and 
$N_c=2,3,\infty$ for $k^2=m_b^2/2$. 
}
\label{cpv1p}
\begin{tabular} {|l|c|c|r|r|r|}
\hline
Channel & Class&   CP-Class&  $N_c=2$ &  $N_c=3$ & $N_c=\infty$ \\
\hline
$B^0/\bar{B}^0 \to\rho^-\pi^+/\rho^+\pi^- $ &I & (iv)&$ 5.3$ & $ 5.2$ & $
5.1$ \\
$B^0/\bar{B}^0 \to \rho^+  \pi^-/\rho^-\pi^+ $&I & (iv) &$ 5.4$ & $ 5.4$ & $
5.4$\\

$\optbar{B^0} \to \rho^0 \pi^0$  &II & (ii)  &$-14.8$ & $-44.9$ & $ 17.8$\\
$\optbar{B^0} \to \omega \pi^0 $  &II & (ii) &$ 63.7$ & $ 51.1$ & $ 19.5$\\

$\optbar{B^0} \to \rho^0 \eta $ &II & (ii) & $-56.5$ & $ 1.3$ & $ 57.7$ \\
$\optbar{B^0} \to \rho^0 \eta^\prime$ &II & (ii)& $-10.2$ & $-70.8$ & $
31.0$ \\
$\optbar{B^0} \to \omega \eta $ &II & (ii) & $ 36.9$ & $ 61.8$ & $ 6.1$ \\
$\optbar{B^0} \to \omega \eta^\prime$&II & (ii) & $ 26.1$ & $ 43.0$ & $
15.7$ \\

$B^\pm \to \rho^0\pi^\pm$ & III & (i)&$-2.7$ & $-3.7$ & $-8.3$ \\
$B^\pm\to \rho^\pm \pi^0$ & III & (i)&$ 1.9$ & $ 2.1$ & $ 2.6$ \\
$B^\pm \to \omega\pi^\pm$ & III & (i)&$ 7.0$ & $ 5.6$ & $-1.3$ \\

$B^\pm \to \rho^\pm \eta$ & III & (i)&$ 2.7$ & $ 3.1$ & $ 4.0$ \\
$B^\pm \to \rho^\pm\eta^\prime$ & III & (i)&$ 2.7$ & $ 3.0$ & $ 3.9$ \\

$B^0/\bar{B}^0 \to \bar K^{*0}  K_S^0/ K^{*0} K_S^0 $&IV & (iv)& $ 22.3$ & 
$21.4$ & $ 20.1$\\
$B^0/\bar{B}^0 \to  K^{*0}   K_S^0/\bar K^{*0} K_S^0$&V & (iv) & $-17.0$ & 
$-14.9$ &
$-11.5$\\

$ B^\pm \to K^\pm  \optbar{K^{*0}}$ &IV & (i)& $ 21.3$ & $ 20.3$ & $ 18.8$\\

$B^\pm \to K^{*\pm}  K_S^0$ & V & (i)& $-1.6$ & $ 54.6$ & $ 62.5$\\
$ B^\pm \to \phi \pi^\pm$  & V & (i)& $ 22.7$ & $ 1.4$ & $ 14.8$ \\
$ \optbar{B^0} \to \phi \pi^0$  & V & (ii)& $ 27.5$ & $ 2.0$ & $ 19.0$\\ 
$\optbar{B^0} \to \phi \eta$  & V & (ii)&    $ 27.5$ & $ 2.0$ & $ 19.0$\\ 
$\optbar{B^0} \to \phi \eta^\prime$  & V & (ii)&   $ 27.5$ & $ 2.0$ & $
19.0$\\ 
\hline
\end{tabular}\end{center}
\end{table}

\begin{table}[htb]\begin{center}
\caption{CP-violating asymmetries $A_{CP}$ in $\protect\optbar{B}\to PV$
decays ($b\to s$ transition) (in percent) 
 using $\rho=0.12$, $\eta=0.34$ and $N_c=2,3,\infty$ 
for $k^2=m_b^2/2\pm 2$ GeV$^2$. 
 }\label{cpv3}
\begin{tabular} {|l|c|c|r|r|r|}
\hline
Channel & Class& CP-Class & $N_c=2$ &  $N_c=3$ & $N_c=\infty$ \\
\hline
$\optbar{B^0} \to \rho^\mp K^\pm$  &I & (i)& $-16.4^{-3.9}_{+9.7}$ 
&$-16.4^{-3.9}_{+9.8}$ & $-16.3^{-3.9}_{+9.7}$\\
$B^\pm \to K^{*\pm}\eta^\prime$ &III & (i)& $-72.6 ^{-5.1}_{+35.9}$ 
&$-84.3^{-8.6}_{+44.0}$ &$-61.5^{-16.1}_{+36.5}$\\

$\optbar{B^0} \to K^{*\pm} \pi^\mp$  &IV & (i)& $-15.5^{-5.0}_{+8.9}$ 
&$-15.9^{-5.2}_{+9.2}$ & $-16.6^{-5.4}_{+9.6}$\\

$\optbar{B^0} \to K^{*0} \pi^0$     &V & (i)& $ 1.4^{+1.2}_{-2.2}$ 
&$-1.3^{+0.1}_{-0.4}$ &$-4.9^{-1.3}_{+2.0}$\\

$B^\pm \to K^{*\pm}\pi^0$ &IV & (i)&$-12.8^{-3.9}_{+7.3}$
&$-12.0^{-3.7}_{+6.8}$
&$-10.5^{-3.2}_{+5.8}$\\
$ B^\pm \to\rho^0 K^\pm $ &IV & (i)& $-13.2^{-3.2}_{+6.8}$
&$-12.8^{-3.2}_{+6.7}$
& $-7.5^{-2.0}_{+3.8}$\\

$ B^\pm \to K^{*\pm} \eta$ &IV & (i)&  $-9.1^{-2.7}_{+5.1}$ 
&$-9.3^{-2.8} _{+5.2}$  &$-9.6^{-2.9}_{+5.3}$\\
$ \optbar{B^0} \to\optbar{ K^{*0}} \eta$   &V & (i)  &  $-2.4^{-0.5}_{+0.9}$ 
&$-1.4^{-0.1}_{+0.2}$ &$ 0.6^{+0.5}_{-1.1}$\\

$B^\pm \to \optbar {K^{*0}} \pi^\pm$ &IV & (i)& $-1.7\mp0.1$ &$-1.6\mp0.1$
& $-1.5^{-0.1}_{+0.0}$\\

$\optbar{B^0} \to \rho^0 K_S^0$        &V & (ii)& $ 10.2^{+0.7}_{-1.3}$ 
&$ 28.2^{+0.1}_{-0.3}$ &$ 45.5^{-0.5}_{+1.4}$\\
$ B^\pm \to \rho^\pm K_S^0$  &V & (i) &$ 3.0^{-0.4}_{+0.8}$ & 
$ 4.6^{-1.1}_{+4.8}$& $-4.4^{+0.6}_{-0.5}$\\
$\optbar{ B^0} \to \optbar{K^{*0}} \eta^\prime$  &V & (i)&
$-44.0^{+11.9}_{-48.0}$ 
&$-13.3^{+2.1}_{+0.5}$ & $ 8.0^{+4.6}_{-7.5}$\\
$ B^\pm \to \phi K^\pm$   &V & (i)& $-1.7\mp0.1$ &$-1.8\mp0.1$
&$-2.7^{-0.1}_{+0.1}$\\
$\optbar{B^0} \to \phi K_S^0$    &V & (ii)    &$ 31.0^{-0.1}_{+0.0}$ 
&$ 30.9^{-0.0}_{+0.0}$ &$ 30.5^{+0.0}_{-0.1}$\\
$\optbar{B^0} \to \omega K_S^0$  &V & (ii)& $ 20.6^{-1.2}_{+1.9}$ 
&$-6.6^{-5.2}_{+9.2}$ &$ 23.6^{-0.8}_{+1.3}$\\
$ B^\pm \to \omega K^\pm $ &V & (i)&  $-13.1^{-4.1}_{+7.4} $ 
&$-19.6^{-4.7}_{+11.1}$ &$ 0.9^{+0.7}_{-1.3}$\\
\hline
\end{tabular}\end{center}
\end{table}

\begin{table}[htb]\begin{center}
\caption{CP-violating asymmetries $A_{CP}$ in $\protect\optbar{B} \to PV$
decays ($b\to s$ transition) (in percent)  
using $\rho=0.23$, $\eta=0.42$ and $N_c=2,3,\infty$ for $k^2=m_b^2/2$. 
 }\label{cpv3p}
\begin{tabular} {|l|c|c|r|r|r|}
\hline
Channel & Class& CP-Class & $N_c=2$ &  $N_c=3$ & $N_c=\infty$ \\
\hline
$\optbar{B^0} \to \rho^\mp K^{\pm}$  &I & (i)& $-11.5$ & $-11.5$ & $-11.4$\\
$B^\pm \to K^{*\pm}\eta^\prime$ &III & (i)& $-55.2$ & $-71.7$ & $-75.2$\\

$\optbar{B^0} \to K^{*\pm} \pi^\mp$ &IV & (i)& $-22.1$ & $-22.6$ & $-23.6$\\

$\optbar{B^0} \to \optbar{K^{*0}} \pi^0$  &V & (i)& $ 1.6$ & $-1.6$ &
$-6.3$\\

$B^\pm \to K^{*\pm}\pi^0$ &IV & (i)& $-18.2$ & $-17.1$ & $-14.8$\\
$ B^\pm \to\rho^0 K^\pm $ &IV & (i)& $-14.5$ & $-15.9$ & $-10.7$\\

$ B^\pm \to K^{*\pm} \eta$ &IV & (i)& $-13.0$ & $-13.3$ & $-13.5$\\
$ \optbar{B^0} \to \optbar{K^{*0}} \eta$ &V & (i)& $-3.1$ & $-1.7$ & $
0.7$\\

$B^\pm \to \optbar{K^{*0}} \pi^\pm$ &IV & (i)& $-2.1$ & $-2.0$ & $-1.9$\\

$\optbar{B^0} \to \rho^0 K_S^0$   &V & (ii)& $ 14.4$ & $ 36.4$ & $ 45.6$\\
$ B^\pm \to \rho^\pm K_S^0$  &V & (i) &$ 3.7$ & $ 5.6$ & $-5.3$\\
$ \optbar{B^0} \to \optbar{K^{*0}} \eta^\prime$ &V & (i)& $-47.7$ & $-16.3$
& $ 9.0$\\
$ B^\pm \to \phi K^\pm$    &V & (i)& $-2.1$ & $-2.2$ & $-3.4$\\
$ \optbar{B^0} \to \phi K_S^0$ &V & (ii)    &$ 38.7$ & $ 38.7$ & $ 38.0$\\
$\optbar{B^0} \to \omega K_S^0$  &V & (ii)& $ 27.3$ & $-9.1$ & $ 30.9$\\
$ B^\pm \to \omega K^\pm $ &V & (i)&  $-18.6$ & $-15.1$ & $ 1.1$\\
\hline
\end{tabular}\end{center}
\end{table}

\subsection{Parametric Dependence of CP-violating Parameters and $A_{CP}$}

We now discuss the CP-asymmetries given in Tables~\ref{cppa} - \ref{cvv2}
and their parametric dependence.
\begin{itemize}
\item {\bf Form factor dependence of $\mathbf A_{CP}$}:
The CP-violating asymmetries depend very weakly on the form factors. We
have calculated the CP-violating asymmetries for the form factors based on
both the BSW \cite{BSW87} and the 
hybrid Lattice-QCD/QCD-SR models. The form factor values
used  are given in \cite{akl98-1}.
However, the dependence of $A_{CP}$ on the form
factors is weak. Hence, we show results only for the former case.
\item{\bf $\mathbf N_c$-dependence of $\mathbf A_{CP}$}: 
The classification of the $B \to h_1 h_2$ decays using $N_c$-dependence of 
the rates that we introduced in \cite{akl98-1} is also very useful in
discussing $A_{CP}$.
 We see that the CP-asymmetries in the class-I 
and class-IV decays have mild dependence on $N_c$,
reflecting very much the characteristics of the decay rates. As already 
remarked, this 
can be traced back to the $N_c$-dependence of the effective coefficients.
However, in some decays classified as class-IV decays in \cite{akl98-1}, we 
have found that $A_{CP}$ shows a marked $N_c$ dependence. All these cases
are on the borderline as far as the $N_c$-sensitivity of the decay rates is 
concerned due to the presence of several competing amplitudes.
The decays, which were classified in \cite{akl98-1} as class-IV decays but 
are now classified as class-V decays,  are as follows:

\begin{table}[htb]\begin{center}
\caption{CP-violating asymmetries $A_{CP}$ in $\protect\optbar{B} \to VV$
decays (in percent)  
 using $\rho=0.12$, $\eta=0.34$ and $N_c=2,3,\infty$ 
for $k^2=m_b^2/2\pm 2$ GeV$^2$. 
}\label{cvv1}
\begin{tabular} {|l|c|c|r|r|r|}
\hline
Channel &Class&CP-Class& $N_c=2$ &  $N_c=3$ &  $N_c=\infty$\\
\hline
$\optbar{B^0}\to\rho^+ \rho^-$ &I & (iii)& $ 10.8^{+0.2}_{-0.7}$ 
& $ 10.8^{+0.1}_{-0.8}$ &$ 10.6^{+0.1}_{-0.8}$\\
$\optbar{B^0} \to \rho^0 \rho^0$   &II & (iii)& $-51.4^{-1.9}_{+5.0}$ 
&$-18.5^{-3.8}_{+7.1}$ & $ 49.9^{+0.8}_{-2.2}$\\
$\optbar{ B^0} \to \omega \omega$  &II & (iii)& $ 58.9^{+1.6}_{-3.6}$
& $ 51.8^{+0.9}_{-1.2}$& $ 14.7^{+0.2}_{-0.9}$\\
$B^\pm \to \rho^\pm\rho^0$ &III & (i)& $ 0.2^{+0.1}_{-0.1}$ 
&$ 0.2^{+0.1}_{-0.0}$ &$ 0.3^{+0.0}_{-0.1}$\\
$B^\pm \to \rho^\pm\omega$ &III & (i) & $ 8.9^{+1.9}_{-4.4}$ 
&$ 7.7^{+1.7}_{-3.9}$ &$ 4.0^{+1.0}_{-2.2}$\\

$\optbar{B^0}\to K^{*\pm} \rho^\mp$ &IV & (i)&  $-15.5^{-5.0}_{+8.9}$ 
&$-15.9^{-5.2}_{+9.2}$ &$-16.6^{-5.4}_{+9.6}$\\
$\optbar{B^0} \to \optbar{K^{*0}} \rho^0$   &V & (i)  & $ 5.1^{+2.8}_{-4.8}$ 
&$-0.8^{+0.5}_{-0.9}$ &$-9.2^{-2.8}_{+4.8}$ \\
$B^\pm\to K^{*\pm }\rho^0$ &IV & (i)& $-11.8^{-3.6}_{+6.6}$& 
$-10.3^{-3.1}_{+5.7}$& $-7.3^{-2.1}_{+3.8}$\\

$B^\pm \to \rho^\pm \optbar{K^{*0}}$ &IV & (i)& $-1.7\mp0.1$ &$-1.6\mp0.1$
&$-1.5^{-0.1}_{+0.0}$\\
$B^\pm \to K^{*\pm} \optbar{ K^{*0}}$ &IV & (i)& $ 15.2^{-3.3}_{+6.1}$ 
&$ 14.5^{-3.2}_{+5.9}$ &$ 13.4^{-3.1}_{+5.7}$\\

$\optbar{B^0}\to K^{*0}\bar K^{*0}$&IV & (iii)& $ 18.6^{-3.5}_{+6.2}$ 
&$ 17.8^{-3.4}_{+6.0}$ &$ 16.6^{-3.2}_{+5.7}$\\

$\optbar{ B^0} \to \rho^0 \omega$ &V & (iii) & $-4.8^{-6.4}_{+11.2}$ 
& $ 13.8^{-0.8}_{+1.5}$ &$ 4.4^{-3.5}_{+6.2}$\\
$\optbar{ B^0} \to \optbar{K^{*0}}\omega $  &V & (i) & $-3.1^{-0.7}_{+1.1}$ 
&$-2.1^{-0.3}_{+0.4} $&$-11.7^{-3.7}_{+6.8}$\\
$B^\pm \to K^{*\pm}\omega$ &V & (i)&$-9.6^{-2.9}_{+5.2}$&
$-14.3^{-4.6}_{+8.2}$&
$ 7.2^{+2.6}_{-5.1}$\\
$B^\pm \to K^{*\pm} \phi$ &V & (i)&  $-1.7\mp0.1$  &$-1.8\mp0.1$
&$-2.7^{-0.1}_{+0.1}$\\
$\optbar{B^0} \to \optbar{K^{*0}} \phi$   &V & (i) &  $-1.7\mp0.1$
&$-1.8\mp0.1$
&$-2.7^{-0.1}_{+0.1}$\\ 
$B^\pm \to\rho^\pm \phi$ &V & (i)& $ 16.2^{-3.4}_{+6.2}$ 
&$ 1.0^{-0.4}_{+0.7}$  &$ 10.5^{-2.6}_{+5.1}$\\
$\optbar{B^0} \to \rho^{0} \phi$  &V & (iii) & $ 19.6^{-3.6}_{+6.5}$ 
&$ 1.4^{-0.3}_{+0.7}$  &$ 13.4^{-2.7}_{+5.0}$\\
$\optbar{B^0} \to \omega \phi$   &V & (iii)  & $ 19.6^{-3.6}_{+6.5}$ 
&$ 1.4^{-0.3}_{+0.7}$  &$ 13.4^{-2.7}_{+5.0}$\\
\hline
\end{tabular}\end{center}
\end{table}
\begin{table}[htb]\begin{center}
\caption{CP-violating asymmetries $A_{CP}$ in $\protect\optbar{B} \to VV$
decays (in percent)  
using $\rho=0.23$, $\eta=0.42$ and $N_c=2,3,\infty$
for $k^2=m_b^2/2$. 
}\label{cvv2}
\begin{tabular} {|l|c|c|r|r|r|}
\hline
Channel &Class&CP-Class& $N_c=2$ &  $N_c=3$ &  $N_c=\infty$\\
\hline
$\optbar{B^0}\to\rho^+ \rho^-$ &I & (iii)& $ 9.7$ & $ 9.6$ & $ 9.5$\\
$\optbar{B^0} \to \rho^0 \rho^0$   &II & (iii)& $-43.9$ & $-25.3$ & $
41.6$\\
$\optbar{ B^0} \to \omega \omega$  &II & (iii)& $ 52.0$ & $ 60.2$ & $
12.4$\\
$B^\pm \to \rho^\pm\rho^0$ &III & (i)& $ 0.2$ & $ 0.2$ & $ 0.2$\\
$B^\pm \to \rho^\pm\omega$ &III & (i) & $ 6.3$ & $ 5.4$ & $ 2.8$\\
$\optbar{B^0}\to K^{*\pm} \rho^\mp$ &IV & (i)&  $-22.1$ & $-22.6$ &
$-23.6$\\
$\optbar{B^0} \to\optbar{ K^{*0}} \rho^0$   &V & (i)  & $ 5.7$ & $-1.0$ &
$-12.3$ \\
$B^\pm\to K^{*\pm}\rho^0$ &IV & (i)& $-16.8$ & $-14.6$ & $-10.1$\\
$B^\pm \to \rho^\pm \optbar{K^{*0}}$ &IV & (i)& $-2.1$ & $-2.0$ & $-1.9
$\\
$B^\pm \to K^{*\pm} \optbar{ K^{*0}}$ &IV & (i)& $ 21.3$ & $ 20.3$ & $
18.8$\\
$\optbar{B^0}\to K^{*0}\bar K^{*0}$&IV & (iii)& $ 26.1$ & $ 25.0$ & $
23.4$\\
$\optbar{ B^0} \to \rho^0 \omega$ &V & (iii) & $-6.8$ & $ 19.4$ & $ 6.3$\\
$\optbar{ B^0} \to\optbar{ K^{*0}}\omega $  &V & (i) & $-4.0$ & $-2.6$ &
$-16.7$\\
$B^\pm \to K^{*\pm}\omega$ &V & (i)&$-13.3$ & $-20.4$ & $ 6.6$\\
$B^\pm \to K^{*\pm} \phi$ &V & (i)&  $-2.1$ & $-2.2$ & $-3.4$\\
$\optbar{B^0} \to \optbar{K^{*0}} \phi$   &V & (i) &  $-2.1$ & $-2.2$ &
$-3.4$\\ 
$B^\pm \to\rho^{\pm} \phi$ &V & (i)& $ 22.7$ & $ 1.4$ & $ 14.8$\\
$\optbar{B^0} \to \rho^{0} \phi$  &V & (iii) & $ 27.5$ & $ 2.0$ & $ 19.0$\\
$\optbar{B^0} \to \omega \phi$   &V & (iii)  & $ 27.5$ & $ 2.0$ & $ 19.0$\\
\hline
\end{tabular}\end{center}
\end{table}

$B \to PP$ decays: $B^0 \to \pi^0 \eta^{(\prime)}$.

$B \to PV$ decays involving $b \to s$ transitions: $B^0 \to K^{*0} 
\eta$. (The decay mode $B^0 \to K^{*0} \eta^\prime$ was already 
classified in \cite{akl98-1} as a class-V decay.)

$B \to VV$ decays: $B^0 \to K^{*0} \rho^0$.

With this, we note that the $N_c$-dependence of $A_{CP}$ in 
the class-I and class-IV decays is at most $\pm 20\%$, as one varies
$N_c$ in the range $N_c=2$ to $N_c=\infty$. 

Concerning class-III decays, in most cases $A_{CP}$ is   
found to vary typically by a factor 2 as  $N_c$ is varied, with one 
exception: $B^+ \to \omega \pi^+$, in which case the estimate for 
$A_{CP}$ is
uncertain by an order of magnitude. This is in line with the observation
made on the decay rate for this process in \cite{akl98-1}.
Both CP-violating asymmetries and decay rates for the class-II and class-V
decays are 
generally strongly $N_c$-dependent. We had shown this for the decay rates
in \cite{akl98-1} and for the CP-violating asymmetries this feature can be
seen in the 
tables presented here.

\item{\bf $\mathbf k^2$-dependence of $\mathbf A_{CP}$}:
The CP-violating asymmetries depend on the  
value of $k^2$, as discussed in the literature \cite{kps}. The value of $k^2$ 
relative to the charm threshold, i.e., $k^2 \leq (\geq) ~4 m_c^2$, 
plays a central role here.  For the choice $k^2 \leq 4 m_c^2$, the 
$c\bar{c}$ loop  will not generate a strong phase.
We treat $k^2$ as a variable parameter
and have studied the sensitivity of the CP-asymmetries by varying it in the 
range $k^2=m_b^2/2 \pm 2~\mbox{GeV}^2$. This range may appear somewhat
arbitrary, however, it is large enough to test which
of the asymmetries are sensitive to the choice of $k^2$. 
 One sees from the tables, that $A_{CP}$ as well as the CP-violating
parameters are sensitive to $k^2$ in most cases. Fortunately,
there are some decays which have large $A_{CP}$ with only moderate
theoretical errors from the $k^2$-dependence.

\item{\bf $\mathbf \mu$-dependence of $\mathbf A_{CP}$}:
It should be remarked that the CP-asymmetries depend on the renormalization 
scale $\mu$. Part of this dependence is due to the fact that the strong 
phases are generated only by the explicit ${\cal O}(\alpha_s)$ corrections.
This can be seen in the numerator $A^-$ given in eq.~(\ref{aminus}). In
other words, NLO corrections to $A_{CP}$, which are of of $O(\alpha_s^2)$  
remain to be
calculated. Despite this, 
the scale-dependence  of $A_{CP}$ in $B \to h_1 h_2$ decays is not very
marked, except for a few decays for which the relevant Wilson coefficients
do show some scale dependence. We give a list of these decays in Table~13.
This feature is quantitatively different from the scale-dependence of
$A_{CP}$ in the inclusive radiative decays
$B \to X_s \gamma$ and $B \to X_d \gamma$ \cite{aag98}, for which  the
$\mu$-dependence of the Wilson coefficient in the electromagnetic penguin
operator introduces quite significant scale dependence in the 
CP-asymmetries. In contrast, the Wilson
coefficients contributing to $A_{CP}$ in the decays $B \to h_1 h_2$ are less 
$\mu$-dependent. Of course, there is still some residual scale dependence
in the quark masses. For all numbers and figures shown here, we use
$\mu=m_b/2$, a scale suggested by NLO corrections in the decay rates for 
$B \to X_s \gamma$ and $B \to X_d \gamma$ \cite{aag98}, for which NLO
corrections are small. 
\end{itemize}

\begin{table}[htb]
\begin{center}
\caption{CP-violating asymmetries $A_{CP}$ in $\protect\optbar{B} \to h_1h_2$
decays (in percent) using 
$\rho=0.12$, $\eta=0.34$ and $N_c=2,3,\infty$, $k^2=m_b^2/2$ 
for $\mu=m_b/2$ and $\mu=m_b$.}
\label{cppu}
\begin{tabular}{|l|r|r|r|r|r|r|}
\hline
 &   \multicolumn{2}{|c|}{$N_c=2$} &\multicolumn{2}{|c|}{  $N_c=3$} &
\multicolumn{2}{|c|}{ $N_c=\infty$} \\
\hline
Channel& $\mu=m_b/2$ & $\mu=m_b$& $\mu=m_b/2$ & $\mu=m_b$& $\mu=m_b/2$ 
& $\mu=m_b$\\
\hline
$\optbar{B^0} \to \pi^0 \pi^0$     &$-42.0$ &$-37.8$ & $-15.1 $
& $-32.2$ &$ 43.9$ &$ 45.5$\\

$ B^\pm \to  K^\pm \eta$ &$ 8.5$ & $ 12.2$  &$ 6.2$ & $ 9.0$ 
& $ 2.8$ & $ 4.4$\\

\hline
$\optbar{B^0} \to \rho^0 \pi^0$   &$-23.5$ & $-18.3$  
& $-49.4$ & $-49.9$ & $ 22.2$ & $ 20.1$\\
$\optbar{B^0} \to \omega \pi^0 $    &$ 57.5$ & $ 61.5$ 
&$ 39.2$ & $ 48.4$  &$ 24.3$ & $ 23.4$\\

$\optbar{ B^0} \to \rho^0 \eta $   & $-59.5$ & $-61.0$ 
&$ 0.9$ & $-21.4$ &$ 64.4$& $ 64.7$ \\
$\optbar{ B^0} \to \rho^0 \eta^\prime$  & $-16.5$ & $-10.0 $
&$-56.7$ & $-59.3$ &$ 40.2$& $ 35.8$ \\

$\optbar{ B^0} \to \omega \eta^\prime$  & $ 33.5$ & $ 29.1$ & 
$ 54.9$& $ 41.3$ & $ 19.2$& $ 17.9$\\

$ B^\pm \to K^{*\pm}  K_S^0$ &   $-1.2$   & $-0.9$
&$ 46.8$ &$ 35.0$  &$ 48.1$ & $ 46.8$\\
\hline
$\optbar{B^0} \to \optbar{K^{*0}} \pi^0$     & $ 1.4$ & $ 3.5$   
&$-1.3$ & $-0.3$  &$-4.9$ & $-5.3$\\

$\optbar{B^0} \to \omega K_S^0$  & $ 20.6$ & $ 18.3$ 
&$-6.6$ & $-14.6$ &$ 23.6$ & $ 23.3$\\
\hline
$\optbar{B^0} \to \rho^0 \rho^0$   & $-51.4$ & $-49.5$  
&$-18.5$ & $-33.5$ & $ 49.9$ & $ 50.5$\\

$\optbar{B^0} \to \optbar{K^{*0}} \rho^0$     & $ 5.1$ & $ 10.0$
&$-0.8$ & $ 1.3$ &$-9.2$& $-11.2$ \\

$\optbar{ B^0} \to \rho^0 \omega$  & $-4.8$ & $-14.8$ 
& $ 13.8$ & $ 18.5$ &$ 4.4$ & $ 2.8$\\
\hline
\end{tabular}\end{center}
\end{table}

\subsection{Decay Modes with Stable $A_{CP}$ in the Factorization Approach}

We use the parametric dependence of $A_{CP}$ just discussed to pick out the
decay modes which are stable against the variation of $N_c$, $k^2$ and 
the scale $\mu$. 
As only class-I and class-IV (and some class III) decays are stable against 
$N_c$, we need concentrate only on decays in these classes. With the help 
of the entries in Tables 5-13, showing the $k^2$ and $\mu$ dependence, we 
find that the following decay modes have measurably large asymmetries, i.e., 
$|A_{CP}| \ge 20\%$ (except for $A_{CP}(\rho^+\rho^-)$ which is estimated 
to be more
like $O(10\%)$) with
large branching ratios, typically $O(10^{-5})$ (except for $B^0 \to K_S^0 
\eta$, which is
estimated to be of $O(10^{-6})$ \cite{akl98-1}).
\begin{itemize}
\item $\optbar{B^0}\to \pi^+\pi^-$, $\optbar{B^0}\to K^0_S \pi^0$,
 $\optbar{B^0}\to K^0_S \eta^\prime$,  
$\optbar{B^0} \to K^0_S \eta$ and $\optbar{B^0} \to \rho^+ \rho^-$.
\end{itemize}
We discuss these cases in detail showing the CKM-parametric 
dependence of $A_{CP}$ in each case. Since these decays measure, ideally,
one of the phases in the unitarity triangle, we shall also plot $A_{CP}$
 as a function of the relevant phase, which is  $\sin 2 \alpha$
for $A_{CP}(\pi^+\pi^-)$, and  $\sin 2 \beta$ for  $A_{CP}(K_S^0 \pi^0)$,
$A_{CP}(K_S^0 \eta)$ and 
$A_{CP}(K_S^0 \eta^\prime)$.

%
\begin{figure}
    \epsfig{file=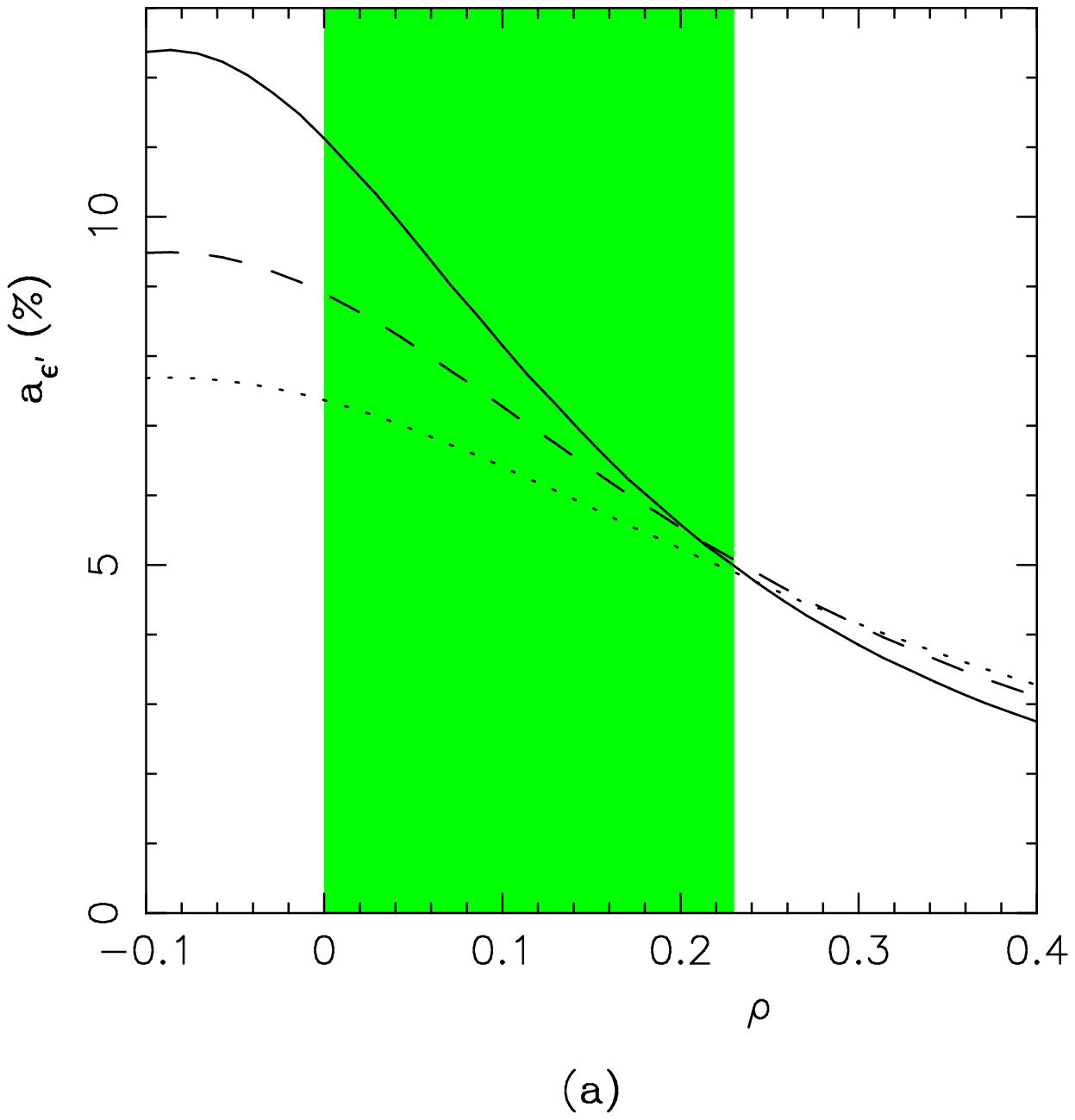,bbllx=5cm,bblly=7cm,bburx=18cm,bbury=19cm,%
width=8cm,height=6.5cm,angle=0}
    \epsfig{file=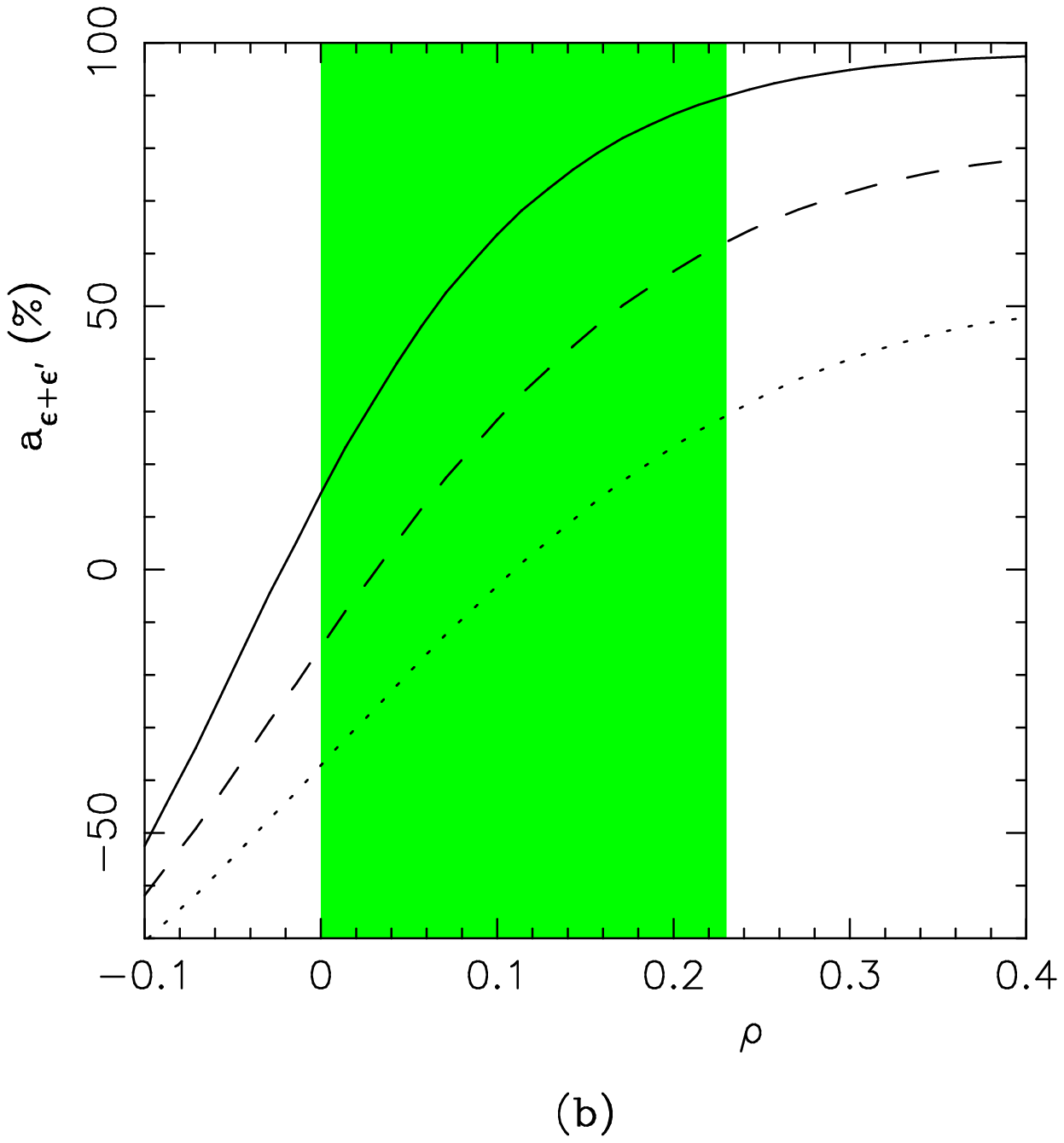,bbllx=5cm,bblly=7cm,bburx=18cm,bbury=19cm,%
width=8cm,height=6.5cm,angle=0}
\caption{CP-violating asymmetry parameters $a_{\epsilon^\prime}$ (a) and 
$a_{\epsilon+\epsilon^\prime}$ (b)
for the decay $\protect\optbar{B^0}  \to \pi^{+}\pi^-$ as a function of 
the CKM parameter $\rho$ with $k^2=m_b^2/2$.  
The dotted, dashed and solid curves correspond to 
the CKM parameter values $\eta=0.42$, $\eta=0.34$ and $\eta=0.26$, 
respectively.}
 \label{cppp}
\end{figure}

\begin{itemize}
\item \underline{CP-violating asymmetry in  $\optbar{B^0} \to \pi^+\pi^-$}

We show in Fig.~\ref{cppp}(a) and \ref{cppp}(b) the CP-asymmetry parameters 
$a_{\epsilon'}$ and 
$a_{\epsilon+\epsilon'}$, defined in eq.~(\ref{a1}) and (\ref{a2}),
 respectively, plotted as a 
function of the CKM-Wolfenstein parameter $\rho$ with the indicated values 
of $\eta$.
The shadowed area in this and all subsequent figures
showing the $\rho$-dependence corresponds to the 
range $0<\rho<0.23$, which is the $\pm 1\sigma$ allowed values of this 
parameter from the unitarity fits \cite{aliapctp97}.
The three curves in Fig.~\ref{cppp}(a) and \ref{cppp}(b)
 represent three different values of the
CKM-Wolfenstein parameter: $\eta=0.26$ (solid curve), $\eta=0.34$ (dashed
curve) and
$\eta=0.42$ (dotted curve).  
The time-integrated asymmetry $A_{CP}$ calculated with the help of 
eq.~(\ref{cpint}) is shown for three values of $\eta$ ($\eta=0.42$, 
$0.34$, $0.26$)
with $k^2=m_b^2/2$ in Fig.~\ref{cc6}(a).
One notes that the CKM-dependence of $A_{CP}$ is very significant.
The $k^2$-dependence of $A_{CP}(\pi^+\pi^-)$ is found to be very weak as
shown is 
Fig.~\ref{cc6}(b), where we plot this quantity as a function of $\rho$ for 
$\eta =0.34$ by varying $k^2$ in the range $k^2=m_b^2/2 \pm 2$ GeV$^2$.
Hence, there is a good case for $A_{CP}(\pi^+ \pi^-)$ yielding information
on the CKM parameters. 
%
%
%
\begin{figure}
    \epsfig{file=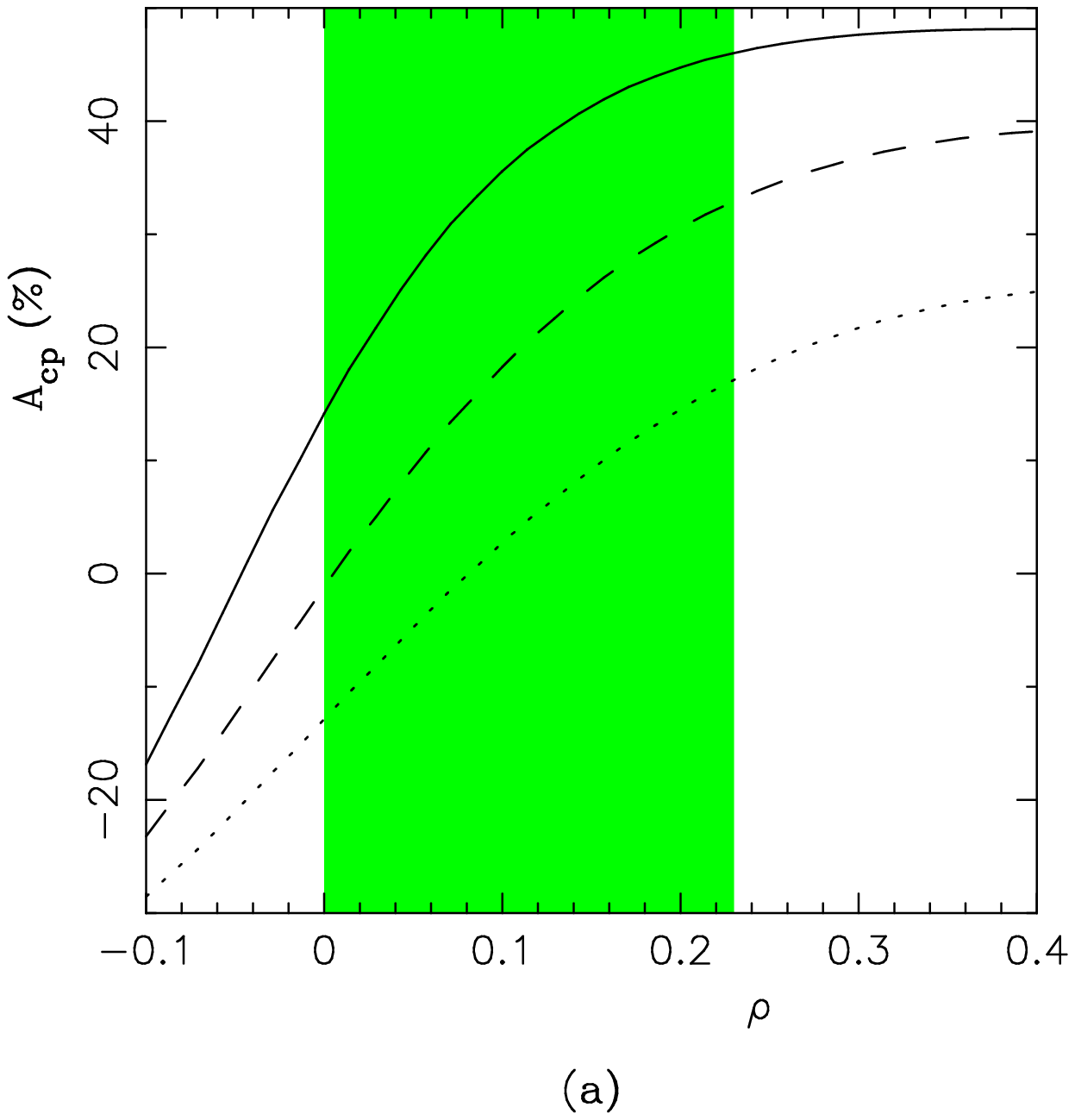,bbllx=5cm,bblly=7cm,bburx=18cm,bbury=19cm,%
width=8cm,height=6.5cm,angle=0}
    \epsfig{file=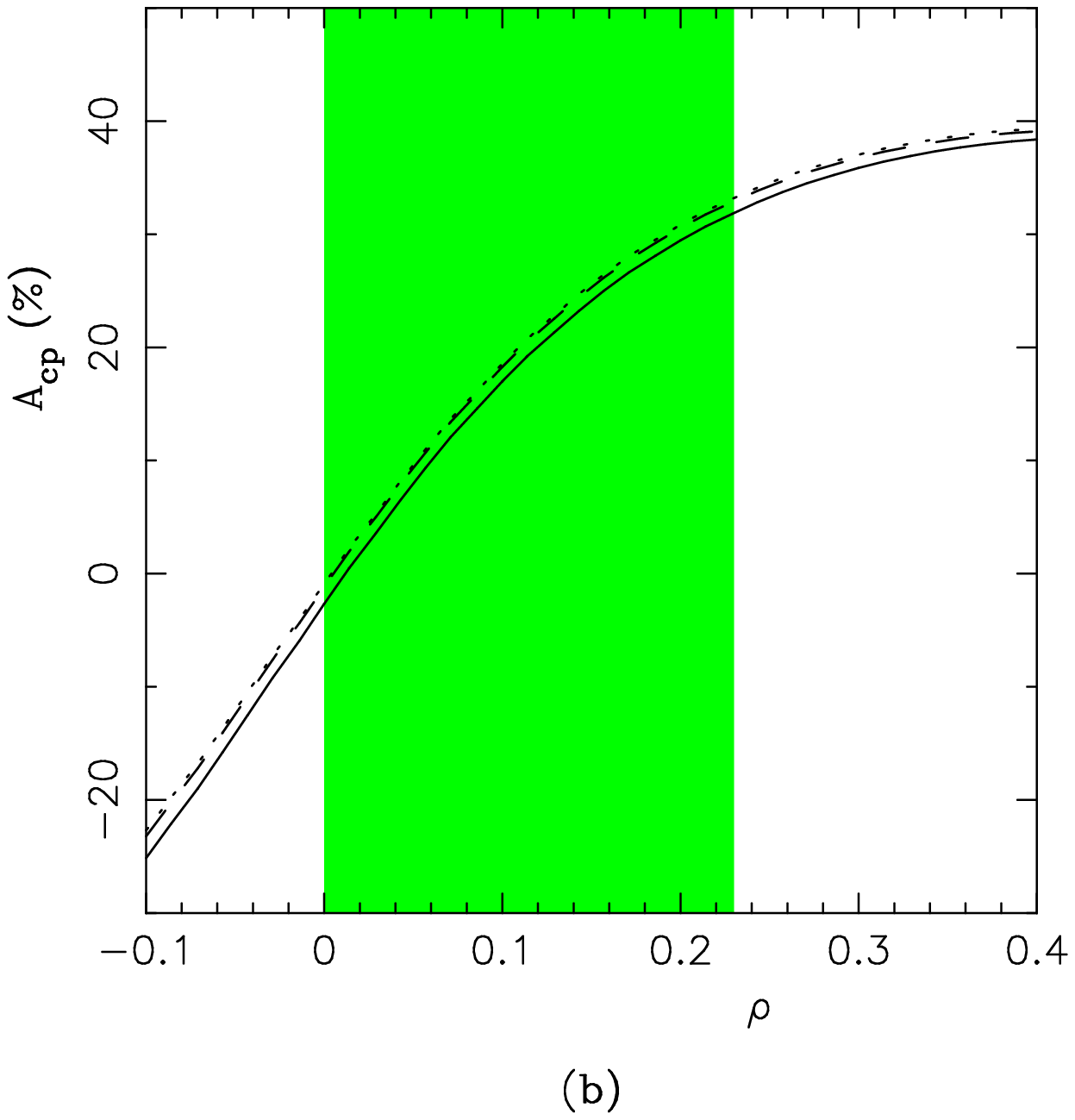,bbllx=5cm,bblly=7cm,bburx=18cm,bbury=19cm,%
width=8cm,height=6.5cm,angle=0}
\caption{CP-violating asymmetry $ A_{CP}$ in
 $\protect\optbar{B^0} \to \pi^{+}\pi^-$ as a function of
the CKM parameter $\rho$.
(a) $k^2=m_b^2/2$. The dotted, dashed and solid curves correspond to
the CKM parameter values $\eta=0.42$, $\eta=0.34$ and $\eta=0.26$,
respectively;
(b)  $\eta=0.34$. The dotted, dashed and solid lines correspond to
$k^2=m_b^2/2+2$ GeV$^2$, $k^2=m_b^2/2$ and $k^2=m_b^2/2-2$ GeV$^2$, 
respectively.} \label{cc6}
\end{figure}

%
%
%
%
%
%
\begin{figure}
  \begin{center}
    \epsfig{file=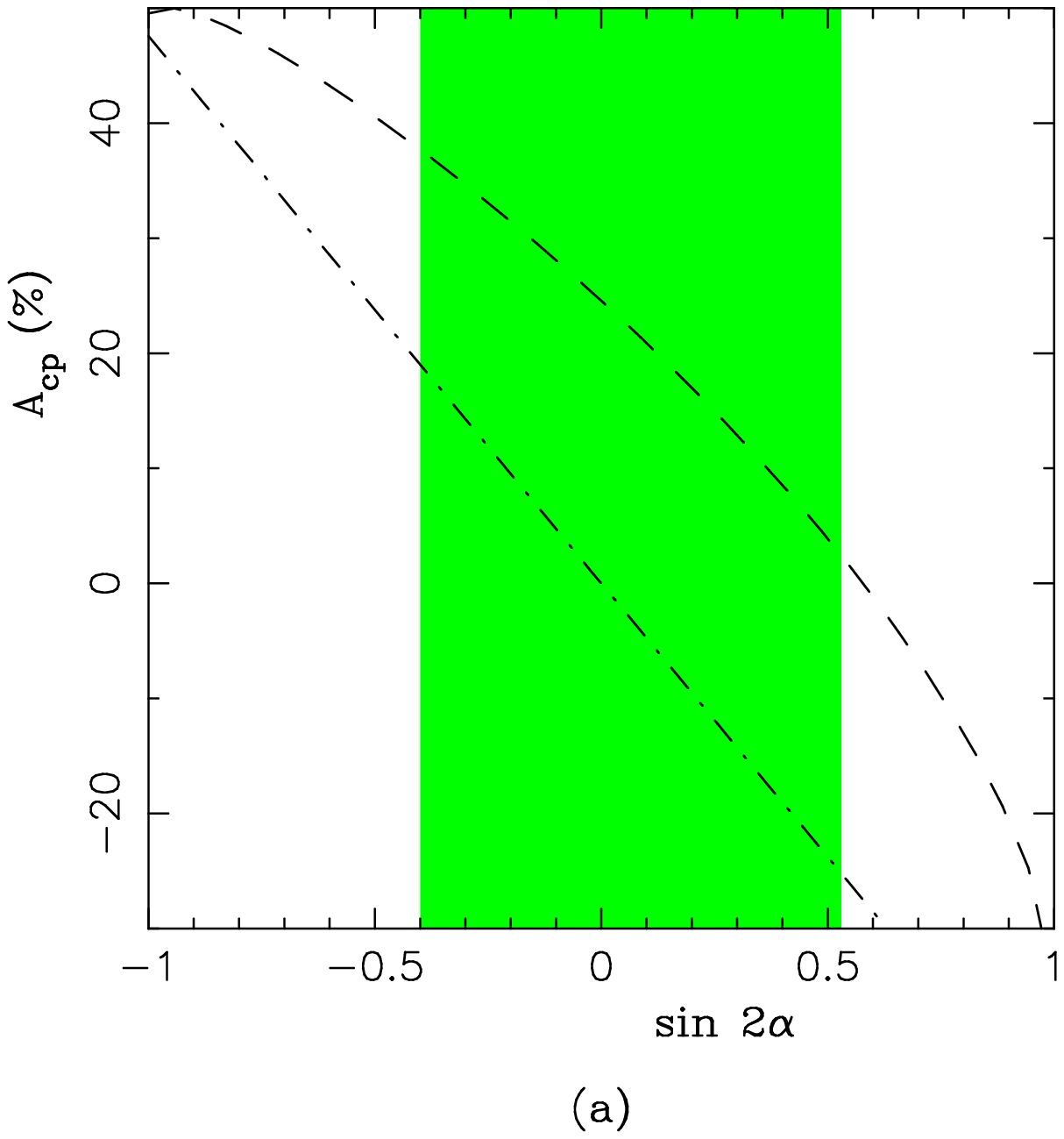,bbllx=5cm,bblly=7cm,bburx=18cm,bbury=19cm,%
width=8cm,height=6.5cm,angle=0}
    \epsfig{file=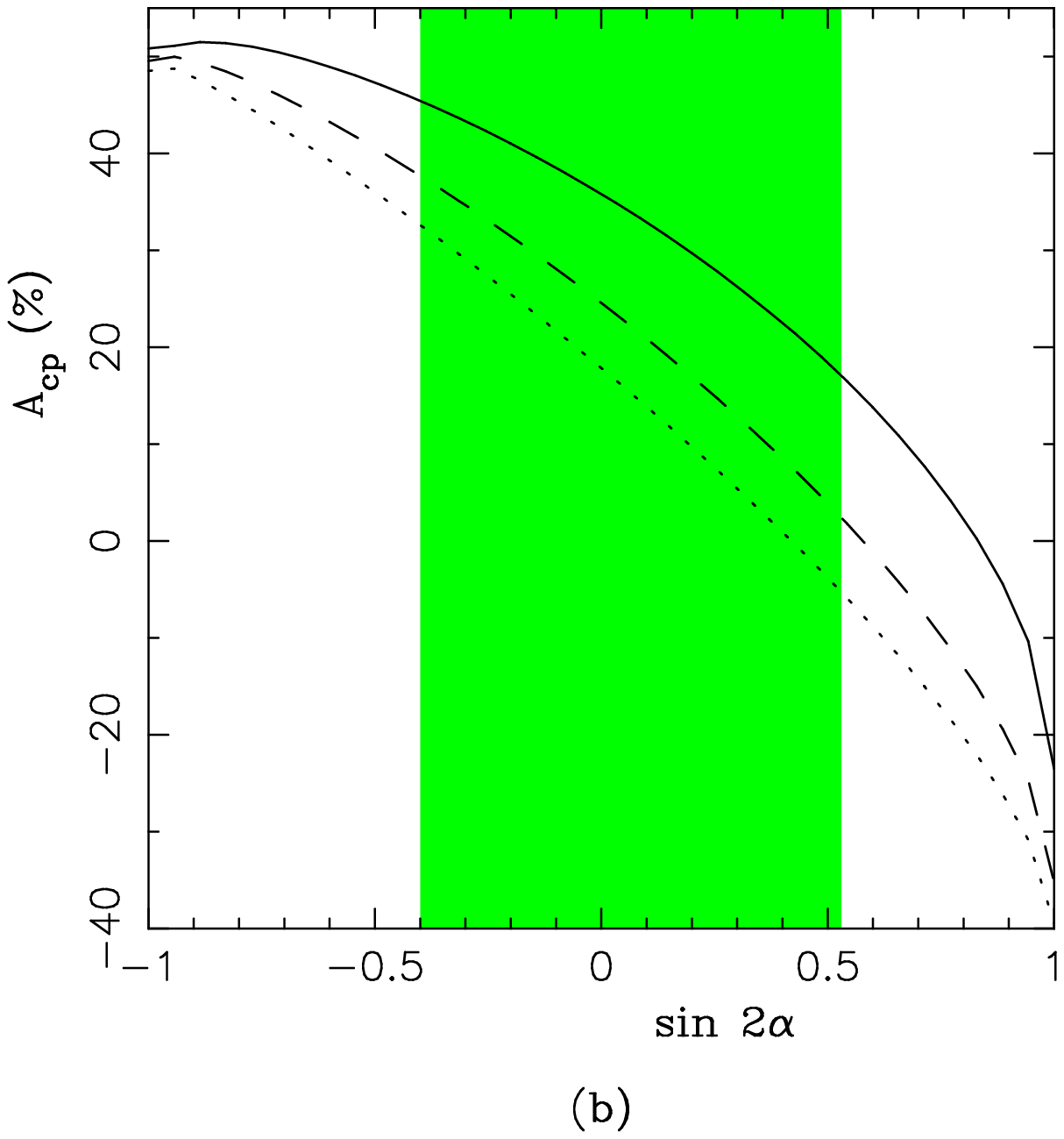,bbllx=5cm,bblly=7cm,bburx=18cm,bbury=19cm,%
width=8cm,height=6.5cm,angle=0}
\caption{CP-violating asymmetry $ A_{CP}$ in
 $\protect\optbar{B^0} \to \pi^{+}\pi^-$ as a function of
$\sin 2 \alpha$ for $k^2=m_b^2/2$. (a)
Effect of the ``penguin pollution":
the lower (upper)  curve corresponds to keeping only the tree contribution
(the complete amplitude, tree + penguin). Note that $\vert V_{ub} 
\vert =0.003$.
(b) Dependence on $\vert V_{ub} \vert$: $\vert V_{ub} \vert = 0.002$ 
(solid curve),
$\vert V_{ub}\vert = 0.003$ (dashed curve),
$\vert V_{ub}\vert = 0.004$ (dotted curve)}
 \label{bcc6}
 \end{center}
\end{figure}

To have a closer look at this, we plot in Fig.~\ref{bcc6}(a) and
\ref{bcc6}(b), the asymmetry $A_{CP}(\pi^+ \pi^-)$ as a function of
$\sin 2 \alpha$ to study the effect of the penguin contribution
(called in the jargon ``penguin pollution") and the dependence on
$\vert V_{ub} \vert$, respectively. The lower (upper) curve in 
Fig.~\ref{bcc6}(a) 
 corresponds to keeping only the tree contribution in the decays
$\optbar{B^0} \to \pi^+ \pi^-$ (tree + penguin). We see that in the entire
$\pm 1 \sigma$ expected range of $\sin 2 \alpha$, depicted as a shadowed 
region, the ``penguin pollution" is quite significant, changing both
$A_{CP}(\pi^+ \pi^-)$ and its functional dependence on $\sin 2 \alpha$. 
Based on \ref{bcc6}(b), we estimate $-10 \% \leq A_{CP}(\pi^+ \pi^-)
\leq +45\%$, with $A_{CP}(\pi^+ \pi^-)=0$ as an allowed solution,
 varying $\sin 2 \alpha$ in the $\pm 1 \sigma$ range:
$-0.40 \leq \sin 2 \alpha \leq 0.53$ \cite{aliapctp97}.

\item \underline{CP-violating asymmetry in $\optbar{B^0} \to K^0_S \eta 
^\prime$}
%
\begin{figure}
    \epsfig{file=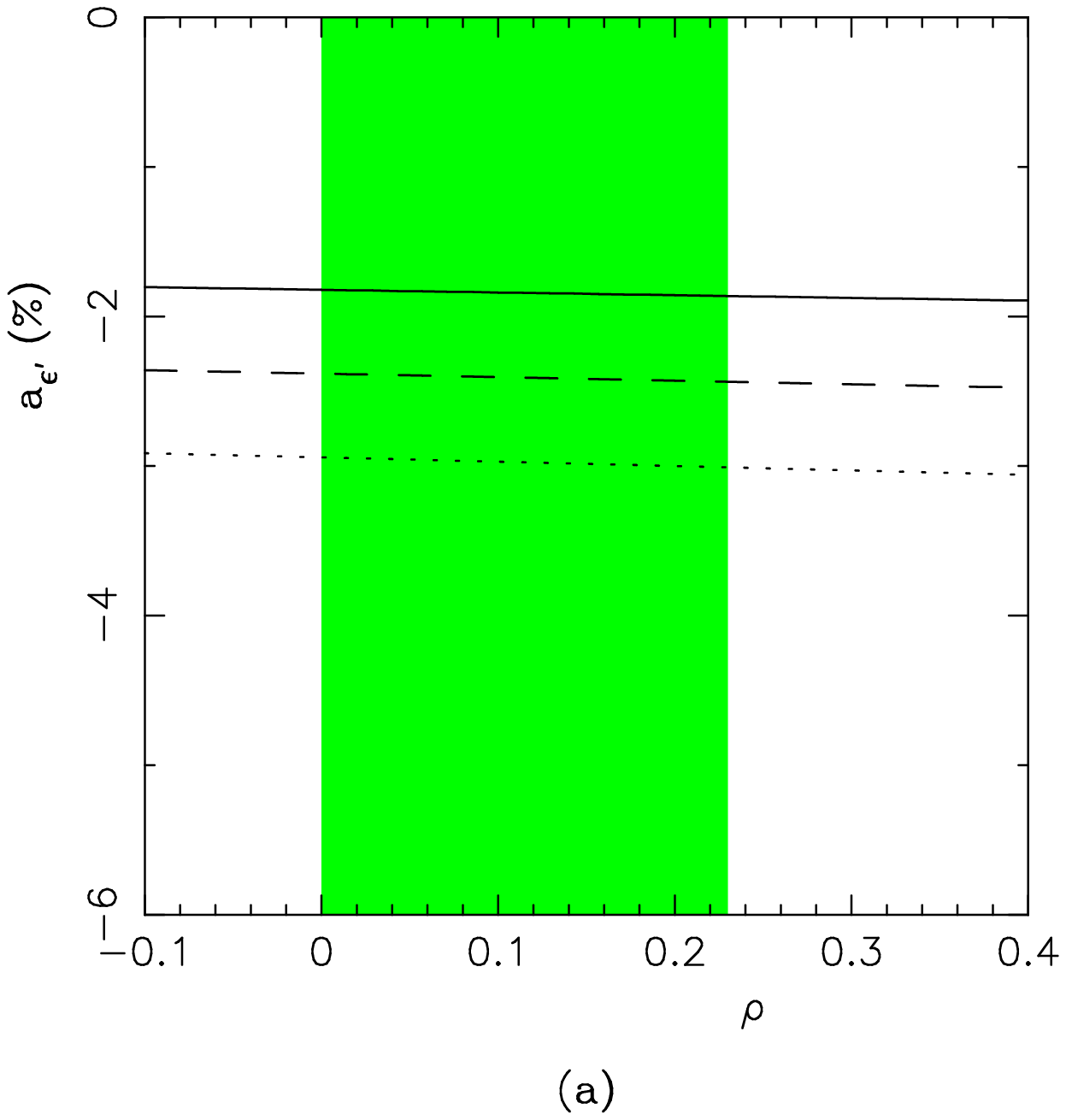,bbllx=5cm,bblly=7cm,bburx=18cm,bbury=19cm,%
width=8cm,height=6.5cm,angle=0}
    \epsfig{file=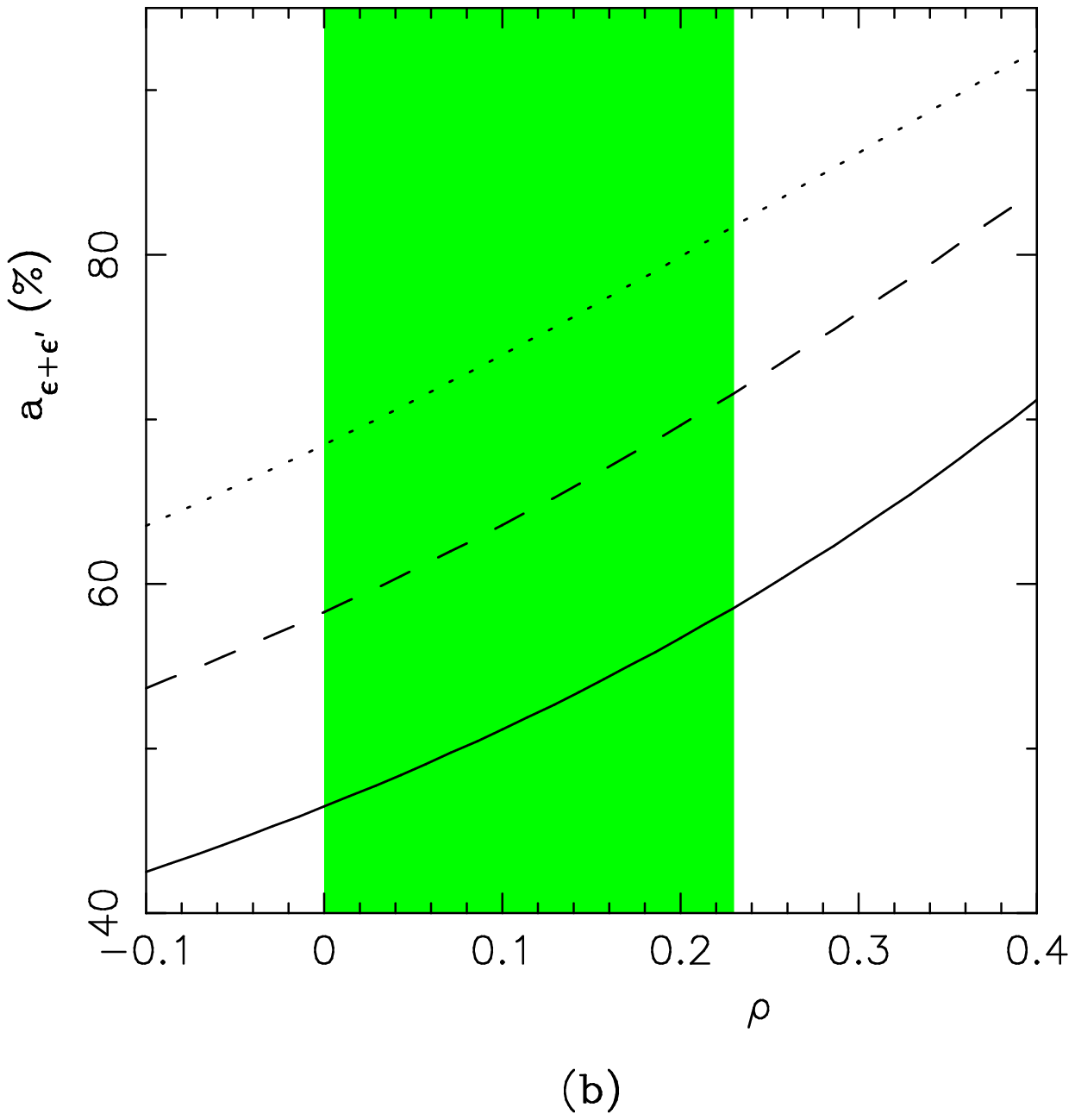,bbllx=5cm,bblly=7cm,bburx=18cm,bbury=19cm,%
width=8cm,height=6.5cm,angle=0}
\caption{CP-asymmetry parameters $a_{\epsilon^\prime}$ (a) and 
$a_{\epsilon+\epsilon^\prime}$ (b) for
$ \protect\optbar{B^0} \to K_S^0\eta^\prime$ as a 
function of the CKM parameter $\rho$.  
 The dotted, dashed and solid curves correspond to 
the CKM parameter values $\eta=0.42$, $\eta=0.34$ and $\eta=0.26$, 
respectively.}
\label{cpke}
\end{figure}
The parameters $a_{\epsilon^\prime}$ and  $a_{\epsilon+\epsilon^\prime}$ 
for the decays
$\optbar{B^0} \to K^0_S \eta ^\prime$ are shown in Fig.~\ref{cpke}(a) and
\ref{cpke}(b), respectively, for $\eta=0.42$, $0.34$, $0.26$
 with fixed $k^2=m_b^2/2$. As can be seen from these figures,
the time-integrated CP-violating asymmetry $A_{CP} (B^0\to K^0_S \eta
^\prime)$ is
completely dominated by the $a_{\epsilon + \epsilon^\prime}$ term. The
CP-violating asymmetry $A_{CP}(K_S^0 \eta^\prime)$ is  shown in
Fig.~\ref{cpketap}(a) for three values of $\eta$ ($\eta=0.42$, $0.34$, $0.26$).
The upper curve for each value of $\eta$ is obtained by neglecting the tree
contribution in $\optbar{B^0} \to K_S^0 \eta^\prime$ and the
lower curves represent the corresponding full (tree + penguin)
contribution.
Fig.~\ref{cpketap}(b) shows the $k^2$-dependence of $A_{CP} (K^0_S \eta')$
with the three (almost) overlapping curves corresponding to $k^2=m_b^2/2$ and
 $k^2=m_b^2/2 \pm 2 $ GeV$^2$ for fixed value, $\eta=0.34$.
As we see from this set of figures, the CKM-parametric dependence of 
$A_{CP}(K_S^0
\eta^\prime)$ is marked and the effect of the ``tree shadow'' is relatively
small. To illustrate this further, we plot in Figs.~\ref{bcc7}(a) and
 \ref{bcc7}(b) this asymmetry as a function of $\sin 2 \beta$, showing the
effect of the ``tree-shadowing" and dependence of $A_{CP}(K_S^0 \eta^\prime)$
on $\vert V_{td} \vert$, respectively. Restricting to the range $ 0.48 \leq 
\sin 2 \beta
\leq 0.78$, which is the $\pm 1 \sigma$ range for this quantity from the 
unitarity fits \cite{aliapctp97}, we find
that $A_{CP} (K^0_S \eta^\prime)$ has  a value in the range 
$ 20\% < A_{CP} < 36\%$. 
This decay has been measured by the CLEO Collaboration with a branching ratio 
${\cal B} (B^0 \to K^0_S \eta ^\prime)= (4.7 ^{+2.7}_{-2.0}\pm 0.9)\times 
10^{-5}$ and is well accounted for in the factorization-based approach 
\citer{akl98-1,acgk}.
As the ``tree shadow' is small in the decay $B^0\to K^0_S \eta ^\prime$
and the electroweak penguin contribution is also small \cite{akl98-1},
$A_{CP} (K^0_S \eta ^\prime)$ is a good measure of $\sin 2\beta$. This was
anticipated by London and Soni \cite{LS97}, who also advocated $A_{CP}
(K_S^0\phi)$ as a measure of the angle $\beta$, following the earlier 
suggestion of the same in
ref. \cite{LP89}. The CP-asymmetry for this decay, like $A_{CP}(K_S^0 
\eta^\prime)$, is
dominated by the $a_{\epsilon + \epsilon^\prime}$ term.
The quantity $A_{CP}(K_S^0\phi)$ is found to
be stable against variation in $N_c$ and $k^2$ (see Tables 9 and 10). However,
being a class-V decay, the branching ratio for $B^0 \to K_S^0 \phi$ 
(and its charged
conjugate) is very sensitively dependent on $N_c$, with ${\cal B}(B^0 \to
 K_S^0 \phi) =
(0.2 - 9) \times 10^{-6}$, with the lower (higher) range corresponding to 
$N_c=\infty$
($N_c=2)$ \cite{akl98-1}. Moreover, the electroweak penguin effect in this
 decay is
estimated to be rather substantial. The present upper bound on this decay is
${\cal B}(B^0 \to K_S^0 \phi) < 6.2 \times 10^{-5}$ \cite{cleobok}. 
Depending on $N_c$,
the above experimental bound is between one and two orders of magnitude away
from the expected rate.
Despite the large and stable value of $A_{CP}(K_S^0 \phi)$, it may turn out 
not to be measurable in the first generation of B factory experiments.  
%
%
%
%
\begin{figure}
    \epsfig{file=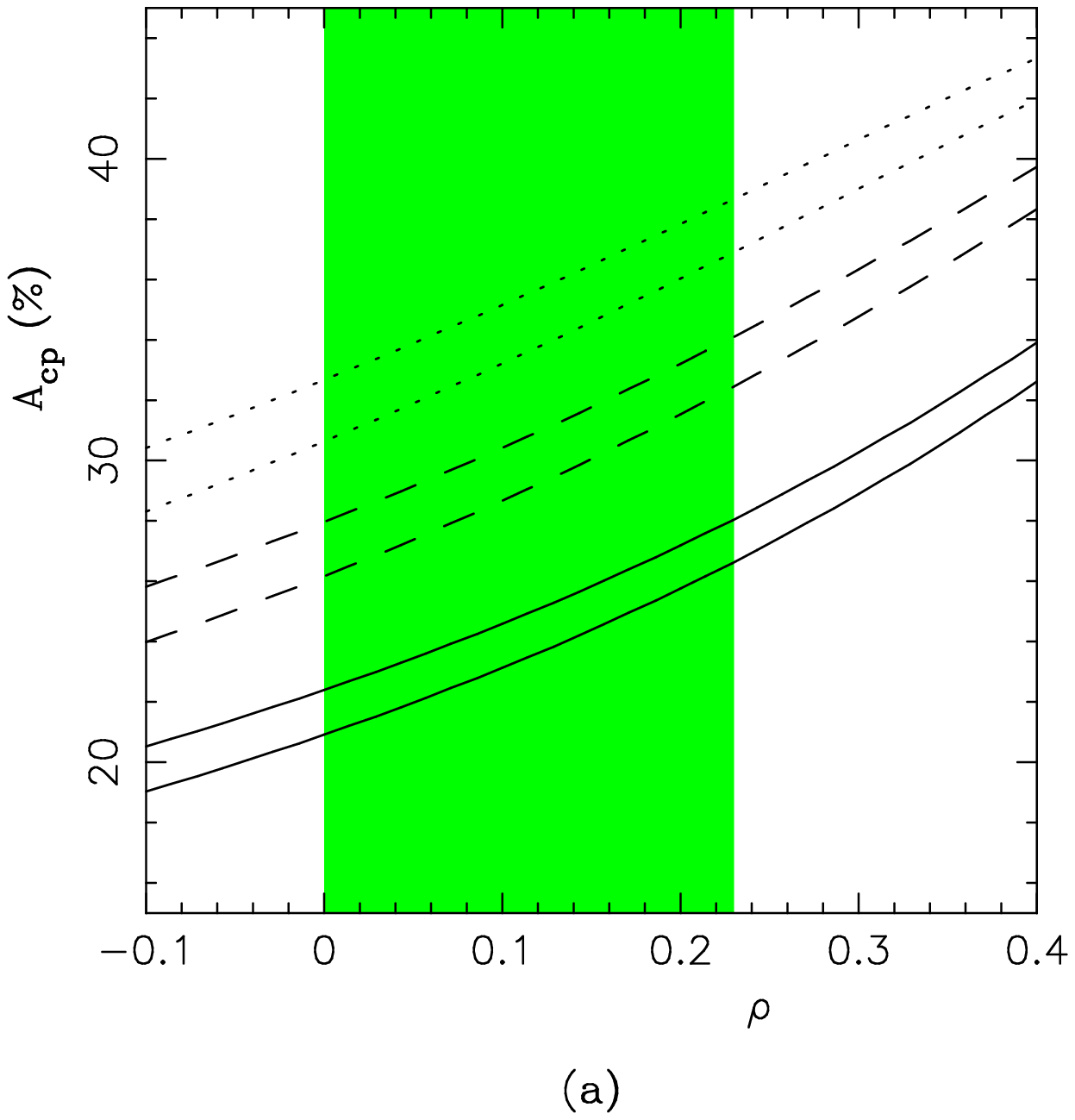,bbllx=5cm,bblly=7cm,bburx=18cm,bbury=19cm,%
width=8cm,height=6.5cm,angle=0}
    \epsfig{file=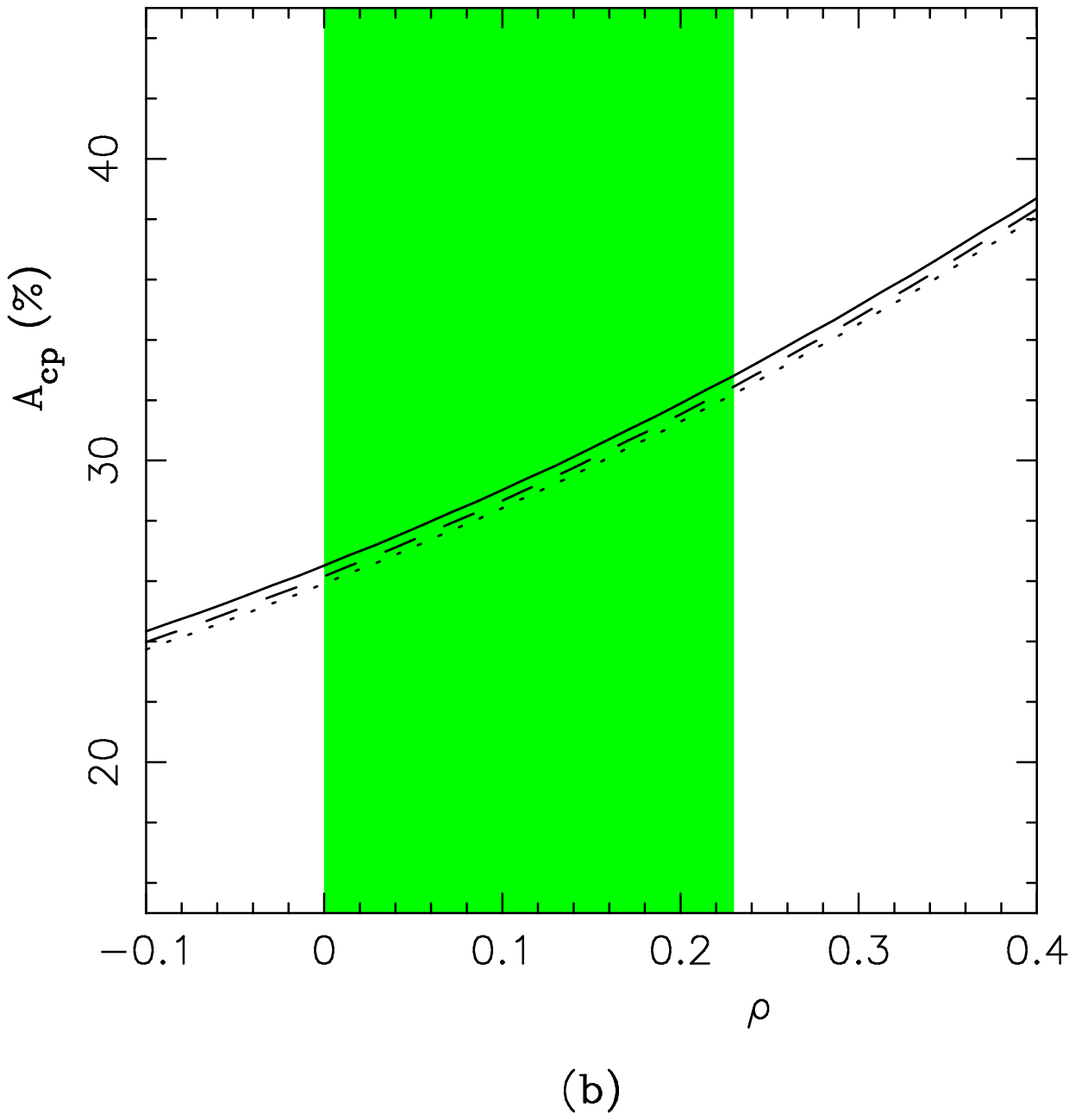,bbllx=5cm,bblly=7cm,bburx=18cm,bbury=19cm,%
width=8cm,height=6.5cm,angle=0}
\caption{CP-violating asymmetry $A_{CP}$ in 
$\protect\optbar{B^0} \to K_S^0\eta^\prime $ decays as a
function of the CKM parameter $\rho$.
(a) $k^2=m_b^2/2$. The dotted, dashed and solid curves correspond to
the CKM parameter values $\eta=0.42$, $\eta=0.34$ and $\eta=0.26$,
respectively; In all three groups, the upper (lower) curve   
corresponds to
neglecting the tree contributions (with the complete amplitude).
(b)  $\eta=0.34$. The dotted, dashed and solid curves correspond to
$k^2=m_b^2/2+2$ GeV$^2$, $k^2=m_b^2/2$ and $k^2=m_b^2/2-2$ GeV$^2$, 
respectively.}
\label{cpketap}
\end{figure}
%
%
%
\begin{figure}
 \begin{center}
    \epsfig{file=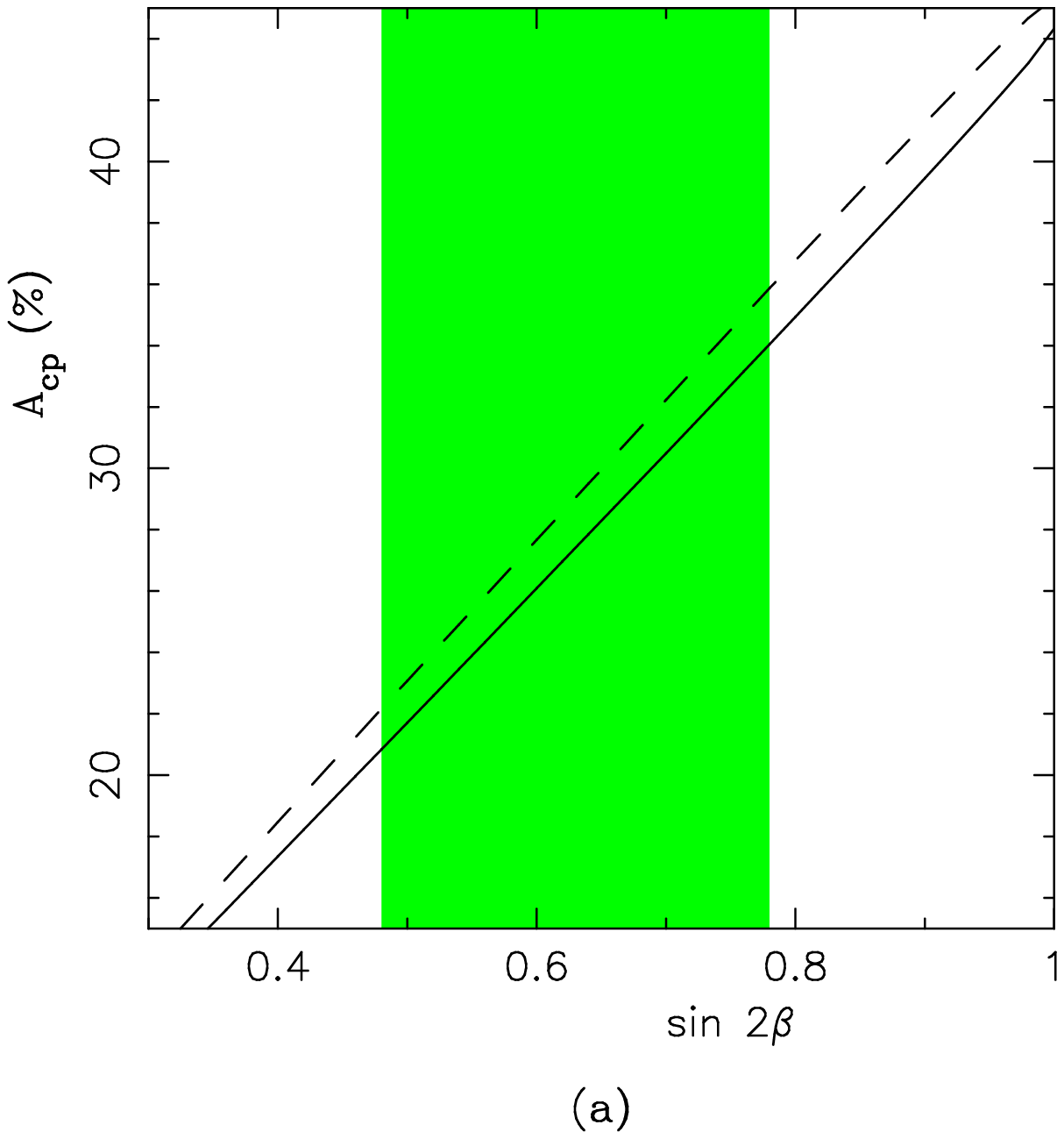,bbllx=5cm,bblly=7cm,bburx=18cm,bbury=19cm,%
width=8cm,height=6.5cm,angle=0}
    \epsfig{file=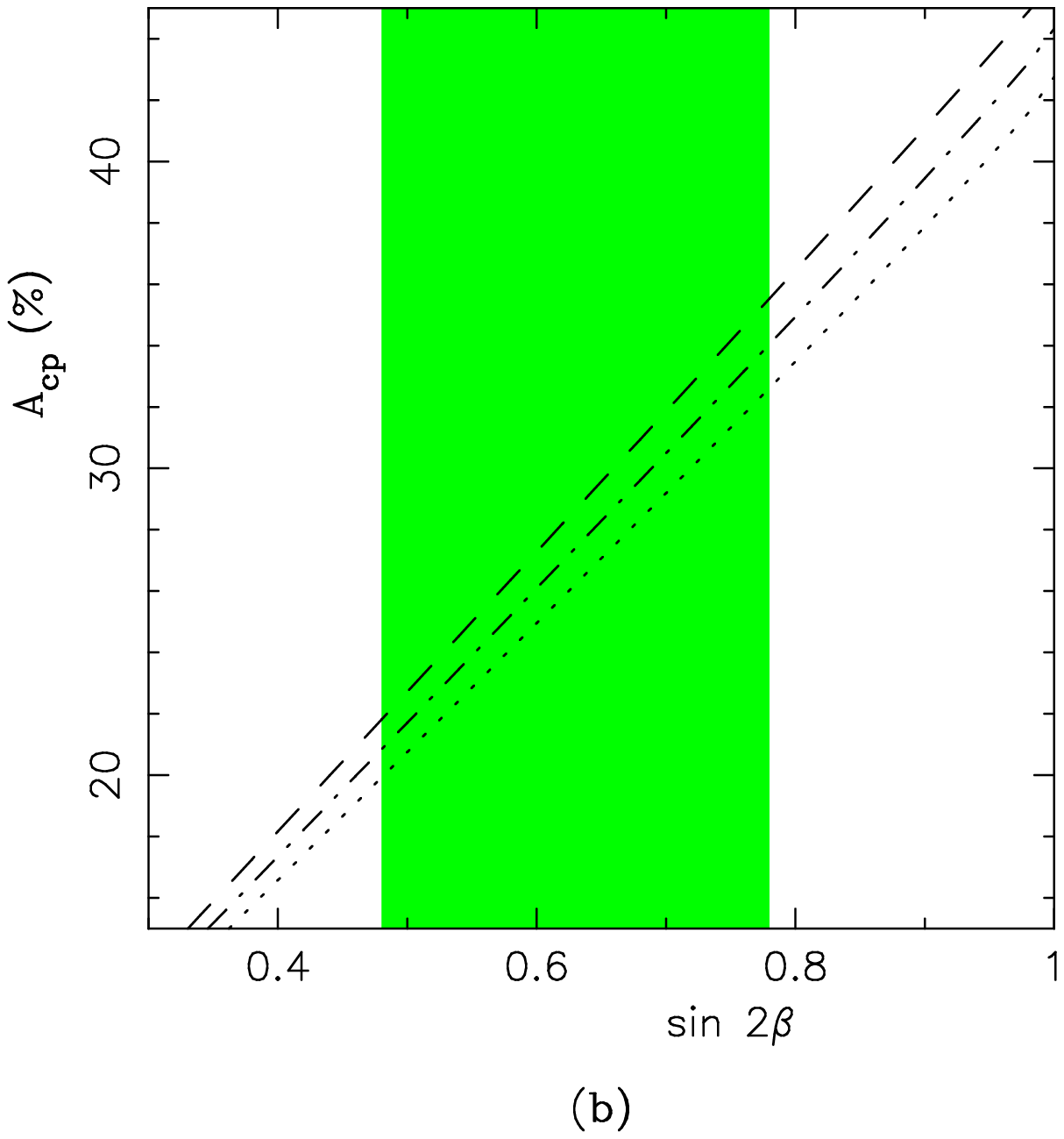,bbllx=5cm,bblly=7cm,bburx=18cm,bbury=19cm,%
width=8cm,height=6.5cm,angle=0}
\caption{CP-violating asymmetry $ A_{CP}$ in
 $\protect\optbar{B^0} \to K_S^0 \eta^\prime$ as a function of
$\sin 2 \beta$ for $k^2=m_b^2/2$.
(a) ``Tree shadow'': The solid (dashed) curve correspond to the full 
amplitude (neglecting the tree contribution).
(b) $\vert V_{td} \vert$ dependence:  
Dashed curve ($\vert V_{td}\vert =0.004$), dashed-dotted curve 
($\vert V_{td}\vert= 0.008$), dotted curve ($\vert V_{td}\vert = 0.012$).}
 \label{bcc7}
 \end{center}
\end{figure}

\item \underline{CP-violating asymmetry in $\optbar{B^0} \to K_S^0 \pi^0$}
%
%
\begin{figure}
    \epsfig{file=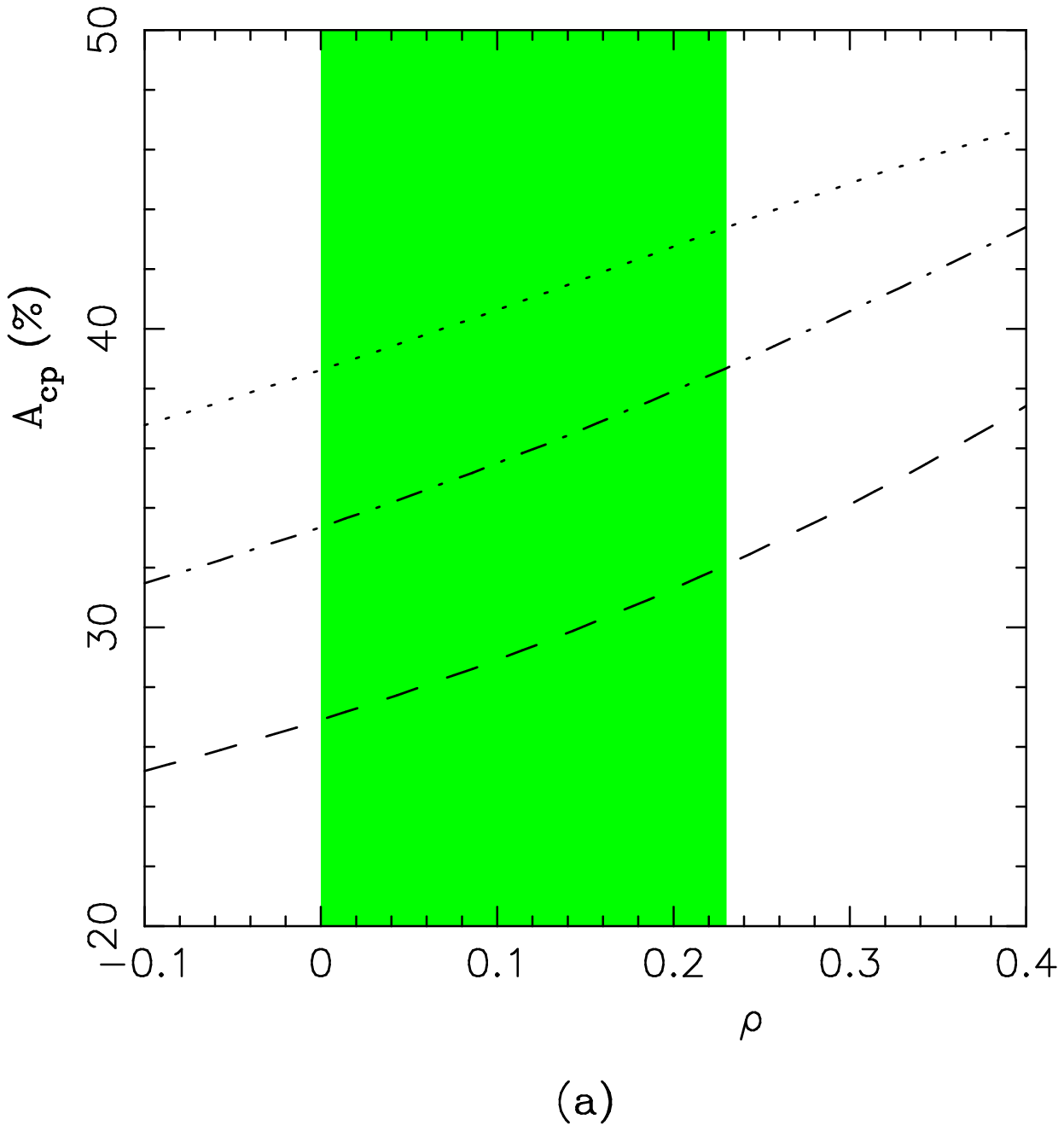,bbllx=5cm,bblly=7cm,bburx=18cm,bbury=19cm,%
width=8cm,height=6.5cm,angle=0}
    \epsfig{file=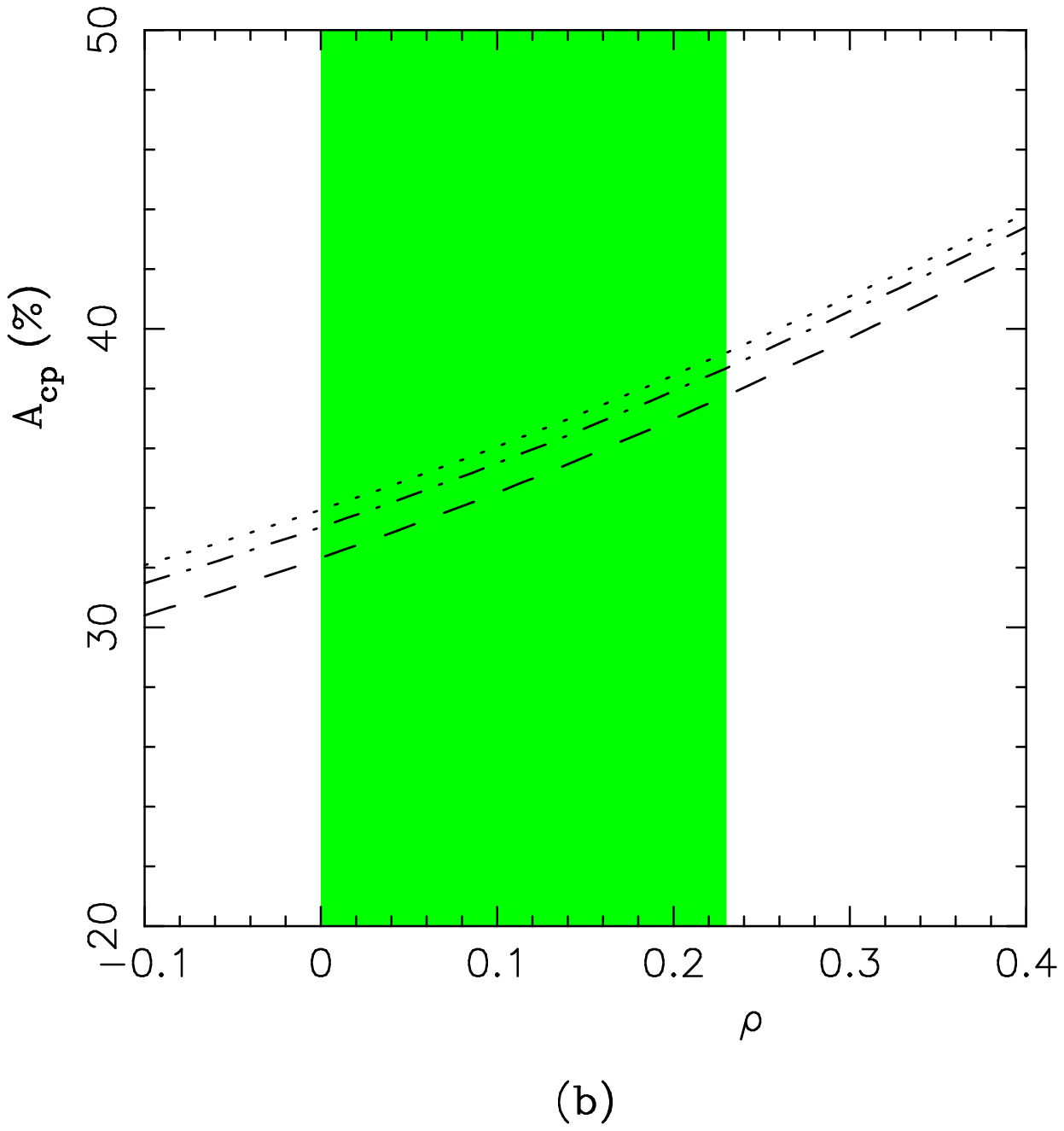,bbllx=5cm,bblly=7cm,bburx=18cm,bbury=19cm,%
width=8cm,height=6.5cm,angle=0}
\caption{CP-violating asymmetry $ A_{CP}$ in
 $\protect\optbar{B^0} \to  K_S^0 \pi^0$ decays as a
function of the CKM parameter $\rho$.
(a) $k^2=m_b^2/2$. The dotted, dashed-dotted and dashed curves correspond to
the CKM parameter values $\eta=0.42$, $\eta=0.34$ and $\eta=0.26$,
respectively;
(b)  $\eta=0.34$. The dotted, dashed-dotted and dashed curves correspond to
$k^2=m_b^2/2+2$ GeV$^2$, $k^2=m_b^2/2$ and $k^2=m_b^2/2-2$ GeV$^2$, 
respectively.} \label{cc8}
\end{figure}

The decay $B^0\to K_S^0 \pi^0$ is dominated by the penguins, with significant
 electroweak penguin contribution \cite{akl98-1}.
The estimated decay rate in the factorization approach is ${\cal B} (
B^0\to K_S^0 \pi^0 ) =(2.5-5) \times 10^{-6}$, with the present experimental 
bound
being ${\cal B} (B^0\to K_S^0 \pi^0 ) < 4.1 \times 10^{-5}$ \cite{cleo}, with
 these
numbers to be understood as averages over the charge conjugated decays.
We expect that with $10^8$ $B\bar B$ events, several hundred $K_S^0\pi^0$ 
decays will be measured. The CP-asymmetry $A_{CP}(K_S^0 \pi^0)$ is dominated by
the $a_{\epsilon + \epsilon^\prime}$ term (see Tables 1 and 2), which is 
large, stable
against variation in $k^2$ and shows only a mild dependence on $N_c$.
The quantities $a_{\epsilon^\prime }$ and $a_{\epsilon + \epsilon^\prime}$ 
for this decay
(together with the others in the $B \to \pi \pi, K\bar{K}$ and $B^\pm \to 
(K\pi)^\pm$
decays)
were worked out by Kramer and Palmer in \cite{kps}. As remarked already, 
there are detailed
differences in the underlying theoretical framework used here and in \cite{kps}
and also in the values of the CKM and other input parameters, but using
identical values of the various input parameters for the sake of comparison,
 the agreement
 between the two is fair. 
We show in Fig.~\ref{cc8}(a), $A_{CP} ( K_S^0 \pi^0)$
as a function of $\rho$ for three values of $\eta$: 
$\eta=0.42$, $0.34$, $0.26$ and note that this dependence is quite marked.
 The $k^2$-dependence of $A_{CP} ( K_S^0 \pi^0)$ is found to be small, 
as shown in Fig.~\ref{cc8}(b). Thus, we expect that $A_{CP}(K_S^0 \pi^0)$ is 
also a good measure of $\sin 2 \beta$. This is shown in Fig.~\ref{bcc8},
 with the three
curves showing the additional  dependence of $A_{CP}(K_S^0 \pi^0)$ 
 on $\vert V_{td} \vert$. Restricting the value of $\sin 2
\beta$ in the $\pm 1 \sigma$ range shown by the shadowed region, we find
$24 \% \leq A_{CP}(K_S^0 \pi^0) \leq 44 \%$. 
%
%
%
\begin{figure}
\begin{center}
    \epsfig{file=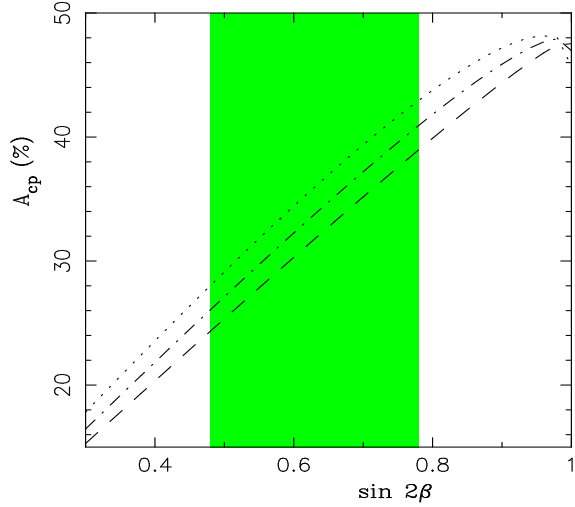,bbllx=5cm,bblly=7cm,bburx=18cm,bbury=19cm,%
width=8cm,height=6.5cm,angle=0}
\caption{CP-violating asymmetry $ A_{CP}$ in  
 $\protect\optbar{B^0} \to K_S^0 \pi^0$ as a function of
$\sin 2 \beta$ for $k^2=m_b^2/2$.  The three curves correspond to the
following values of the CKM matrix element $\vert V_{td}\vert$:
dashed curve ($\vert V_{td}\vert= 0.004$), dashed-dotted curve
($\vert V_{td}\vert =0.008$), dotted curve ($\vert V_{td}\vert =0.012$).}
 \label{bcc8}
\end{center}
\end{figure}

\item \underline{CP-violating asymmetry in $\optbar{B^0} \to K_S^0 \eta$}
%
%
\begin{figure}
    \epsfig{file=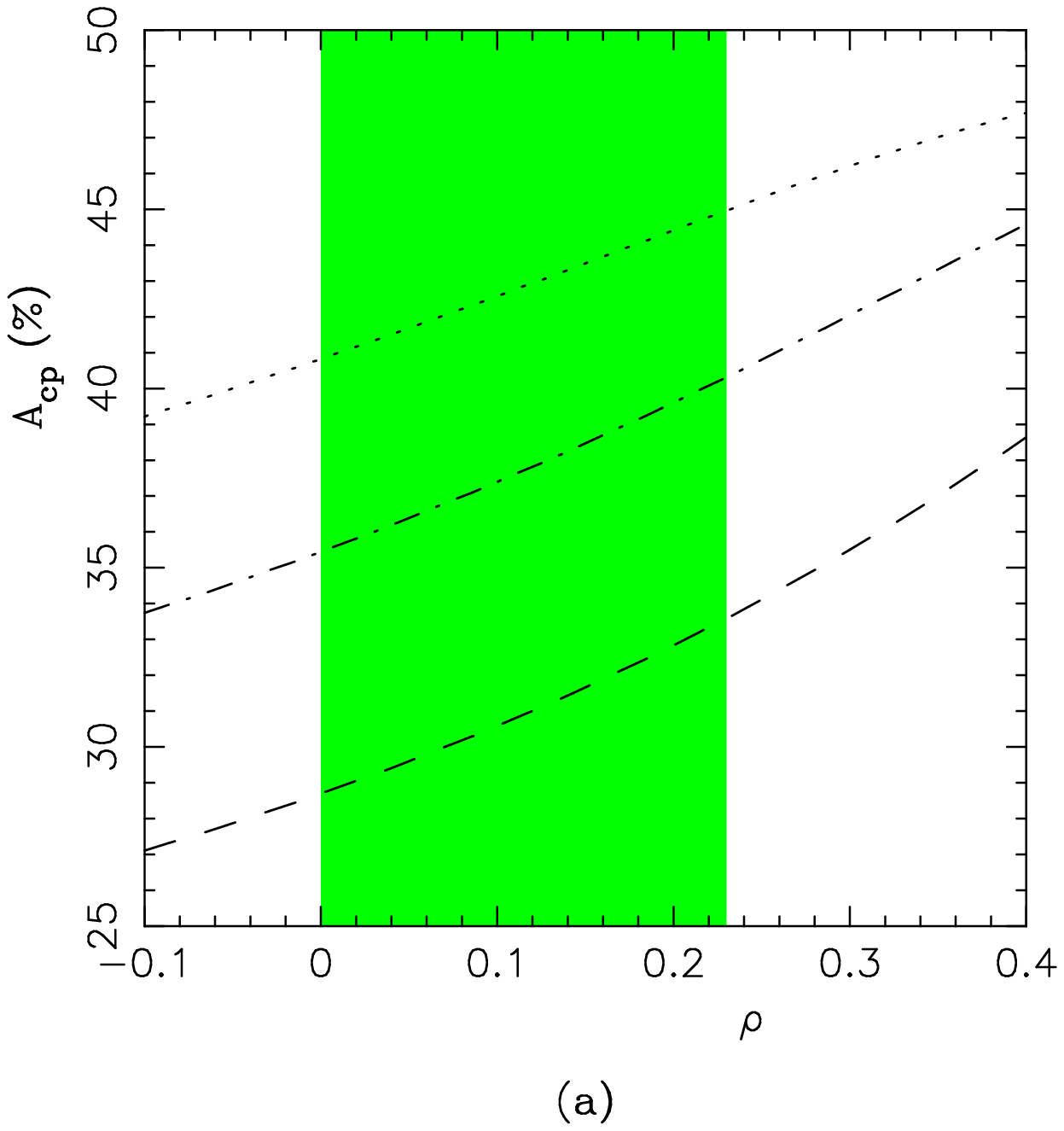,bbllx=5cm,bblly=7cm,bburx=18cm,bbury=19cm,%
width=8cm,height=6.5cm,angle=0}
    \epsfig{file=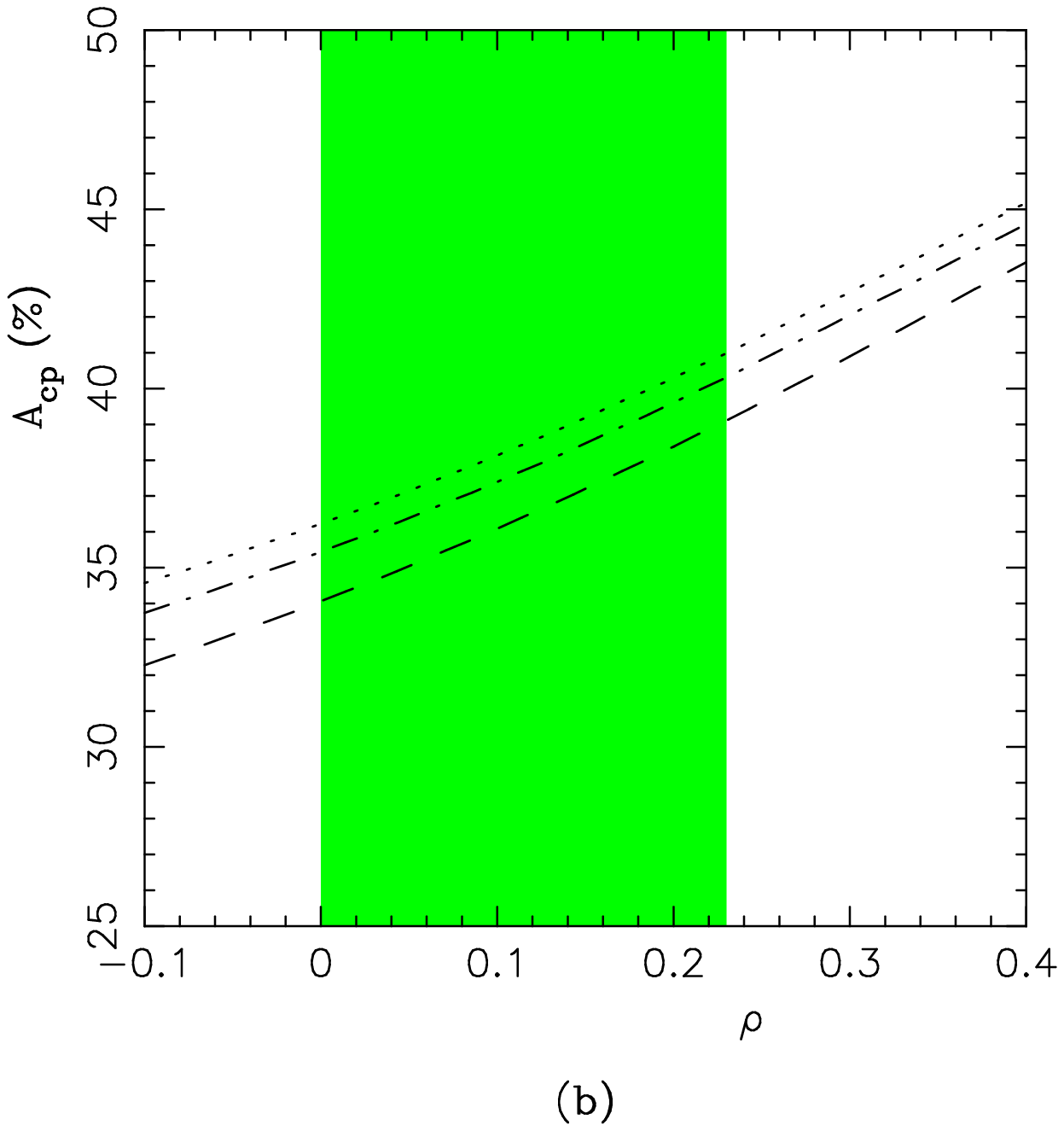,bbllx=5cm,bblly=7cm,bburx=18cm,bbury=19cm,%
width=8cm,height=6.5cm,angle=0}
\caption{CP-violating asymmetry $ A_{CP}$ in 
$\protect\optbar{B^0} \to K_S^0 \eta$ decays as a
function of the CKM parameter $\rho$.
(a) $k^2=m_b^2/2$. The dotted, dashed-dotted and dashed curves correspond to
the CKM parameter values $\eta=0.42$, $\eta=0.34$ and $\eta=0.26$,
respectively;
(b)  $\eta=0.34$. The dotted, dashed-dotted and dashed lines correspond to
$k^2=m_b^2/2+2$  GeV$^2$, $k^2=m_b^2/2$ and $k^2=m_b^2/2-2$ GeV$^2$, 
respectively.} \label{cp2}
\end{figure}

The decay  $B^0\to K_S^0 \eta$, like the preceding decay, is dominated by 
the penguins with significant electroweak penguin contribution \cite{akl98-1}.
The branching ratio for this mode is estimated to be about a factor 3  too
small compared to 
$B^0\to K_S^0 \pi^0$, with ${\cal B} (B^0\to K_S^0 \eta) \simeq 
(1-2)\times 10^{-6}$.
The CP-violating asymmetry $A_{CP} (K_S^0 \eta)$ is, however, found to be
very similar
 to $A_{CP} ( K_S^0 \pi^0)$. 
This is shown in Fig.~\ref{cp2}(a) where we plot  $A_{CP} (K_S^0 \eta)$ as a 
function of $\rho$ for the three indicated values of $\eta$, keeping 
$k^2=m_b^2/2$ fixed.
The $k^2$-dependence of  $A_{CP} (K_S^0 \eta)$ is shown in Fig.~\ref{cp2}(b) 
and is found to be moderately small in the range $k^2=m_b^2/2\pm 2$ GeV$^2$.
We show in Fig.~\ref{bcp2} $A_{CP} (K_S^0 \eta)$ as a function of $\sin 2 
\beta$, with the
three curves showing three different values of $\vert V_{td} \vert$. 
Restricting again
to the $\pm 1 \sigma$ range of $\sin2 \beta$, we estimate: $24\% \leq A_{CP}
 (K_S^0 \eta)
\leq 46\%$.

%
%
%
\begin{figure}
\begin{center}
    \epsfig{file=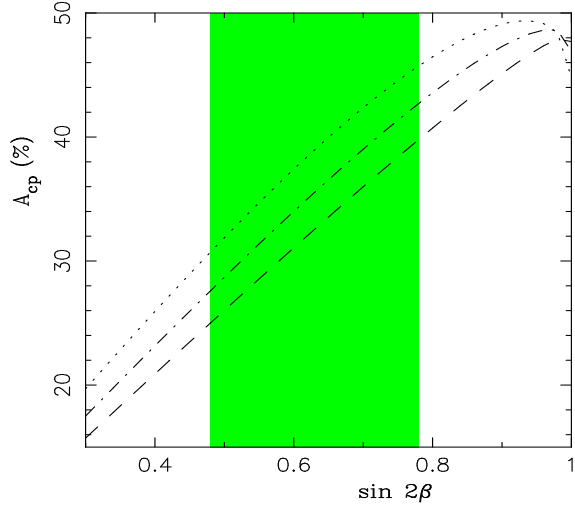,bbllx=5cm,bblly=7cm,bburx=18cm,bbury=19cm,%
width=8cm,height=6.5cm,angle=0}
\caption{CP-violating asymmetry $ A_{CP}$ in
 $\protect\optbar{B^0} \to K_S^0 \eta$ as a function of
$\sin 2 \beta$ for $k^2=m_b^2/2$.  The three curves correspond to the
following values of the CKM matrix element $\vert V_{td}\vert$:
$\vert V_{td}\vert =0.004$ (dashed curve),
$\vert V_{td}\vert =0.008$ (dashed-dotted curve),
$\vert V_{td}\vert= 0.012$ (dotted curve).}
 \label{bcp2}
\end{center}
\end{figure}

\item \underline{CP-violating asymmetry in
$\optbar{B^0} \to K_S^0 h^0$, with $h^0=\pi^0,K_S^0,\eta,\eta^\prime$}
 
%
\begin{figure}
    \epsfig{file=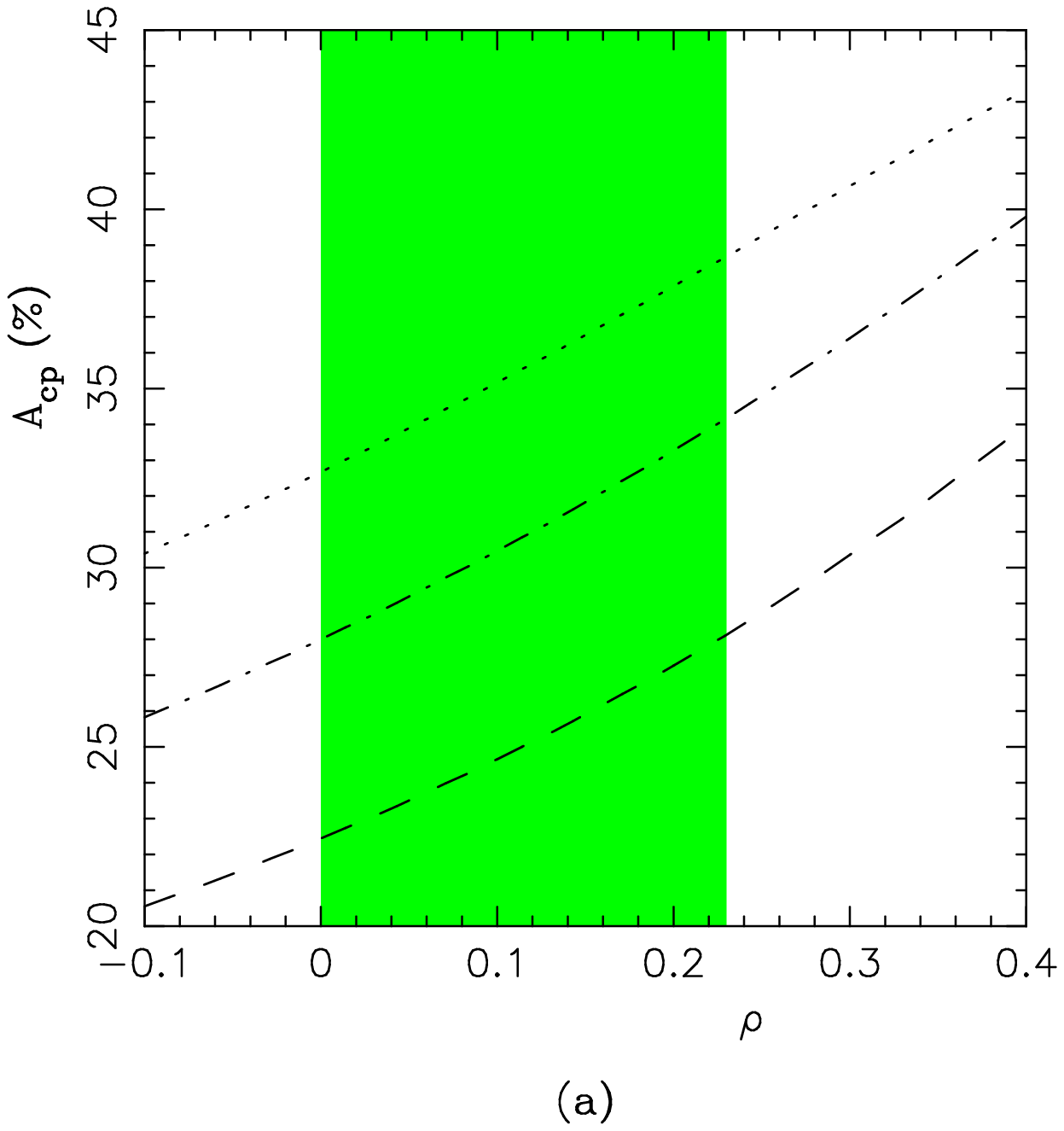,bbllx=5cm,bblly=7cm,bburx=18cm,bbury=19cm,%
width=8cm,height=6.5cm,angle=0}
    \epsfig{file=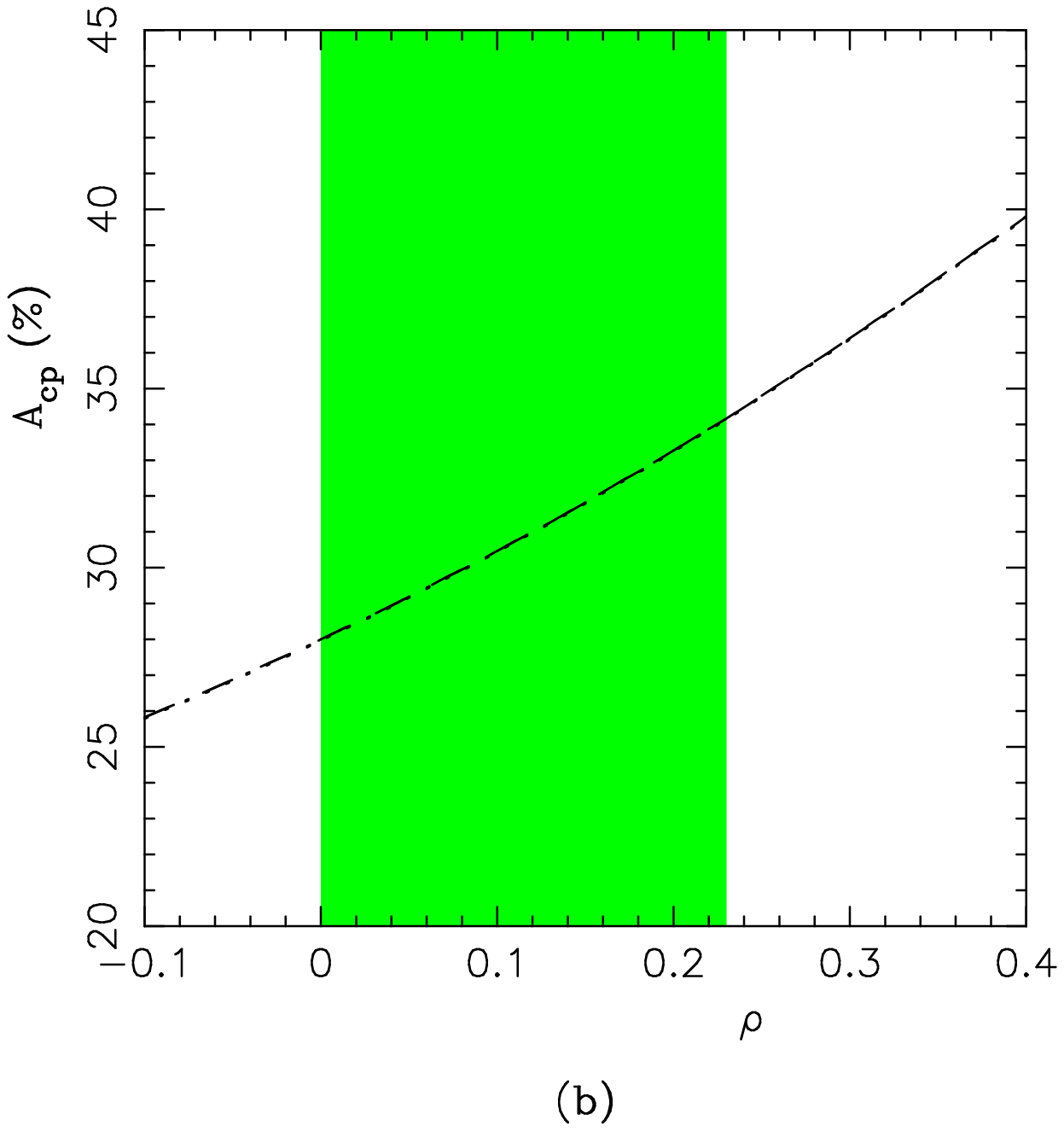,bbllx=5cm,bblly=7cm,bburx=18cm,bbury=19cm,%
width=8cm,height=6.5cm,angle=0}
\caption{CP-violating asymmetry $ A_{CP}$ in
 $\protect\optbar{B^0} \to K_S^0 h^0 $ decays with
$h^0=\pi^0, K_S^0, 
\eta, \eta^\prime$ as a function of the CKM parameter $\rho$.
(a) $k^2=m_b^2/2$. The dotted, dashed-dotted and dashed curves correspond to
the CKM parameter values $\eta=0.42$, $\eta=0.34$ and $\eta=0.26$,
respectively.
(b)  $\eta=0.34$. The overlapping curves correspond to
$k^2=m_b^2/2+2$ GeV$^2$, $k^2=m_b^2/2$ and $k^2=m_b^2/2-2$ GeV$^2$.}
\label{cp6}
\end{figure}

As the CKM-parametric dependence of the CP-violating asymmetries  $A_{CP}
(K_S^0 \pi^0)$,
 $A_{CP} (K_S^0 \eta)$,  $A_{CP}
 (K_S^0 \eta^\prime)$ are very similar, one could combine these asymmetries.
We estimate ${\cal B} ( B^0\to K^0_S h^0) \simeq (2.7-4.6) \times 10^{-5}$,
with  $A_{CP} ( K^0_S h^0)\simeq (22-36)\%$, for $h^0=\pi ^0 , \eta$ and 
$\eta ^\prime$. 
As the branching ratio for the decay $B^0 \to K^0_S  \bar K^0$ is estimated
 to be
small, typically ${\cal B} ( B^0\to K^0_S \bar K^0) \simeq 5 \times 10^{-7}$,
the above estimates of 
${\cal B} ( B^0\to K^0_S h^0)$ and  $A_{CP} ( K^0_S h^0)$ hold also to a 
very good approximation if we now also include $K_S^0$ in $h^0$. 
The  dependence of $A_{CP} ( K^0_S h^0)$ on the CKM parameters $\rho$ and 
$\eta$ is shown in Fig.~\ref{cp6}(a) and the $k^2$-dependence in 
Fig.~\ref{cp6}(b). Interestingly, the $k^2$-dependence in various components
which is already small gets almost canceled in the sum, yielding
$A_{CP}(K_S^0h^0)$ which is practically independent of $k^2$. We show the 
dependence of $A_{CP}(K_S^0 h^0)$ on $\sin 2 \beta$ in Fig.~\ref{bcp6}, 
with the three curves
representing each a different value of $\vert V_{td} \vert$.
Thus, we predict  $A_{CP} ( K^0_S h^0)\simeq (22-36)\%$,
for $h^0=\pi ^0 $, $K_S^0$, $\eta$ and $\eta ^\prime$ for the
$\pm 1 \sigma$ range of $\sin2 \beta$.
 %
%
%
%
\begin{figure}
\begin{center}
    \epsfig{file=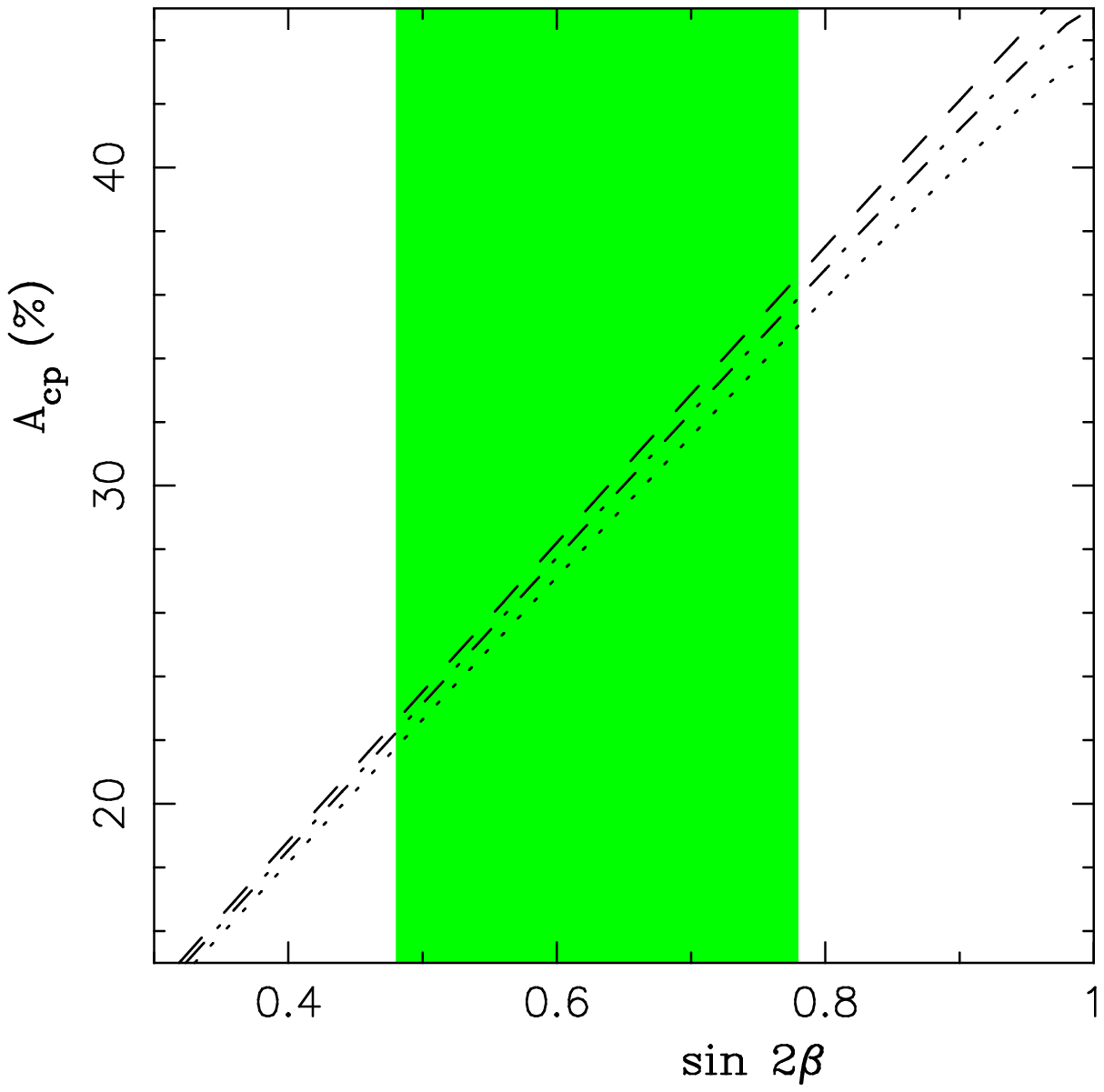,bbllx=5cm,bblly=7cm,bburx=18cm,bbury=19cm,%
width=8cm,height=6.5cm,angle=0}
\caption{CP-violating asymmetry $ A_{CP}$ in
 $\protect\optbar{B^0} \to K_S^0 h^0$ decays with
$h^0=\pi^0, K_S^0,\eta, \eta^\prime$ 
 as a function of
$\sin 2 \beta$ for $k^2=m_b^2/2$.  The three curves correspond to the
following values of the CKM matrix element $\vert V_{td}\vert$:
dashed curve ($\vert V_{td}\vert =0.004$), dashed-dotted curve
($\vert V_{td}\vert =0.008$), dotted curve ($\vert V_{td}\vert =0.012$).}
 \label{bcp6}
\end{center}
\end{figure}

\item \underline{CP-violating asymmetry in $\optbar{B^0} \to \rho^+ \rho^-$}

As another example of the decay whose $A_{CP}$ is stable against variation 
in $N_c$ and $k^2$, we remark that the decay mode $B^0\to \rho^+\rho^-$ is 
estimated to have an asymmetry $A_{CP} \simeq 10\%$, as can be seen in 
Table 11 and 12. 
This decay mode is dominated by the tree amplitudes (like $B^0\to \pi^+\pi^-$)
and belongs to the CP~class (iii) decays.
Estimated branching ratio for this mode is  ${\cal B} ( B^0\to \rho^+ \rho^-)
 \simeq (2-3) \times 10^{-5}$.
\end{itemize}
    
\subsection{The Decays $B^0 \to \rho^+ \pi^-$, $B^0 \to \rho^- \pi^+$ and 
CP-Violating
Asymmetries}

Next, we discuss decay modes which belong to the CP class (iv) decays. 
There are four of them $B^0 \to \bar K^{*0} K_S^0$,  $B^0 \to K^{*0} K_S^0$, 
$B^0 \to \rho^+\pi^-$ and $B^0 \to \rho^-\pi^+$. Of these the decay 
$B^0 \to K^{*0} K_S^0$ belongs to the Class-V decay and is estimated to 
have a very small branching ratio in the factorization approach
 ${\cal B} (B^0 \to K^{*0} K_S^0)\simeq O(10^{-9})$ \cite{akl98-1}. 
The other $B^0 \to \bar K^{*0} K_S^0$ is a Class-IV decay but is expected
 to have also a small branching ratio, with  ${\cal B} (B^0 \to\bar  K^{*0}
 K_S^0)\simeq (2-3)\times 10^{-7}$.
In view of this, we concentrate on the decays $B^0 \to \rho^+\pi^-$ and 
$B^0 \to \rho^-\pi^+$.
%
%
\begin{figure}
    \epsfig{file=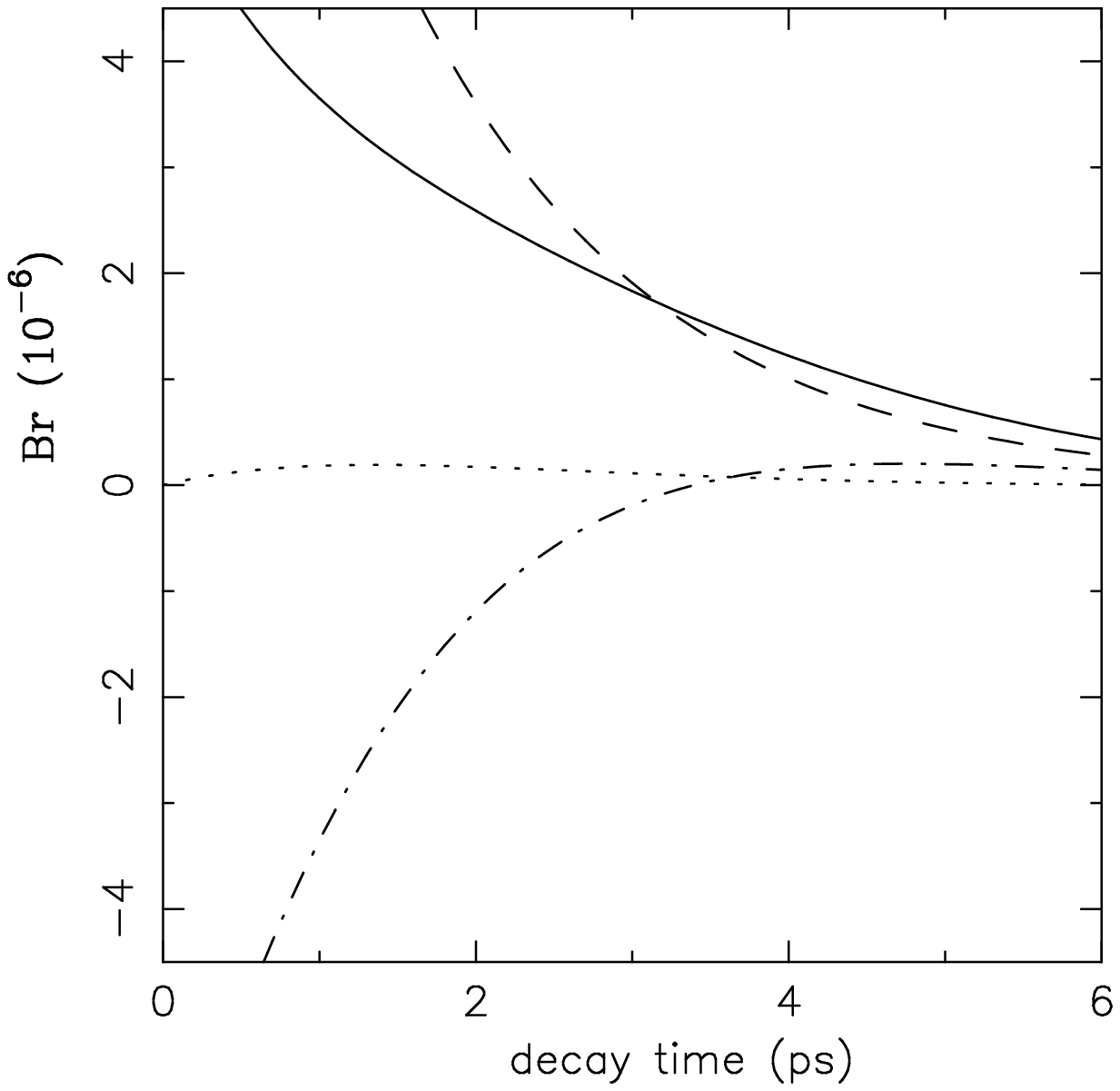,bbllx=5cm,bblly=7cm,bburx=18cm,bbury=19cm,%
width=8cm,height=6.5cm,angle=0}
  \epsfig{file=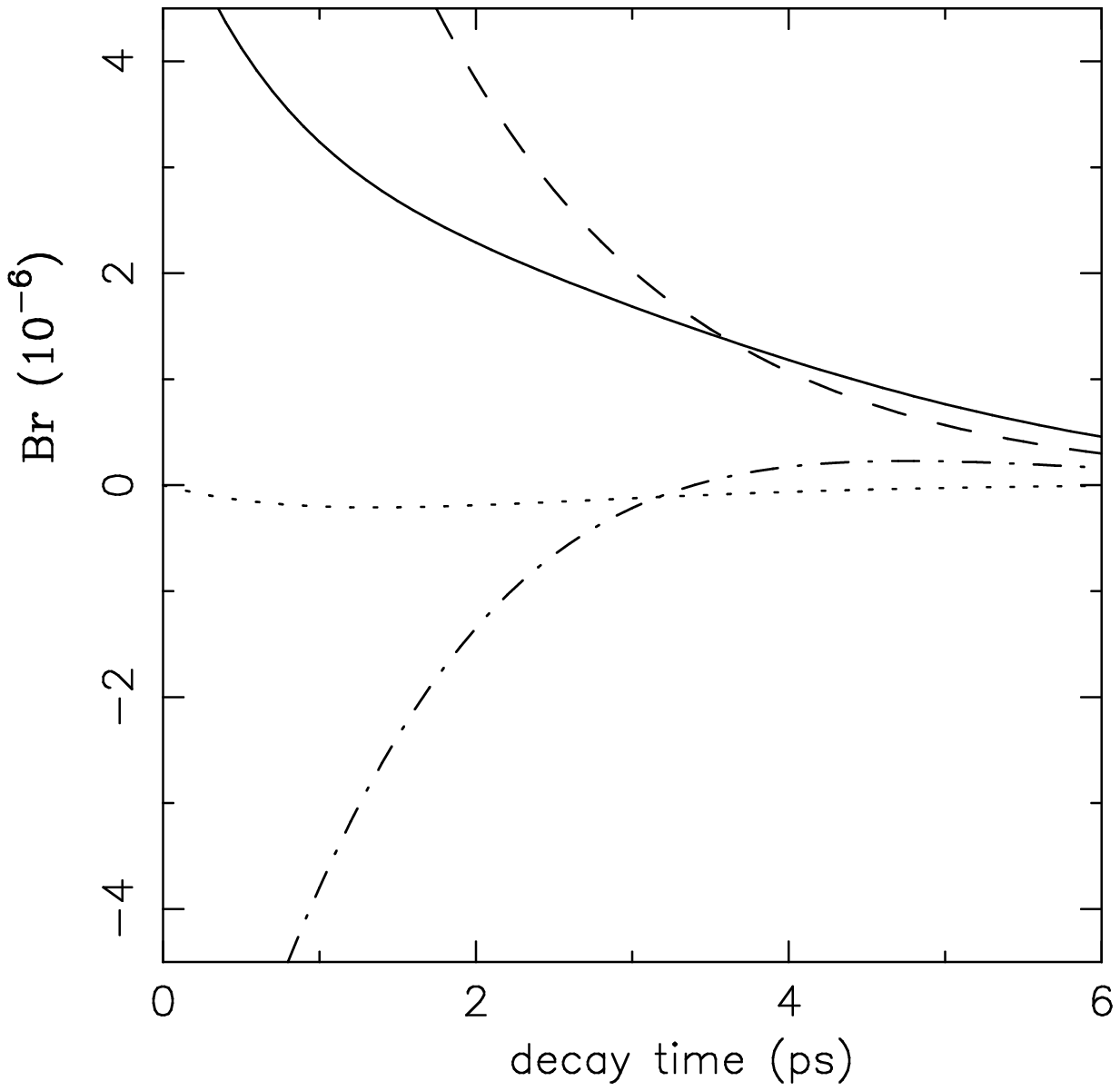,bbllx=5cm,bblly=7cm,bburx=18cm,bbury=19cm,%
width=8cm,height=6.5cm,angle=0}
\caption{ Time-dependent branching ratio 
for the decays $ B^0 \to \rho^-\pi^+$ (left)
and $\bar B^0 \to \rho^+\pi^-$ (right) as a function of the decay time.
The dashed, dashed-dotted and dotted curves correspond to 
the contributions from the exponential decay term $e^{-\Gamma t}$,
$e^{-\Gamma t}\cos\Delta m t$ and
$ e^{-\Gamma t} \sin \Delta m t$ in eq.~(\ref{rate}), respectively.
The solid curve is the sum of the three contributions.}
 \label{bra1}
\end{figure}

\begin{figure}
    \epsfig{file=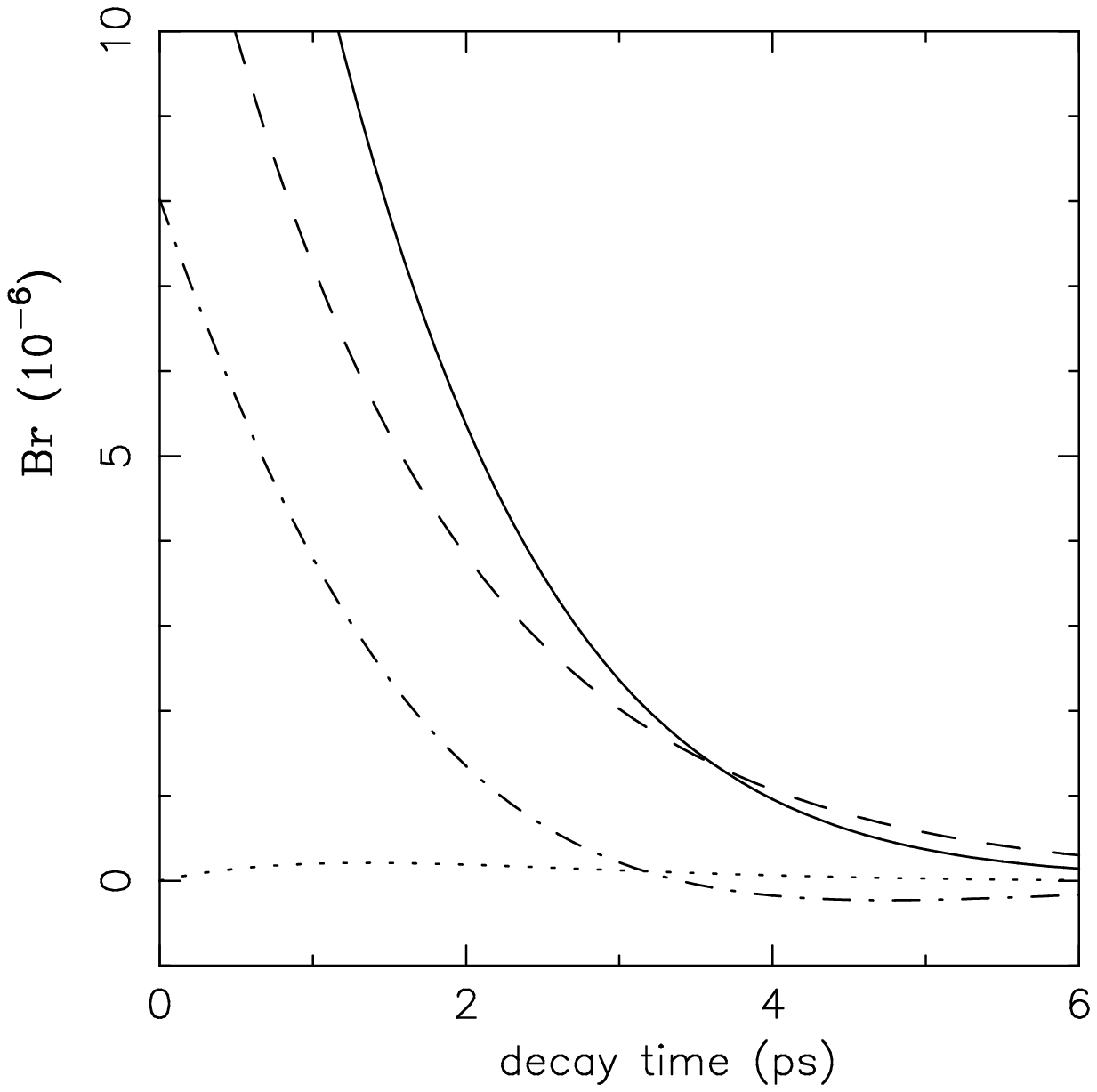,bbllx=5cm,bblly=7cm,bburx=18cm,bbury=19cm,%
width=8cm,height=6.5cm,angle=0}
  \epsfig{file=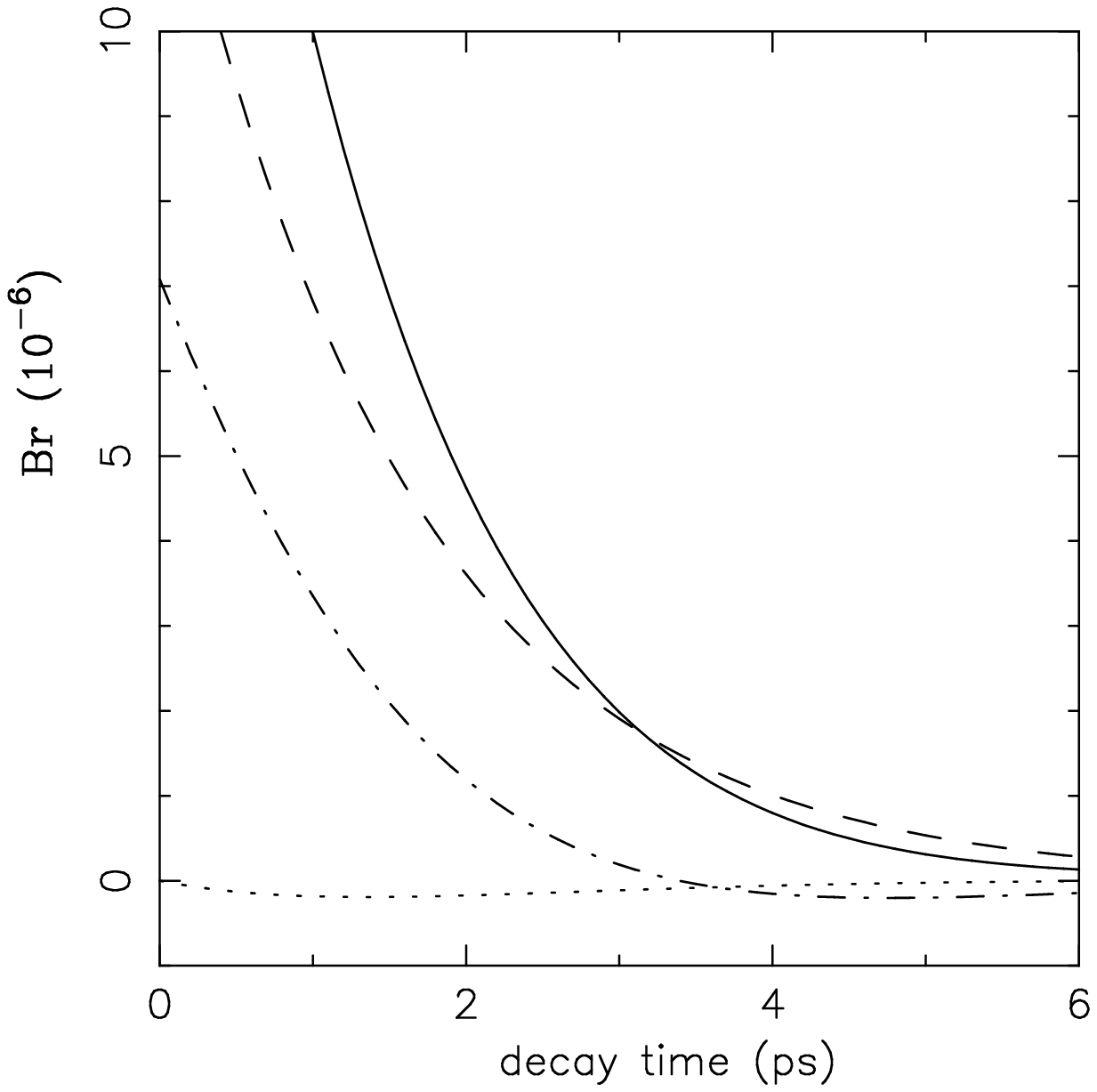,bbllx=5cm,bblly=7cm,bburx=18cm,bbury=19cm,%
width=8cm,height=6.5cm,angle=0}
\caption{ Time-dependent branching ratio for the
decays $ B^0 \to \rho^+\pi^-$ (left)
 and  $\bar B^0 \to \rho^-\pi^+$ (right) as a function of the decay time. 
The dashed, dashed-dotted and dotted curves correspond to
the contributions from the exponential decay term $e^{-\Gamma t}$,
$e^{-\Gamma t}\cos\Delta m t$ and
$ e^{-\Gamma t} \sin \Delta m t$ in eq.~(\ref{rate}), respectively.
The solid curve is the sum of the three contributions.}
\label{bra3}
\end{figure}

 With $f=\rho^+ \pi^-$ and $\bar{f}=\rho^-\pi^+$, the time evolution of the
 four
branching ratios is given in eq.~(\ref{aepsilon}). They have each three 
components with
characteristic time-dependences proportional to $e ^{-\Gamma t}$,
$e ^{-\Gamma t} \cos \Delta mt$ and $e ^{-\Gamma t} \sin \Delta mt$,
 with the relative and overall normalization explicitly stated there.
The time dependence of the branching ratio ${\cal B} (B^0 (t) \to
 \rho^-\pi^+)$ and of the branching ratio for the charge conjugate decay  
${\cal B} (\overline B^0 (t)\to
\rho^+\pi^-)$
is shown in Fig.~\ref{bra1}(a) and \ref{bra1}(b), respectively.
The time dependence of the branching ratio ${\cal B} (B^0 (t) \to \rho^+\pi^-)$
and  of ${\cal B} (\overline B^0 (t)\to \rho^-\pi^+)$
is shown in Fig.~\ref{bra3}(a) and \ref{bra3}(b), respectively.
The three components and the sum are depicted by the four curves.

\begin{figure}
  \begin{center}
    \epsfig{file=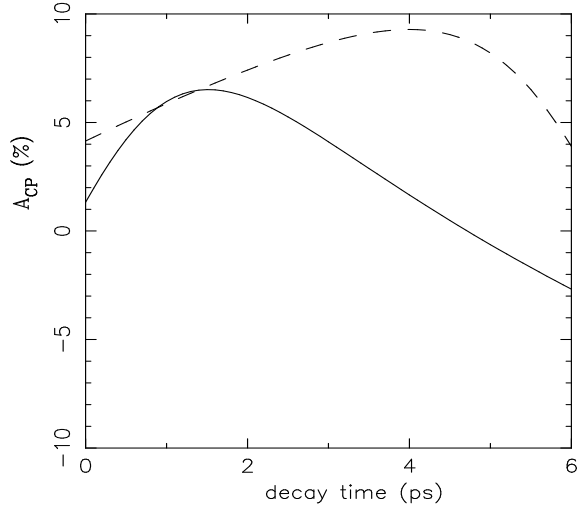,bbllx=5cm,bblly=7cm,bburx=18cm,bbury=19cm,%
width=8cm,height=6.5cm,angle=0}
\caption{Time-dependent CP-violating asymmetry $ A_{CP} (t;\rho^-\pi^+)$
(solid curve) and $A_{CP}(t;  \rho^+\pi^-)$ (dashed curve)
as a function of the decay time, with $\rho=0.12, \eta=0.34$ and 
$k^2=m_b^2/2$.}
 \label{cbrap}
  \end{center}
\end{figure}

The resulting time-dependent CP-violating asymmetry $A_{CP}(t)$ for 
$B^0 \to \rho^-\pi^+$ defined as 
\begin{equation}
  \label{cpt}
  A_{CP} (t;\rho^-\pi^+) \equiv \frac{\Gamma (B^0 (t) \to \rho^-\pi^+)
- \Gamma (\overline B^0 (t)\to \rho^+\pi^-)}
{\Gamma  (B^0 (t) \to \rho^-\pi^+)+ \Gamma(\overline B^0 (t)\to \rho^+\pi^-)
 },
\end{equation}
is shown in Fig.~\ref{cbrap} through the solid curve.
The corresponding asymmetry $ A_{CP} (t;\rho^+\pi^-)$ defined in an 
analogous way as for  $A_{CP} (t;\rho^-\pi^+) $ is given by the dashed curve 
in this figure.

We recall that the decay rate for $B^0 \to \rho^+\pi^-$ averaged over its 
charge conjugated decay $\overline B^0 \to \rho^-\pi^+$ is estimated to have 
a value in the range ${\cal B} ( B^0 \to \rho^+\pi^-) \simeq (2$-$4) \times 
10^{-5}$ \cite{akl98-1}; the time-integrated CP-asymmetry is estimated to be 
$A_{CP}  (\rho^+\pi^-) \simeq  (4$-$7) \%$.
Being a Class-I decay, both the branching ratio ${\cal B} ( B^0 \to \rho^+
\pi^-)$ and $A_{CP}  (\rho^+\pi^-)$ are $N_c$-stable.
In addition,  $A_{CP}  (\rho^+\pi^-)$ is also $k^2$-stable, as shown in 
Table 7.

The branching ratio for the decay $B^0 \to \rho^-\pi^+$, averaged over 
its charge conjugate decay $\overline B^0\to \rho^+\pi^-$, is expected to be 
 ${\cal B} ( B^0 \to \rho^-\pi^+) \simeq (6$-$9) \times 10^{-6}$ 
\cite{akl98-1}, i.e.,
 typically a factor 4 smaller than ${\cal B} ( B^0 \to \rho^+ \pi^-)$.
Also, $A_{CP}  (\rho^-\pi^+)$ is estimated somewhat smaller for the central
 value of the CKM-parameter $\rho=0.12$, $\eta=0.34$.
For these CKM parameter, we estimate $A_{CP}  (\rho^-\pi^+) \simeq (3$-$4)\%$.
For $\rho=0.23$, $\eta=0.42$,  $A_{CP}  (\rho^-\pi^+) \simeq 
A_{CP}  (\rho^+\pi^-) \simeq O(5\%)$ (see Table 8).

We note that our estimate of the ratio ${\cal B} ( B^0 \to \rho^+ \pi^-)/
{\cal B} ( B^0 \to \pi^+ \pi^-) \simeq 2.3$ derived in \cite{akl98-1} is 
in reasonable agreement with the corresponding ratio estimated in \cite{adkd91}
but we also find ${\cal B} ( B^0 \to \rho^- \pi^+)/ {\cal B} ( B^0 \to 
\rho^+ \pi^-)
\simeq 0.27$, which is drastically different from the estimates presented in 
\cite{adkd91}.

\subsection{Decay Modes with Measurable but $k^2$-dependent $A_{CP}$}

In addition to the decay modes discussed above, 
the following decay modes have $A_{CP}$ which are $N_c$- and
$\mu$- stable but show significant or strong $k^2$-dependence.
However, we think that further theoretical work and/or measurements of
$A_{CP}$ in one or more of the following decay modes will greatly help in
determining $k^2$ and hence in reducing the present theoretical dispersion
on $A_{CP}$.
%
%
%
\begin{figure}
    \epsfig{file=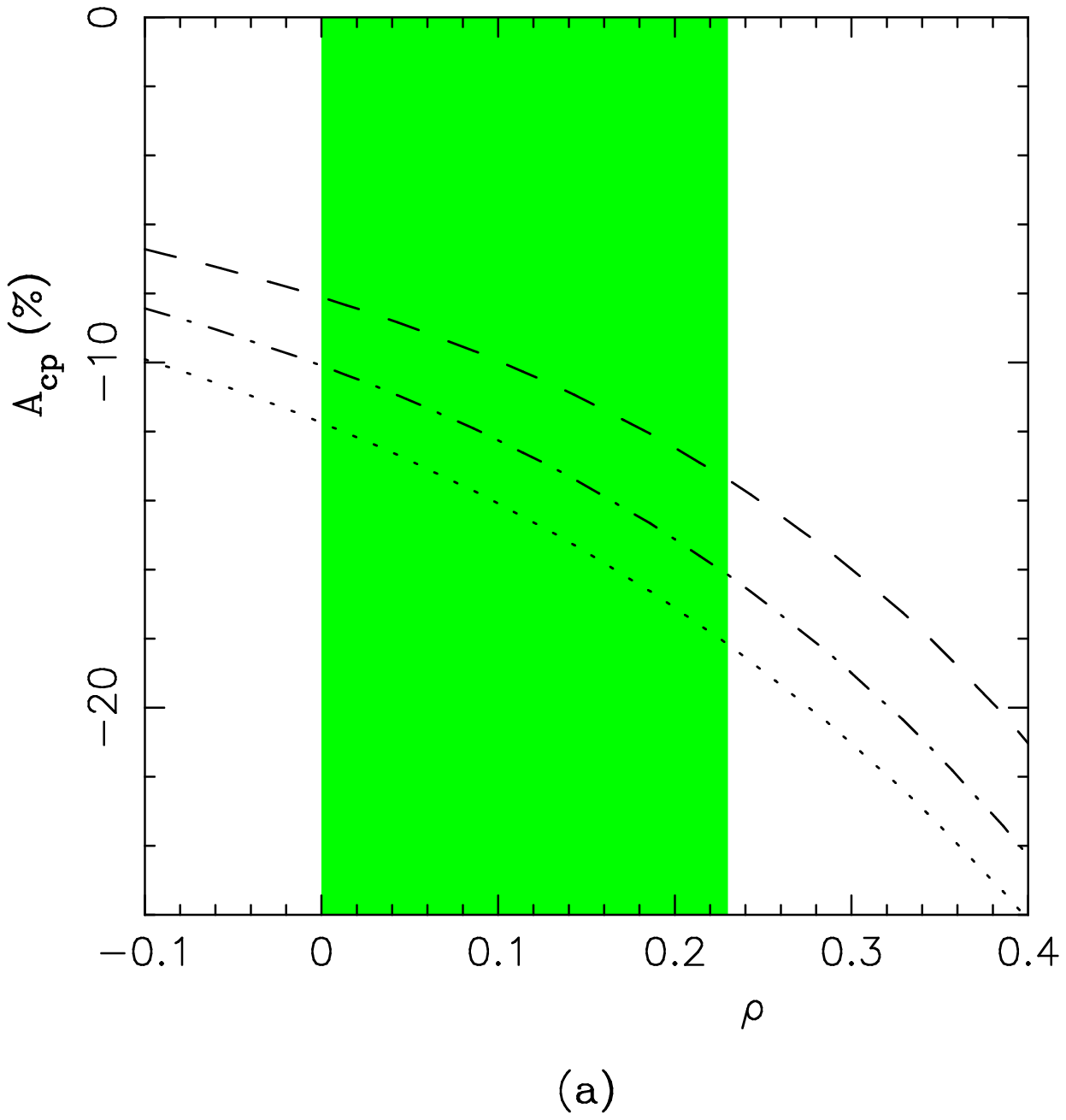,bbllx=5cm,bblly=7cm,bburx=18cm,bbury=19cm,%
width=8cm,height=6.5cm,angle=0}
    \epsfig{file=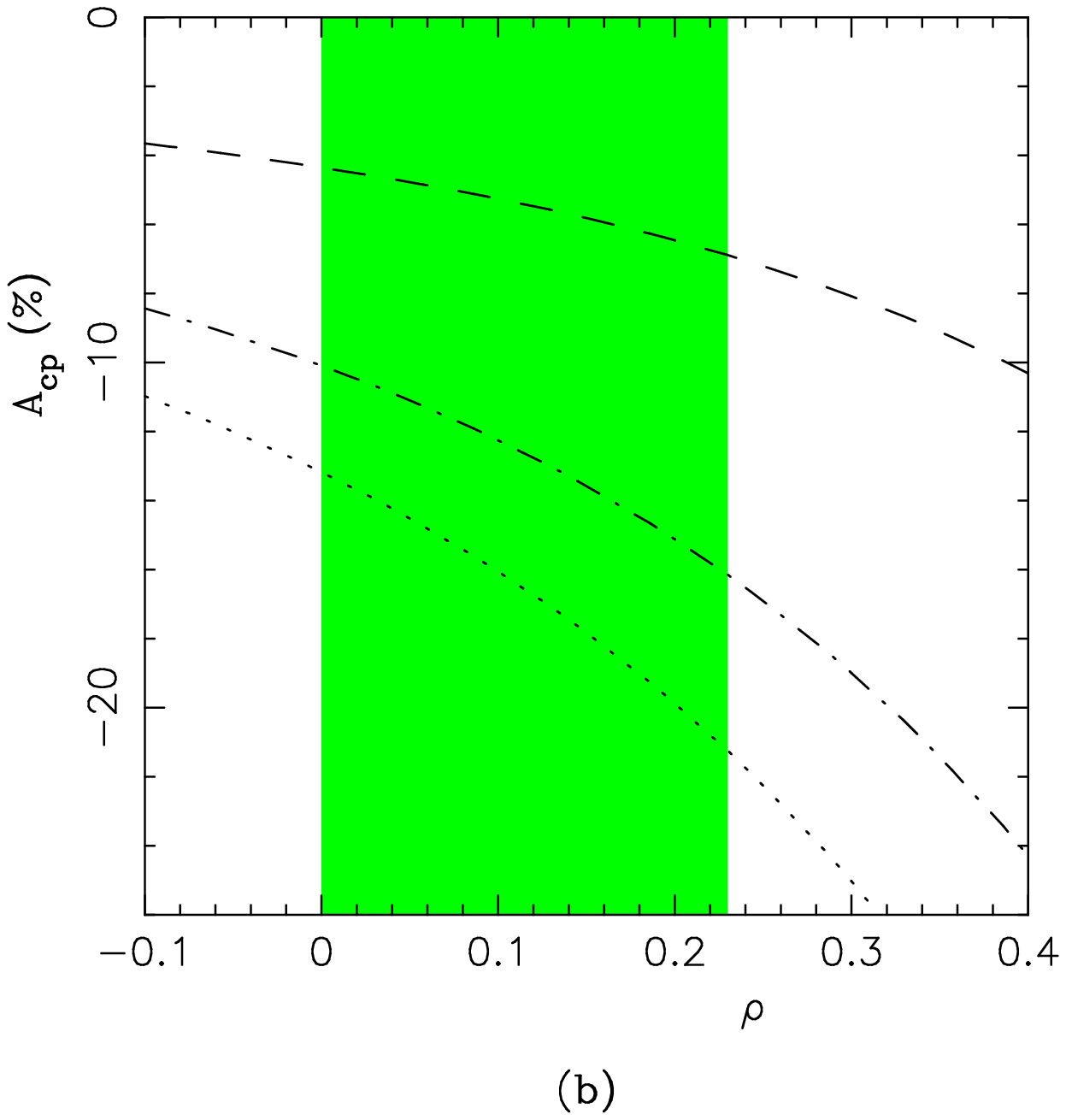,bbllx=5cm,bblly=7cm,bburx=18cm,bbury=19cm,%
width=8cm,height=6.5cm,angle=0}
\caption{CP-violating asymmetry $A_{CP}$ in the decays
$ B^\pm\to K^{*\pm}\pi^0$ as a function of the CKM parameter $\rho$.
(a) $k^2=m_b^2/2$. The dotted, dashed-dotted and dashed curves correspond to
the CKM parameter values $\eta=0.42$, $\eta=0.34$ and $\eta=0.26$,
respectively.
(b)  $\eta=0.34$. The dotted, dashed-dotted and dashed curves correspond to
$k^2=m_b^2/2+2$  GeV$^2$, $k^2=m_b^2/2$ and $k^2=m_b^2/2-2$ GeV$^2$, 
respectively.} \label{cp3}
\end{figure}
\begin{itemize}
\item{$B^\pm\to \pi^\pm\eta^\prime$,  $\optbar{B^0} \to K^{*\pm}\pi^\mp$,
$B^\pm\to K^{*\pm}\pi^0$, $B^\pm \to K^{*\pm} \eta$, $B^\pm \to K^{*\pm} 
\eta^\prime$,
$\optbar{B^0}\to K^{*\pm} \rho^\mp$, $B^\pm\to K^{*\pm} \rho^0$.}
\end{itemize}
 
These decays have branching ratios which are estimated to be several multiples
of $10^{-5}$ to several multiples of $10^{-6}$ and may have $|A_{CP}|$ at
least of $O(5\%)$, but being uncertain due to the $k^2$-dependence may reach
rather large values. The CP-violating asymmetries in these cases belong to the 
class (i), i.e., they are direct CP-violating asymmetries. 
 
\begin{figure}
    \epsfig{file=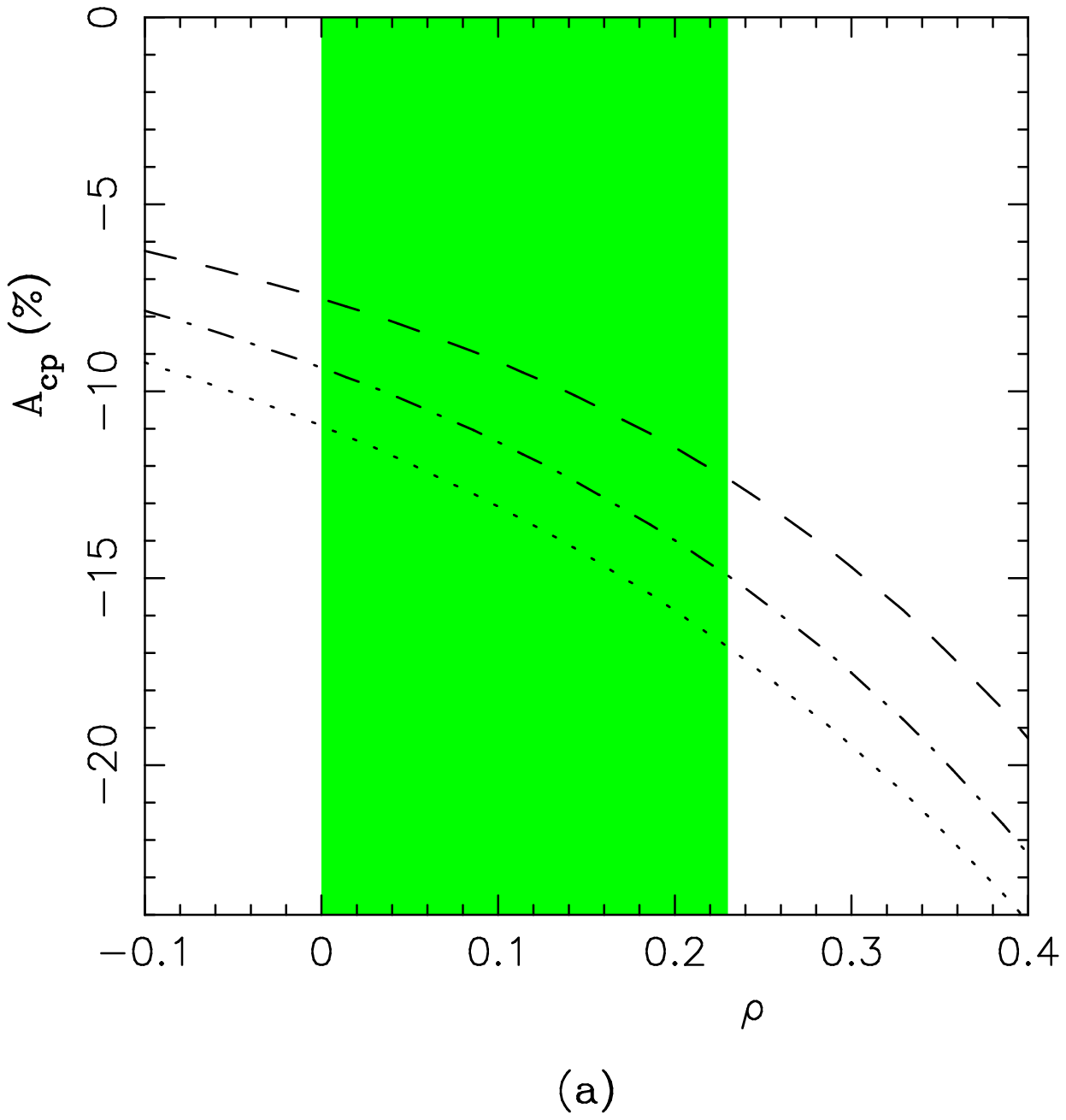,bbllx=5cm,bblly=7cm,bburx=18cm,bbury=19cm,%
width=8cm,height=6.5cm,angle=0}
     \epsfig{file=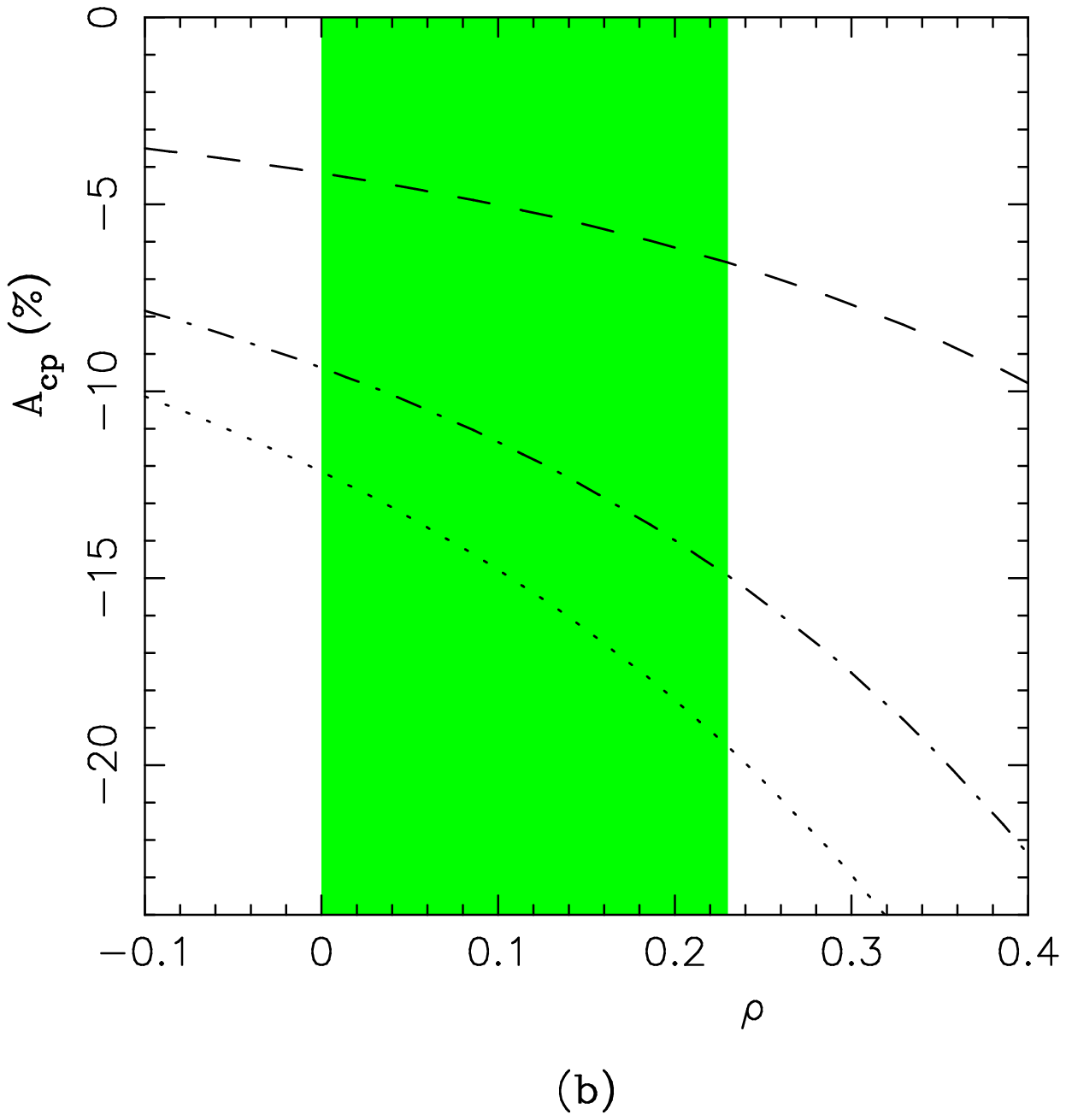,bbllx=5cm,bblly=7cm,bburx=18cm,bbury=19cm,%
width=8cm,height=6.5cm,angle=0}
\caption{CP-violating asymmetry $A_{CP}$ in $B^\pm\to K^{*\pm}\rho^0$
decays  as a function of the CKM parameter $\rho$.
(a) $k^2=m_b^2/2$. The dotted, dashed-dotted and dashed curves correspond to
the CKM parameter values $\eta=0.42$, $\eta=0.34$ and $\eta=0.26$,
respectively.
(b)  $\eta=0.34$. The dotted, dashed-dotted and dashed curves correspond to
$k^2=m_b^2/2+2$ GeV$^2$, $k^2=m_b^2/2$ and $k^2=m_b^2/2-2$ GeV$^2$, 
respectively.}
 \label{cp7}
\end{figure}

In Fig.~\ref{cp3}(a) and \ref{cp3}(b), we show the CP-violating asymmetry
$A_{CP} 
(K^{*\pm}\pi^0)$ as a function of $\rho$. 
The three curves in Fig.~\ref{cp3}(a) correspond to the three choices of 
$\eta$,
with $k^2=m_b^2/2$, whereas the three curves in Fig.~\ref{cp3}(b) correspond
to using $k^2=m_b^2/2+2$ GeV$^2$ (dotted curve), $k^2=m_b^2/2$ (dashed-dotted  
curve), $k^2=m_b^2/2-2$ GeV$^2$ (dashed curve) with $\eta =0.34$.
Depending on the value of $k^2$, $A_{CP} (K^{*\pm}\pi^0)$ could reach a 
value  $-25\%$. 
The branching ratio is estimated to lie in the range  ${\cal B} (B^+ \to
K^{*+}\pi^0)\simeq (4-7)\times 10^{-6}$. 
The decay mode $B^+ \to K^{*+}\rho^0$ has very similar CKM and 
$k^2$-dependence, which is shown in Fig.~\ref{cp7}(a) and (b), respectively,
where we plot the CP-asymmetry $A_{CP} (K^{*\pm}\rho^0)$.
Also, the branching ratio ${\cal B} (B^+ \to
K^{*+}\rho^0)\simeq (5-8)\times 10^{-6}$ estimated in \cite{akl98-1} is 
very similar to $B^+ \to K^{*+}\pi^0$.

%
%

\begin{figure}
    \epsfig{file=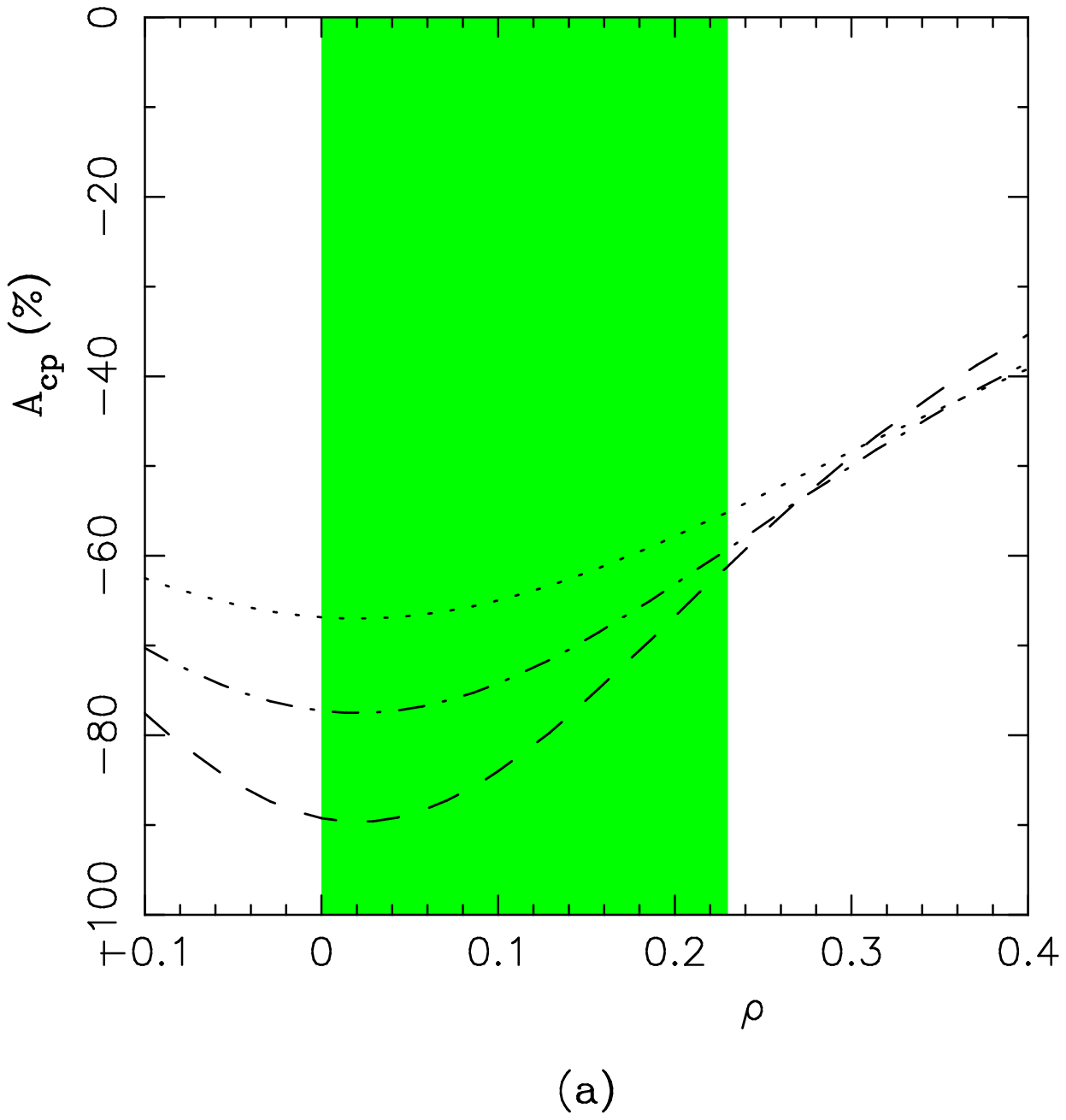,bbllx=5cm,bblly=7cm,bburx=18cm,bbury=19cm,%
width=8cm,height=6.5cm,angle=0}
    \epsfig{file=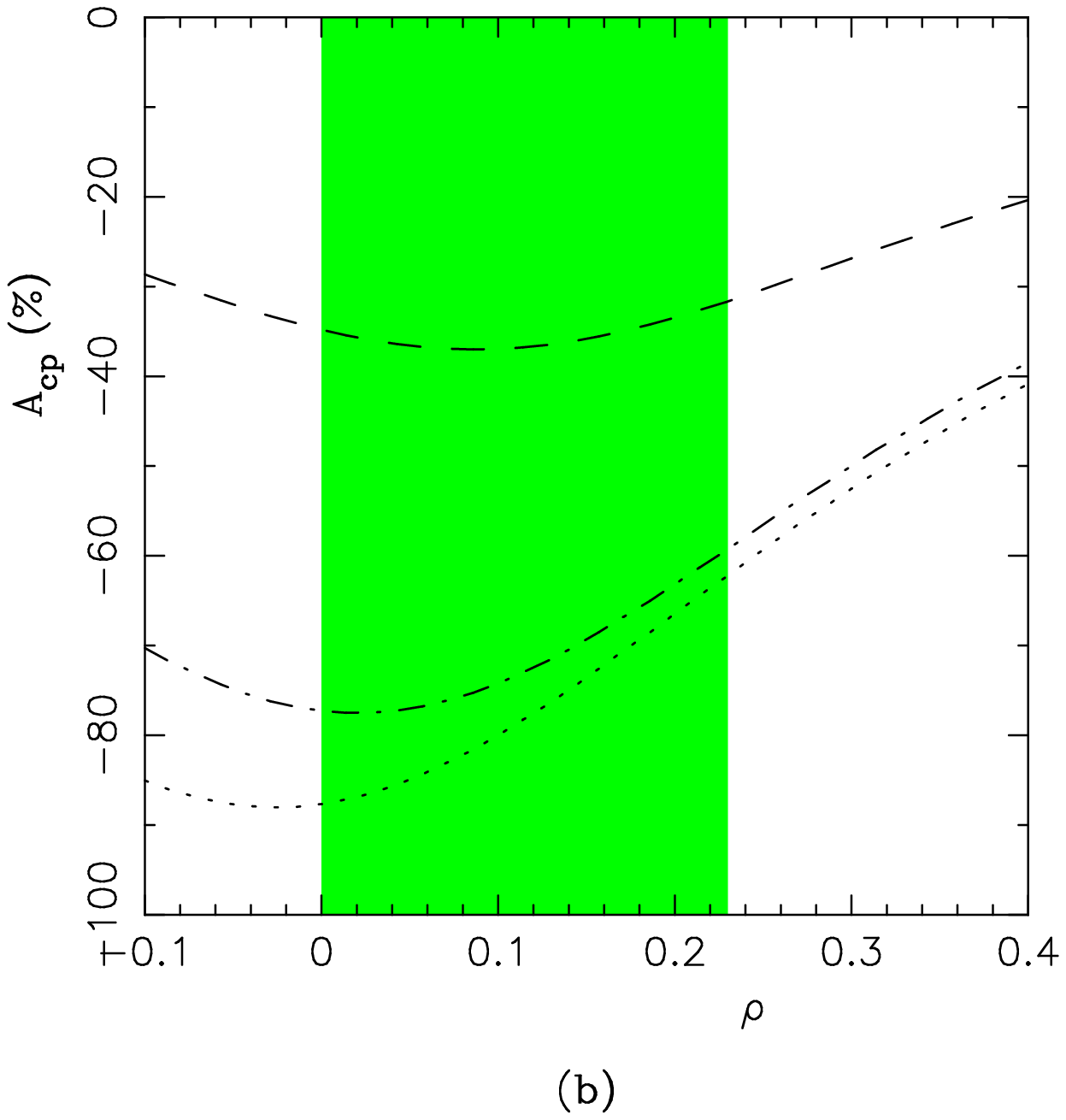,bbllx=5cm,bblly=7cm,bburx=18cm,bbury=19cm,%
width=8cm,height=6.5cm,angle=0}
\caption{CP-violating asymmetry $A_{CP}$ in $B^\pm\to K^{*\pm}\eta^\prime$
decays as a function of the CKM parameter $\rho$.
(a) $k^2=m_b^2/2$. The dotted, dashed-dotted and dashed curves correspond to
the CKM parameter values $\eta=0.42$, $\eta=0.34$ and $\eta=0.26$,
respectively.
(b)  $\eta=0.34$. The dotted, dashed-dotted and dashed curves correspond to
$k^2=m_b^2/2+2$ GeV$^2$, $k^2=m_b^2/2$ and $k^2=m_b^2/2-2$ GeV$^2$, 
respectively.}
    \label{cc4}
\end{figure}

In Fig.~\ref{cc4}(a) and \ref{cc4}(b), 
we show the CP-violating asymmetry $A_{CP}(K^{*\pm}\eta^\prime)$
in the decays  $B^\pm\to K^{*\pm}\eta^\prime$.
This is a Class-III decay dominated by the tree amplitude and is expected 
to have a branching ratio ${\cal B} (B^+ \to
K^{*+}\eta^\prime) \simeq 3\times 10^{-7}$, where an average over the
charge conjugated decays is implied.
However, depending on the value of $k^2$ this decay mode may show a large 
CP-violating asymmetry, reaching  $A_{CP}(K^{*+}\eta^\prime)\simeq -90\%$
for $\rho=0.12$, $\eta=0.34$ and $k^2=m_b^2/2 +2 $ GeV$^2$.
For  $k^2=m_b^2/2 -2 $ GeV$^2$, the CP-asymmetry comes down to a value 
$A_{CP}(K^{*\pm}\eta^\prime)\simeq -20\%$.
All of these values are significantly higher than the ones reported in
 \cite{Petrov97}.
Large but $k^2$-sensitive values of this quantity have also been reported 
earlier in \cite{kps}. 
We also mention here the decay modes $B^\pm\to K^{*\pm} \eta$, whose branching 
ratio is estimated as  ${\cal B} (B^+ \to K^{*+} \eta) \simeq (2-3)\times
 10^{-6}$ \cite{ag,acgk,akl98-1} and which may have CP-violating asymmetry 
in the 
range $A_{CP}(K^{*\pm}\eta)\simeq -(4$-$15)\%$ depending on the CKM 
parameters and $k^2$ (see Tables 9 and 10).

%
%
%

\begin{figure}
    \epsfig{file=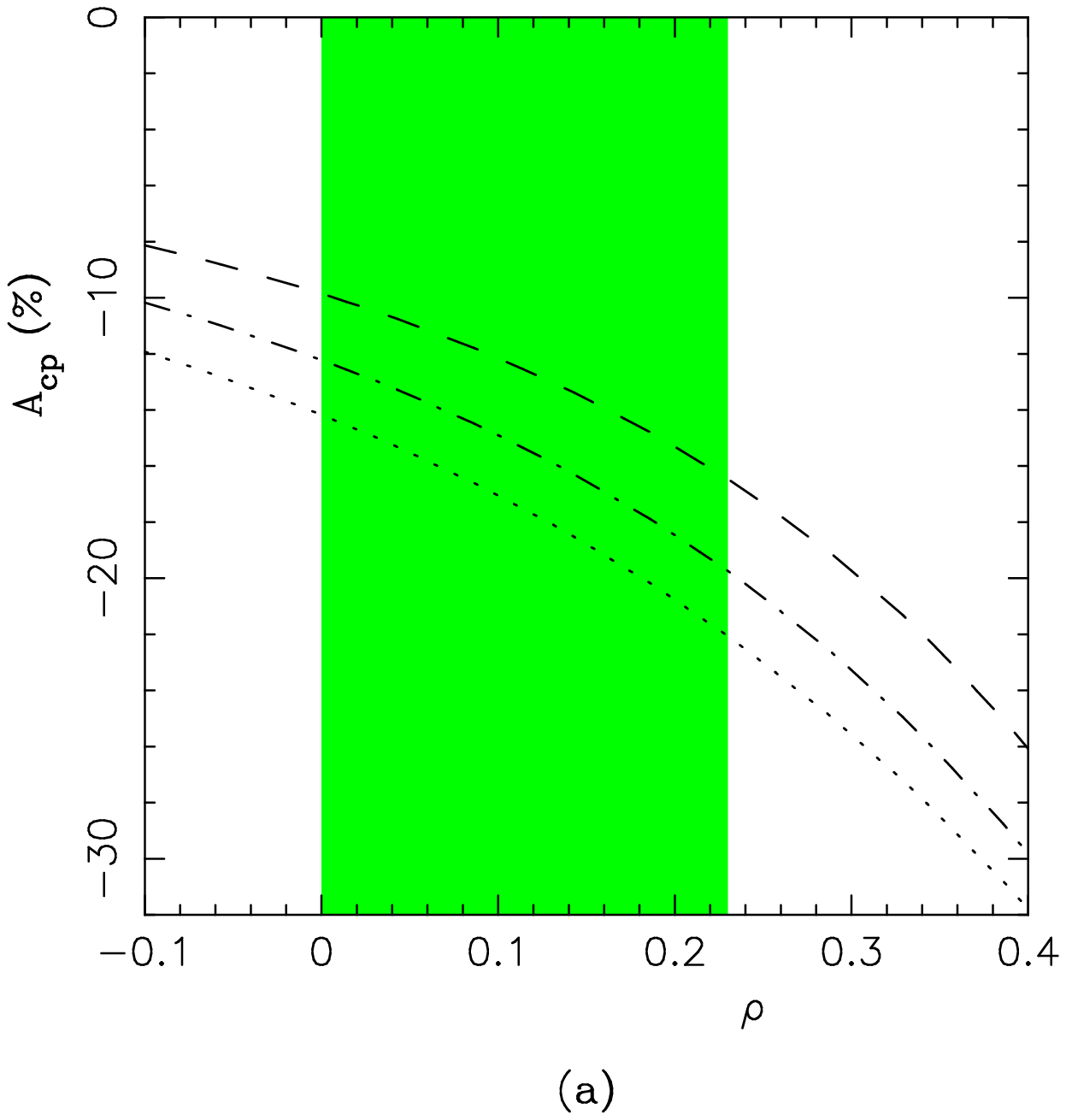,bbllx=5cm,bblly=7cm,bburx=18cm,bbury=19cm,%
width=8cm,height=6.5cm,angle=0}
    \epsfig{file=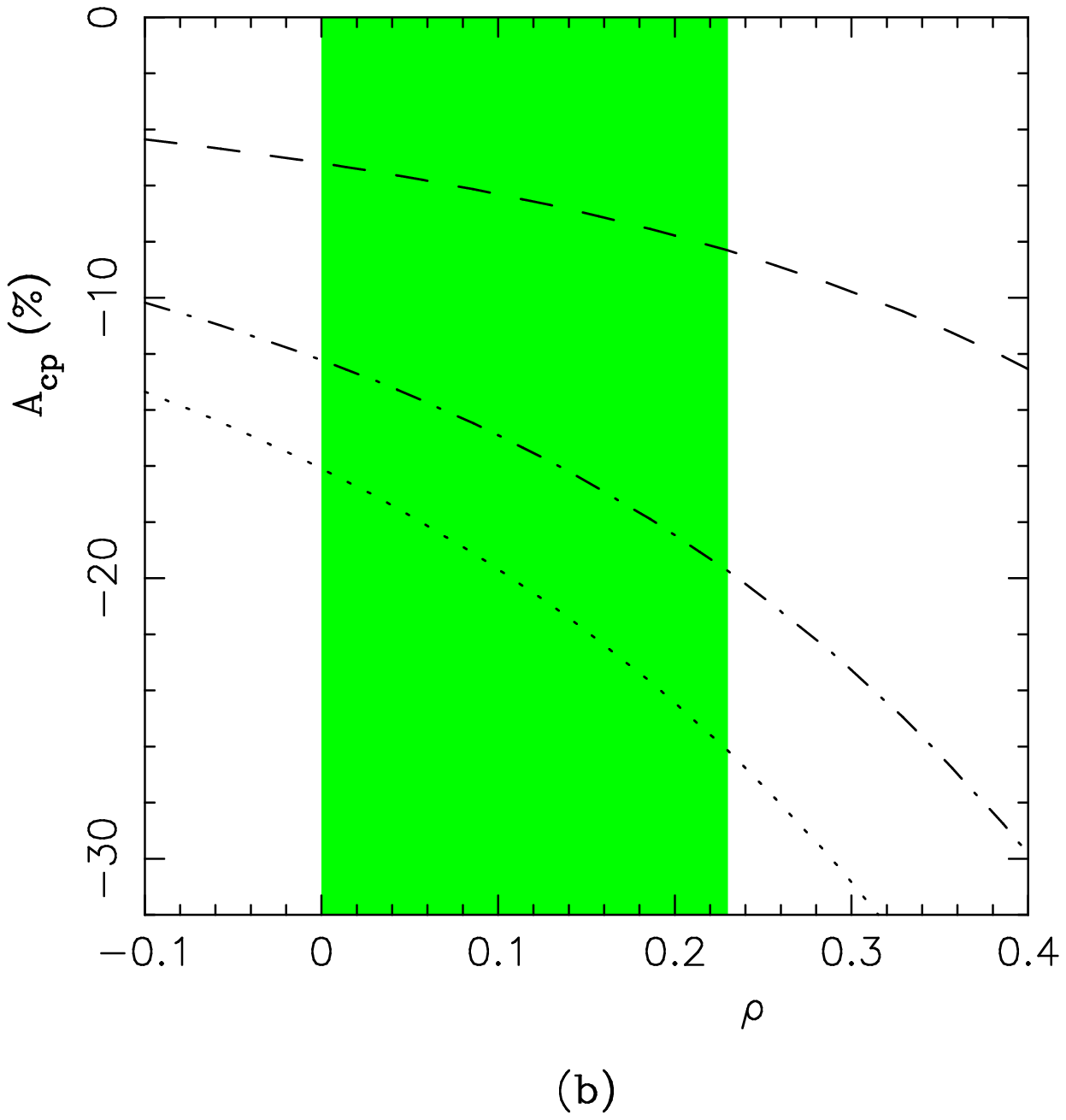,bbllx=5cm,bblly=7cm,bburx=18cm,bbury=19cm,%
width=8cm,height=6.5cm,angle=0}
\caption{CP-violating Asymmetry $A_{CP}$ in $ \protect\optbar{B^0}\to
K^{*\pm}\pi^\mp$
decays  as a function of the CKM parameter $\rho$.
(a) $k^2=m_b^2/2$. The dotted, dashed-dotted and dashed curves correspond to
the CKM parameter values $\eta=0.42$, $\eta=0.34$ and $\eta=0.26$,
respectively.
(b)  $\eta=0.34$. The dotted, dashed-dotted and dashed curves correspond to
$k^2=m_b^2/2+2$  GeV$^2$, $k^2=m_b^2/2$ and $k^2=m_b^2/2-2$ GeV$^2$, 
respectively.} \label{cc2}
\end{figure}

Finally, we mention two more decay modes $B^0 \to K^{*+} \pi^-$ and 
 $B^0 \to K^{*+} \rho^-$ which are both Class-IV decays, with branching ratios 
estimated as   ${\cal B} (B^0 \to K^{*+} \pi^-) \simeq (6-9)\times
 10^{-6}$ and  ${\cal B} (B^0 \to K^{*+} \rho^-) \simeq (5-8)\times
 10^{-6}$ \cite{akl98-1}. The CP-violating asymmetries in these decays are
estimated 
to lie in the range  $A_{CP} (K^{*\pm} \pi^\mp) =
 A_{CP} (K^{*\pm} \rho^\mp) \simeq -(6$-$30) \%$.
In Fig.~\ref{cc2}(a) and \ref{cc2}(b), we show $A_{CP} (K^{*\pm} \pi^\mp)$
as a function of $\rho$ by varying $\eta$ and $k^2$, respectively.
%
%
%

\section{Summary and Conclusions}

  Using the NLO perturbative framework and a generalized factorization 
approach discussed in detail in \cite{akl98-1}, we have calculated the
CP-violating asymmetries
in partial decay rates of all the two-body non-leptonic decays $B \to h_1 
h_2$, where $h_1$ and $h_2$ are the light pseudoscalar and vector
mesons. Our results can be summarized as follows:
\begin{itemize}
\item We find that the decay classification scheme presented in 
\cite{akl98-1} for the branching ratios is also very useful in
discussing the CP-violating asymmetries. In line with this, Class-I and 
class-IV 
decays yield asymmetries which are stable against the variation of $N_c$.
There are two exceptions, $A_{CP}(K^\pm\eta)$ and 
$A_{CP}(K_S^0 \eta)$, which vary by a factor 3 and 1.65, respectively,
for $2 \leq N_c \leq \infty$.  
\item Estimates of CP-violating asymmetries in  Class-II and class-V
decays depend rather sensitively on $N_c$ and hence are very 
unreliable. There is one notable exception $A_{CP}(\phi K_S^0)$, which is
parametrically stable and large. However being a class-V decay, the
branching ratio ${\cal B}(B^0 \to \phi K_S^0)$ is uncertain in the
factorization approach by at least an order of magnitude \cite{akl98-1}.
\item  The CP-asymmetries in Class-III decays vary by approximately a 
factor 2, as one varies $N_c$ in the range $2 \leq N_c \leq \infty$, with
the exception of $A_{CP}(\omega \pi^\pm)$ and $A_{CP}(\rho^0 \pi^\pm)$
which are much more uncertain. The $N_c$-sensitivity of ${\cal B}(B^\pm \to
\omega \pi^\pm)$ was already pointed out in \cite{akl98-1}.
\item The CP-violating asymmetries worked out here are in most cases
relatively insensitive to the scale $\mu$, i.e., this dependence is below
 $\pm 20\%$,
for $m_b/2 \leq \mu \leq m_b$,  
except in some decays which we have listed in Table 13. 

\item As opposed to the branching ratios, asymmetries do not depend in the
first approximation on the form factors and decay constants. However,
in most cases, they depend on the parameter $k^2$, the virtuality of the
$g, \gamma$ and $Z^0$ decaying into $q \bar{q}$ from the penguin 
contributions. This has been already studied in great detail in \cite{kps},
a behavior which we have also confirmed. 

\item Interestingly, we find that a number of $B \to h_1 h_2$ decays
have CP-violating asymmetries which can be predicted within a reasonable 
range in the factorization approach. They include:
$A_{CP}(\pi^+ \pi^-)$, $A_{CP}(K_S^0 \eta^\prime)$, $A_{CP}(K_S^0 \pi^0)$,
$A_{CP}(K_S^0 \eta)$ and $A_{CP}(\rho^+\rho^-)$.
 The decay modes involved have reasonably large branching ratios and the 
CP-violating asymmetries are also measurably large in all these cases.
Hence,
 their measurements can be used to put
constraints on the CKM parameters $\rho$ and $\eta$.
Likewise, these decay modes are well suited to test the hypothesis
that strong phases in these decays are generated dominantly by
perturbative QCD. This, in our opinion, is difficult to test
in class-II and class-V decays.
 Of particular interest
is $A_{CP}(K_S^0 \eta^\prime)$, which is expected to have a value
$A_{CP}(K_S^0 \eta^\prime) \simeq (20 $-$ 36)\%$. This decay mode has already
been measured by the CLEO collaboration \cite{cleo} and estimates of
its branching ratio in the factorization approach are in agreement with
data \citer{akl98-1,acgk}.

\item The CP-asymmetry $A_{CP}(K_S^0 h^0)$, where $h^0=\pi^0, K_S^0,\eta,
\eta^\prime$ is found to be remarkably stable in $k^2$, due to the 
compensation in the various channels. The resulting CP-asymmetry is found 
to be large, with $A_{CP}(K_S^0 h^0) \simeq (20$-$36)\%$, with the range
reflecting the CKM-parametric dependence.
\item We have studied the dependence of $A_{CP}(\pi^+ \pi^-)$ on $\sin 2 
\alpha$, studying
the effect of the ``penguin pollution'', which we find to be significant. 
The effect
of the ``tree-shadowing'' in $A_{CP}(K_S^0\eta^\prime)$ is, however, 
found to be
small. Thus, $A_{CP}(K_S^0\eta^\prime)$, likewise $A_{CP}(K_S^0\pi^0)$, 
$A_{CP}(K_S^0\eta)$ and $A_{CP}(K_S^0h^0)$ are good measures of $\sin 2 \beta$.
\item We have studied time-dependent CP-violating asymmetries $A_{CP}(t; 
\rho^+ \pi^-)$
and $A_{CP}(t; \rho^- \pi^+)$, working out the various characteristic 
components in the time evolution of the individual branching ratios. With the 
branching ratio averaged over the charge-conjugated modes ${\cal B}(B^0 \to 
\rho^+ \pi^-)
=(2-4) \times 10^{-5}$ and time-integrated CP-violating asymmetry $A_{CP}
(\rho^+ \pi^-) =
(4-7)\%$, for the central values $\rho=0.12$ and $\eta=0.34$, it is an 
interesting process
to measure, as stressed in \cite{adkd91}. The branching ratio ${\cal B}
(B^0 \to \rho^-
\pi^+)$ is estimated by us as typically a factor 4 below ${\cal B}(B^0 \to 
\rho^+
\pi^-)$ and hence $A_{CP}(\rho^- \pi^+)$ is a relatively more difficult 
measurement. 
\item There are several class-IV decays whose CP-asymmetries are small
but stable against variation in $N_c$, $k^2$ and $\mu$. They include:
$A_{CP}(K^\pm\eta^\prime)$, $A_{CP}(\pi^\pm K_S^0)$ and $A_{CP}(\rho^\pm 
\optbar{K^{*0}})$.
CP-asymmetries well over $5\%$ in these decay modes can arise through 
SFI and/or new physics. We argue that
the role of SFI can be disentangled already in decay rates
and through the measurements of a number of CP-violating asymmetries which
are predicted to be large. As at this stage it is hard to quantify the
effects of SFI, one can not stress too strongly that a measurement of
CP-violating asymmetry in any of these partial rates significantly above
the estimates presented here will be a sign of new physics.
\item There are quite a few other decay modes which have measurably large
CP-violating asymmetries, though without constraining the 
parameter $k^2$ experimentally, or removing this dependence in an improved
theoretical framework,
 they are at present rather uncertain. A good measurement of
the CP-asymmetry in any one of these could be used to determine $k^2$.
We list these potentially interesting asymmetries below:\\
$A_{CP}(K^{*\pm} \pi^\mp)$, $A_{CP}(K^{*\pm} \pi^0)$,
$A_{CP}(K^{*\pm} \eta)$, $A_{CP}(K^{*\pm} \eta^\prime)$, $A_{CP}(K^{*\pm} 
\rho^\mp)$ and $A_{CP}(K^{*\pm} \rho^0)$.
\end{itemize}

In conclusion, by systematically studying the $B \to h_1 h_2$ decays
in the factorization approach, we hope
that we have found classes of decays where the factorization approach can be
tested as it makes predictions within a reasonable range. If the predictions 
in the rates 
in these decays are borne out by data, then it will strengthen the notion 
based on
color transparency that non-factorization effects in decay rates are small 
and  QCD dynamics in
$B \to h_1 h_2$ decays can be largely described in terms of perturbative
QCD and factorized amplitudes. This will bring in a number of CP-violating 
asymmetries under quantitative control of the factorization-based theory.
If these expectations did not stand the experimental tests, attempts to
quantitatively study two-body
non-leptonic decays would have to wait for a fundamental step in the QCD
technology
enabling a direct computation of the four-quark matrix elements in the decays
$B \to h_1 h_2$.  However, present data on $B \to h_1 h_2$ decays are
rather encouraging and perhaps factorization approach is well poised
to becoming a useful theoretical tool in studying non-leptonic $B$ decays -
at least in class-I and class-IV decays.  
  We look forward to new experimental results where many of the predictions
presented here and in \cite{akl98-1} will be tested in terms of branching
ratios and CP-violating asymmetries in partial decay rates. 

\vspace{1.0cm}
{\bf Acknowledgments}\\
We thank Christoph Greub and Jim Smith for helpful discussions. We would
also like to thank Alexey Petrov for clarifying some points in his
paper \cite{Petrov97}. One of us (A.A.) would like to thank Matthias
Neubert for discussions on the factorization approach, and the CERN-TH
Division for hospitality where this work was completed.

\end{document}